\documentclass[english,11pt]{article}
\pdfoutput=1
\usepackage{amsmath,amsfonts,amssymb,amsbsy,amstext}
\usepackage[top=3cm, bottom=2cm, left=2cm, right=2cm]{geometry}
\usepackage{graphicx}
\usepackage{color}
\usepackage[english]{babel}
\usepackage{subfigure}
\usepackage[numbers,compress]{natbib}

\newcommand{\beq}{\begin{eqnarray}}
\newcommand{\eeq}{\end{eqnarray}}
\newcommand{\non}{\nonumber\\}

\newcommand{\p}{\partial}

\newcommand{\sech}{{\rm sech}}

\newcommand{\Or}{{\rm O}}

\begin{document}

\begin{titlepage}
\def\thefootnote{\fnsymbol{footnote}}
\phantom{.}\vspace{-2cm}
\begin{flushright}
NORDITA-2014-27
\end{flushright}

\bigskip

\begin{center}
{\Large {\bf Domain wall Skyrmions}}

\bigskip
{\large Sven Bjarke Gudnason${}^1$\footnote{\texttt{sbgu(at)kth.se}} 
and Muneto Nitta${}^2$\footnote{\texttt{nitta(at)phys-h.keio.ac.jp}}}
\end{center}

\renewcommand{\thefootnote}{\arabic{footnote}}

\begin{center}
\vspace{0em}
{\em {${}^1$Nordita, KTH Royal Institute of Technology and Stockholm University,
Roslagstullsbacken 23, SE-106 91 Stockholm, Sweden\\
${}^2$Department of Physics, and Research and Education Center for Natural
Sciences, Keio University, Hiyoshi 4-1-1, Yokohama, Kanagawa 223-8521,
Japan 

\vskip .4cm}}

\end{center}

\vspace{1.1cm}

\noindent
\begin{center} {\bf Abstract} \end{center}

Skyrmions of different dimensions are related by domain walls. We
obtain explicit full numerical solutions of various Skyrmion
configurations trapped inside a domain wall. 
We find for the quadratic mass-term that multi-Skyrmions are
ring-shaped, and conjecture for the linear mass-term, that the
lowest-energy state of multi-Skyrmions will consist of charge-2 rings
accommodated in a lattice. 

\vfill

\begin{flushleft}
{\today}
\end{flushleft}
\end{titlepage}

\hfill{}
\setcounter{footnote}{0}

\section{Introduction}

Skyrmions exist in the non-linear sigma model if it is amended by a
higher-derivative term, which in the simplest case can be a 
fourth-order derivative term \cite{Skyrme:1961vq}. 
They are topological solitons supported by the homotopy group
$\pi_3(S^3)=\mathbb{Z}$ and are believed to describe baryons in the
low-energy effective theory of quantum chromodynamics in which the
elementary fields represent mesons (pions) \cite{'tHooft:1973jz}. 
The original Skyrme model is equivalent to the O(4) non-linear sigma
model with the addition of the Skyrme term. The condition of having
finite-energy configurations is tantamount to picking a direction of
the field at infinity and thus spontaneously break O(4) down to O(3)
yielding $S^3$ which when mapped to the $3$-dimensional space with
spatial infinity identified as a point, forms the basis of the
topological nature of the solitons, i.e.~the Skyrmions. 

Skyrmions exist also in other dimensions than $3+1$; in $2+1$
dimensions the O(3) baby-Skyrmion model 
\cite{Piette:1994ug,Piette:1994mh,Kudryavtsev:1997nw,Weidig:1998ii}
is analogous to its
higher-dimensional cousin and it is described by the homotopy group 
$\pi_2(S^2)=\mathbb{Z}$ along the same lines as mentioned
above. Stepping down one dimension, formally, the sine-Gordon kink is
the $(1+1)$-dimensional Skyrmion as known already by Skyrme himself
\cite{Skyrme:1961vr}. By virtue of Derrick's theorem
\cite{Derrick:1964ww}, the $(1+1)$-dimensional soliton does not need 
higher-derivative terms in order to be stable and finite in size and
energy as is the case in higher dimensions. So one might naively think
that the sine-Gordon kink would be modified upon the presence of the
Skyrme term. However, it so happens that along one spatial direction,
the Skyrme term leaves the kink unchanged. 

Skyrmions in $d$ spatial dimensions have been shown in
refs.~\cite{Nitta:2012xq,Nitta:2012wi,Nitta:2012rq} to be related to those
in $d-1$ spatial dimensions 
via exactly a domain wall along one direction; yielding a full 
$d$-dimensional Skyrmion charge in the full $(d+1)$-dimensional
theory (see also refs.~\cite{Kudryavtsev:1997nw,Piette:1997ce}). 
In the low-energy effective theory living on the domain wall
(the kink along one of the spatial directions), the effective soliton
is a $(d-1)$-dimensional Skyrmion enjoying a $(d-1)$-dimensional
Skyrmion charge as well. 
The dynamics of the lowest dimensional case 
\cite{Nitta:2012xq} 
was studied numerically in a related model  
\cite{Jennings:2013aea}. 

In this paper we will study static configurations in $3+1$ dimensions
in which we take one spatial direction (the $x$-direction) to be
always a domain wall. Now we have two options for creating a full
$(3+1)$-dimensional Skyrmion trapped on the wall, 
which we will denote a domain wall Skyrmion. 
The first option is
to make a domain line, say along the $y$-direction and then endow it
with a sine-Gordon kink. 
The adequate potential giving rise to such configuration is of
a hierarchical form, breaking $\Or(4)\to\Or(3)\to\Or(2)$ and finally
either completely, giving a Skyrmion, or down to $\mathbb{Z}_2$
yielding a half-Skyrmion. For this to work out, we need the mass
scales of the symmetry breakings to be of the form 
$m_4\gg m_3\gg m_2$. 
This is the matryoshka construction of ref.~\cite{Nitta:2012rq} in which
three consecutive domain walls form the full Skyrmion which is doubly
trapped. 
An interesting fact about this configuration is that we do not need
higher-derivative terms for creating this configuration and hence its
applicability might be found even in condensed-matter systems. 

The second option at hand is to start again with the domain wall in
the $x$-direction and embed it with a baby-Skyrmion living on the two 
dimensional surface of the domain wall. Pictorially, we can imagine
the domain line of the previous type of configurations to be rolled
up. For this we do need the Skyrme term as well as a mass term for the
baby-Skyrmion, where the first acts as a pressure term and the latter
as an attraction term. This option also does not allow for the
half-Skyrmion which exists in the other branch of solutions. However,
the effective baby-Skyrmion model living on the domain wall can easily
host higher-charged Skyrmions which turn out to be ring-like
configurations, depending on the potential. 
A note of caution is that although the theory living on the domain
wall is effectively like that of the planar baby-Skyrmion, differences
do persist. The two main differences are that the coefficients in the
effective Lagrangian are no longer canonical and the field feels the
derivatives in the transverse direction (these two facts are related). 

We find that for the linear potential in the baby-Skyrmion theory 
induced on a domain wall, the Skyrmions of charge 1,2 and 3 are stable
while that with charge 4 is only meta-stable and will eventually decay
to two charge-2 rings. The solution with charge 5 is unstable and
decays into a bound state consisting of one 2-ring and the remaining
$B=3$ charge. For large $B$, we conjecture that the lowest-energy
solutions form lattices composed of 2-rings. 
For the quadratic potential in the baby-Skyrmion theory all
higher-charged configurations we have found are stable and ring-like
in their energy distributions. 
For hierarchical quadratic and linear potentials in the baby-Skyrmion
theory, we find sine-Gordon kinks placed equidistantly on a ring
inside the domain wall.

This paper is organized as follows. 
In section \ref{sec:skyrmemodel} we introduce the Skyrme model and the
family of potentials, in which we find domain walls with domain lines
with either full Skyrmions or half-Skyrmions in
Sec.~\ref{sec:matryoshka}. In section \ref{sec:trapped_babySkyrmions}
we study baby-Skyrmions trapped on a domain wall with both linear and
quadratic induced potentials and we conclude with 
a summary and discussion in
Sec.~\ref{sec:discussion}. 
We relegate additional technical information to the appendices, such
as the numerical method in app.~\ref{app:lattice}, the domain wall
with full and half-Skyrmions residing on domain lines, but without the
Skyrme term in app.~\ref{app:normal_kinetic} and finally an asymmetric
initial configuration relaxed down to an axially symmetric 2-ring in
app.~\ref{app:relaxation_sequence}.

\section{The Skyrme model\label{sec:skyrmemodel}}

Defining a four-vector of scalar fields 
${\mathbf n} = \{n^a(x) \} 
= \{n^1(x),n^2(x),n^3(x),n^4(x)\}$ ($a=1,2,3,4$) 
subject to the constraint ${\mathbf n}^2=1$,
we consider the O(4) sigma model augmented by the Skyrme term 
\cite{Skyrme:1961vq}
\begin{align}
\mathcal{L} &= 
\frac{1}{2}\p_\mu\mathbf{n}\cdot\p^\mu\mathbf{n}
+\frac{1}{4}\left(\p_\mu\mathbf{n}\cdot\p_\nu\mathbf{n}\right)
\left(\p^\mu\mathbf{n}\cdot\p^\nu\mathbf{n}\right)
-\frac{1}{4}\left(\p_\mu\mathbf{n}\cdot\p^\mu\mathbf{n}\right)^2 
- V(\mathbf{n}) \, , \label{eq:LO4}
\end{align}
and a potential which we will choose according to the specific breed
of matryoshka configuration in sight. 
The static equation of motion is given by
\begin{align}
\p^2 n^a 
+ (\p_i\p_j\mathbf{n}\cdot\p_j\mathbf{n})\p_i n^a
- (\p^2\mathbf{n}\cdot\p_i\mathbf{n})\p_i n^a
+ (\p_i\mathbf{n}\cdot\p_i\mathbf{n})\p^2 n^a
- (\p_i\mathbf{n}\cdot\p_j\mathbf{n})\p_i\p_j n^a
- \frac{\delta V}{\delta n^a} = 0 \, , \label{eq:eom}
\end{align}
where $i,j=1,2,3$ are spatial indices and $a=1,2,3,4$ is the vector
index. 
The Skyrmion or baryon number is given by the number of 3-cycles of
the 3-sphere and reads 
\beq
B = -\frac{1}{12\pi^2}\int d^3x\;
\epsilon_{i j k}\epsilon^{a b c d} \p_i n^a \p_j n^b \p_k n^c n^d \, . 
\eeq
The family of potentials we study is a subset of
\beq
V = -\frac{1}{2}m_2^2 n_2^{a_2}
- \frac{1}{2}m_3^2 n_3^{a_3}
+ \frac{1}{2}m_4^2 (1-n_4^2) \, ,
\eeq
where the powers $a_{2,3}$ can be either 1 or 2. The squared component
of the field allows for a domain wall configuration as both 
$n_4=\pm 1$ are vacua \footnote{In the middle of the $n_4$-domain
  wall, $n_4\approx 0$ which allows for e.g.~the domain line due to the
  ``vacua'' $n_3=\pm 1$ and analogously for $n_2$.}. The linear
potential has no domain wall, but breaks $\mathbb{Z}_2$ allowing for a
full trapped baryon, as we will show later. 

\section{The Matryoshka-wall configurations\label{sec:matryoshka}}

\subsection{The wall}

The simplest and exact solution on which we in the remainder of the
paper will trap a series of configurations is the domain wall. If we
set $\mathbf{n}=\{0,0,\sin f(x),\cos f(x)\}$ and choose the potential 
\beq
V = \frac{1}{2}m_4^2(1 - n_4^2) \, , 
\eeq
yielding
\beq
\mathcal{L} = - \frac{1}{2}(\p_x f)^2 - \frac{1}{2}m_4^2\sin^2 f \, ,
\eeq
which breaks the O(4)-symmetry down to O(3). 
This Lagrangian density gives rise to the equation of motion known as
the sine-Gordon equation and it admits the following exact domain wall
solutions 
\beq
f = 2\tan^{-1}\exp(\pm m_4 x) \, ,
\eeq
or alternatively in the original variables
\beq
n_4 = \mp\tanh(m_4 x) \, , \qquad
n_3 = \sech(m_4 x) \, .
\eeq
This configuration is just a single point on the moduli space, $S^2$,
which we can parametrize as follows
\beq
\mathbf{n} = \{a^1\sin f,a^2\sin f,a^3\sin f,\cos f\} \, ,
\eeq
where $\mathbf{a}^2=1$ and any constant vector $\mathbf{a}$ yields
energetically a wall as the one described above. 

All the different configurations will be trapped objects on this wall, 
by introducing potentials and/or twisting of the moduli $\mathbf{a}$; 
this will however induce some backreaction on the wall. 
In order to keep the host soliton (the wall) alive we need to choose a
hierarchical order of the masses in the potential at hand: 
$m_4\gg m_3\gg m_2$. 
From an energetic point of view, it corresponds to a cascading symmetry
breaking like for instance $\Or(4)\to\Or(3)\to\Or(2)\to\mathbf{1}$,
depending on the specifics of the potential.
Next, we will study a number of different entrapped configurations in
turn. 

\subsection{The wall with a trapped Skyrmion on a domain line}

The simplest entrapment we can deploy is a domain line on the domain
wall. This can be illustrated by choosing 
$\mathbf{n}=\{0,\sin g(y)\sin f(x),\cos g(y)\sin f(x),\cos f(x)\}$
together with the potential 
\beq
V = -\frac{1}{2}m_3^2 n_3^2
+ \frac{1}{2}m_4^2 (1-n_4^2) \, ,
\eeq
which breaks O(4) down to O(3) by means of the wall and in turn down
to O(2) due to the existence of the domain line. The O(2) symmetry
resembles the rotational degree of freedom of the domain line inside
the wall. Plugging into the Lagrangian \eqref{eq:LO4} we obtain
\beq
-\mathcal{L} = \frac{1}{2}f_x^2 
+ \frac{1}{2}(m_4^2 - m_3^2)\sin^2 f
+ \frac{1}{2}\sin^2 f\left[g_y^2 + m_3^2\sin^2 g + f_x^2 g_y^2\right]
\, , 
\eeq
which teaches us two lessons; both $f$ and $g$ are roughly described
by kinks and the effective mass-squared for the $f$-kink is
$(m_4^2-m_3^2)$, which is positive only if we enforce the hierarchical
ordering of mass scales as mentioned above. 
Finally, there is no need for the Skyrme term in this configuration as
its only impact lies in the derivative coupling being the last term in
the bracket. 
The real solution is not simply the kink in $f$ and $g$, as the host
wall receives a backreaction from the presence of the domain line, as
we have mentioned already and $g$ is also a function of $x$ contrary
to what we have shown here for simplicity. Let us stress that in our
numerical solutions we do not impose any factorizing Ans\"atze.

We will solve the system numerically, however before doing so we will
take the next step and inhabit the domain line with a kink. If we
break the O(2)-symmetry completely by using a linear potential for
$n_2$, then the configuration exhibits (a full) baryon charge.
For illustrative purposes, let us consider 
\beq
\mathbf{n}=\{\sin h(z)\sin g(y)\sin f(x),\cos h(z)\sin g(y)\sin
f(x),\cos g(y)\sin f(x),\cos f(x)\} \, , \label{eq:ntriplewall}
\eeq
along with the potential
\beq
V = -\frac{1}{2}m_2^2 n_2
- \frac{1}{2}m_3^2 n_3^2
+ \frac{1}{2}m_4^2 (1-n_4^2) \, . 
\label{eq:VfullSkyrmion}
\eeq
Plugging into the Lagrangian \eqref{eq:LO4} sums up to
\begin{align}
-\mathcal{L} &= \frac{1}{2}f_x^2 
+ \frac{1}{2}(m_4^2 - m_3^2)\sin^2 f
+ \frac{1}{2}\sin^2 f\left[g_y^2 + m_3^2\sin^2 g + f_x^2 g_y^2\right] \\
&\phantom{=\ }
+ \frac{1}{2}\sin^2 f\sin^2 g\left[h_z^2 + f_x^2h_z^2 
  + \sin^2(f) g_y^2 h_y^2\right] 
- \frac{1}{2}m_2^2 \sin f\sin g\cos h \, , \nonumber
\end{align}
from which we can again identify the three kink components of the
system, albeit the terms for the $h$-kink are slightly deformed with
respect to those of the $g$-kink and finally it also enjoys more
mixing terms.
The reason for breaking the symmetry completely (as opposed to leaving
a $\mathbb{Z}_2$ unbroken) is that it forces the
configuration to wind $2\pi$ in the $h$ field yielding a full 3-cycle
on the target space $S^3$ and in turn baryon (Skyrmion) charge unity. 

\begin{figure}[!hpt]
\begin{center}
\mbox{\subfigure[isosurfaces]{\includegraphics[width=0.3\linewidth]{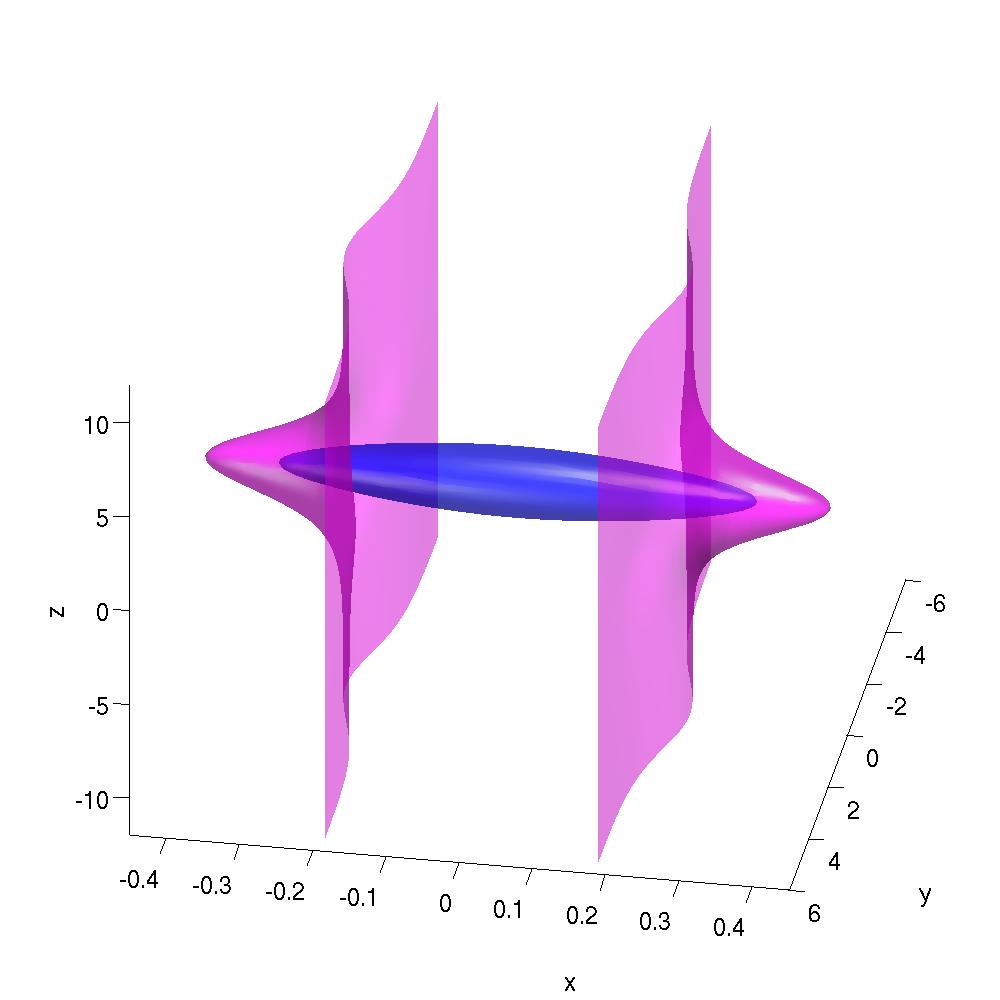}}
\subfigure[energy density]{\includegraphics[width=0.3\linewidth]{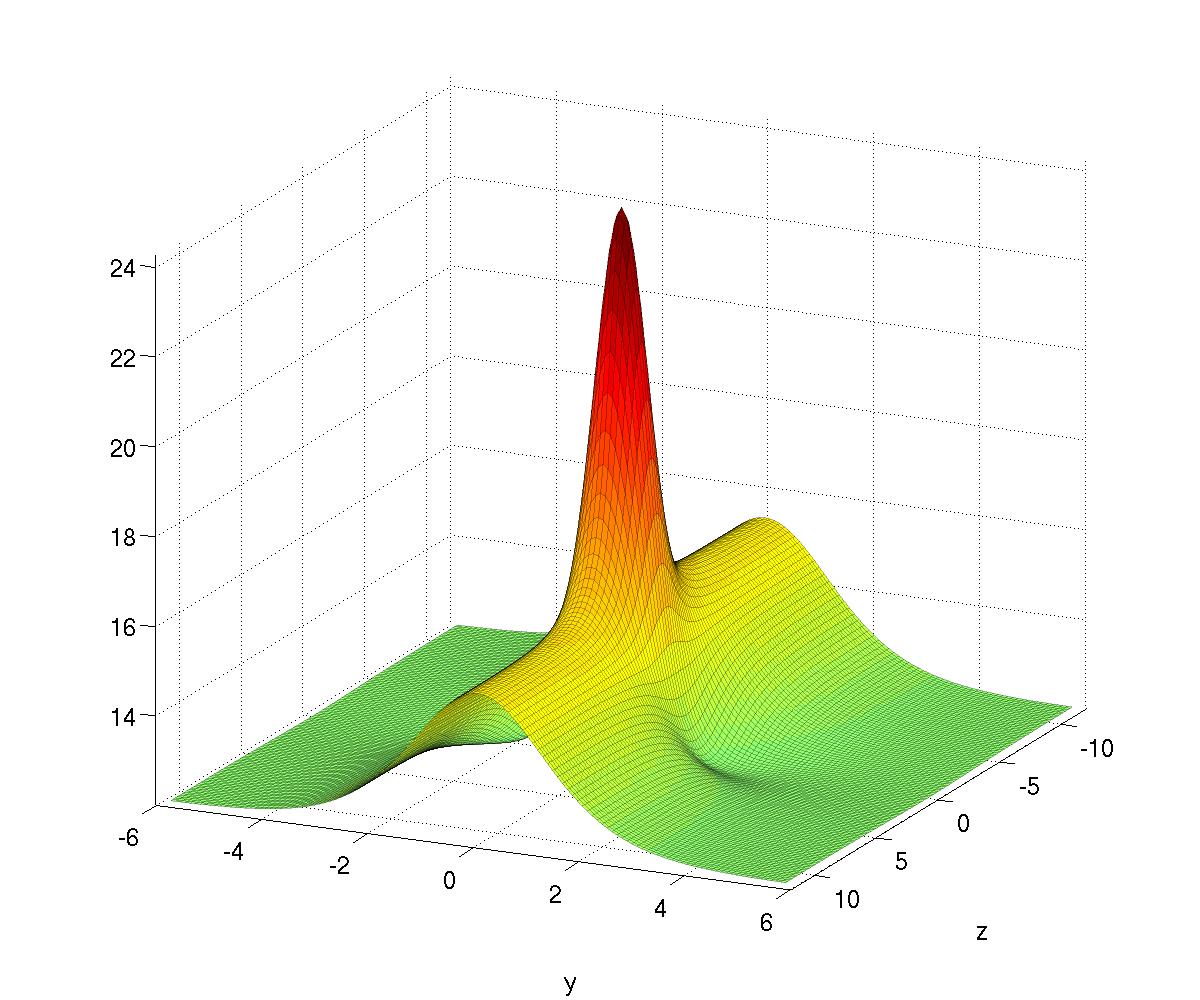}}
\subfigure[baryon charge density]{\includegraphics[width=0.3\linewidth]{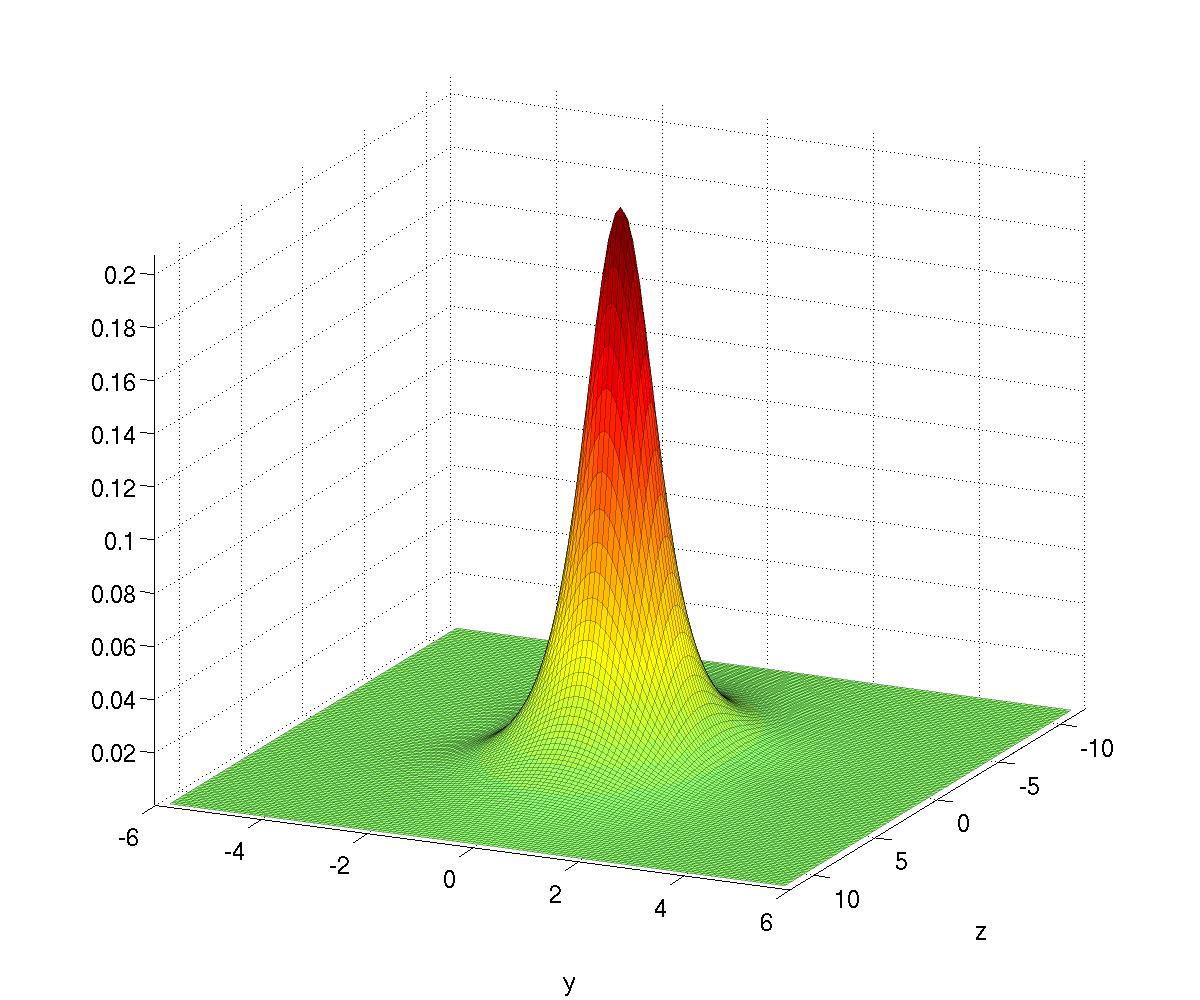}}
}
\caption{(a) 3D view of isosurfaces for the domain wall with a domain
  line on which a Skyrmion resides; the magenta surfaces represent
  the energy isosurfaces at a third of the maximum of the energy and
  the blue surface at the center shows the baryon charge isosurface,
  at half its maximum value. (b) and (c) show respectively the energy
  density and baryon charge density at a $yz$-slice in the middle of
  the domain wall (at $x=0$). The calculation is done on a $129^3$
  cubic lattice, $B^{\rm numerical}=0.992$ and the potential used is
  \eqref{eq:VfullSkyrmion} with $m_4=4,m_3=2,m_2=1$. } 
\label{fig:DW_DL_S}
\end{center}
\end{figure}

We are now ready to find numerically the solutions that we have just 
described. We will not impose any such factorization in the numerical
calculation; that was only for illustrative purposes in the above to
get a feeling for what is going on. The field used is thus
$\mathbf{n}(x,y,z)$ subject to Dirichlet boundary conditions in the
$x$-direction
\beq
\mathbf{n}(\mp\infty,y,z) = \{0,0,0,\pm 1\} \, ,
\label{eq:DBCx}
\eeq
and Neumann boundary conditions on the remaining sides of the cube. 
The equations of motion that we will solve throughout the paper are
those of eq.~\eqref{eq:eom}.
Feeding an appropriate initial configuration as a guess and relaxing
the system numerically on a $129^3$ cubic lattice as described
in more detail in app.~\ref{app:lattice}, we obtain the configuration
shown in fig.~\ref{fig:DW_DL_S}. Notice that the energy density in the
center slice of the $x$-domain wall, shown in fig.~\ref{fig:DW_DL_S}b
has a valley on the domain line on both sides of the Skyrmion
peak. This is a manifestation of the solution not being an effective
theory or factorized function, but a genuinely full-backreacted
solution to the equations of motion.

\subsection{The wall with a trapped half-Skyrmion on a domain line}

With respect to the configuration in the previous (sub)section, we
will now change the potential to
\beq
V = -\frac{1}{2}m_2^2 n_2^2
- \frac{1}{2}m_3^2 n_3^2
+ \frac{1}{2}m_4^2 (1-n_4^2) \, , \label{eq:VhalfSkyrmion}
\eeq
i.e.~the $n_2$-component goes from linear to a quadratic
potential. In this case, we are not forced to make a full winding of
the kink on the domain line, i.e.~in terms of the illustrative
vector \eqref{eq:ntriplewall}, $h$ needs only to wind from 0 to
$\pi$. This in turn halves the baryon charge of the configuration,
leaving us with a ``half Skyrmion'' trapped on the domain line
residing in the wall. Plugging the Ansatz and the above potential into
\eqref{eq:LO4}, we obtain 
\begin{align}
-\mathcal{L} &= \frac{1}{2}f_x^2 
+ \frac{1}{2}(m_4^2 - m_3^2)\sin^2 f
+ \frac{1}{2}\sin^2 f\left[g_y^2 + (m_3^2-m_2^2)\sin^2 g + f_x^2
  g_y^2\right]\non
&\phantom{=\ }
+ \frac{1}{2}\sin^2 f\sin^2 g\left[h_z^2 + m_2^2\sin^2 h + f_x^2h_z^2
  + \sin^2(f) g_y^2 h_y^2\right] \, , 
\end{align}
where we can identify three consecutive domain walls inside each other
with effective mass-squareds $(m_4^2-m_3^2)$, $(m_3^2-m_2^2)$ and
$m_2^2$ for $f$, $g$ and $h$, respectively.

\begin{figure}[!hpt]
\begin{center}
\mbox{\subfigure[isosurfaces]{\includegraphics[width=0.3\linewidth]{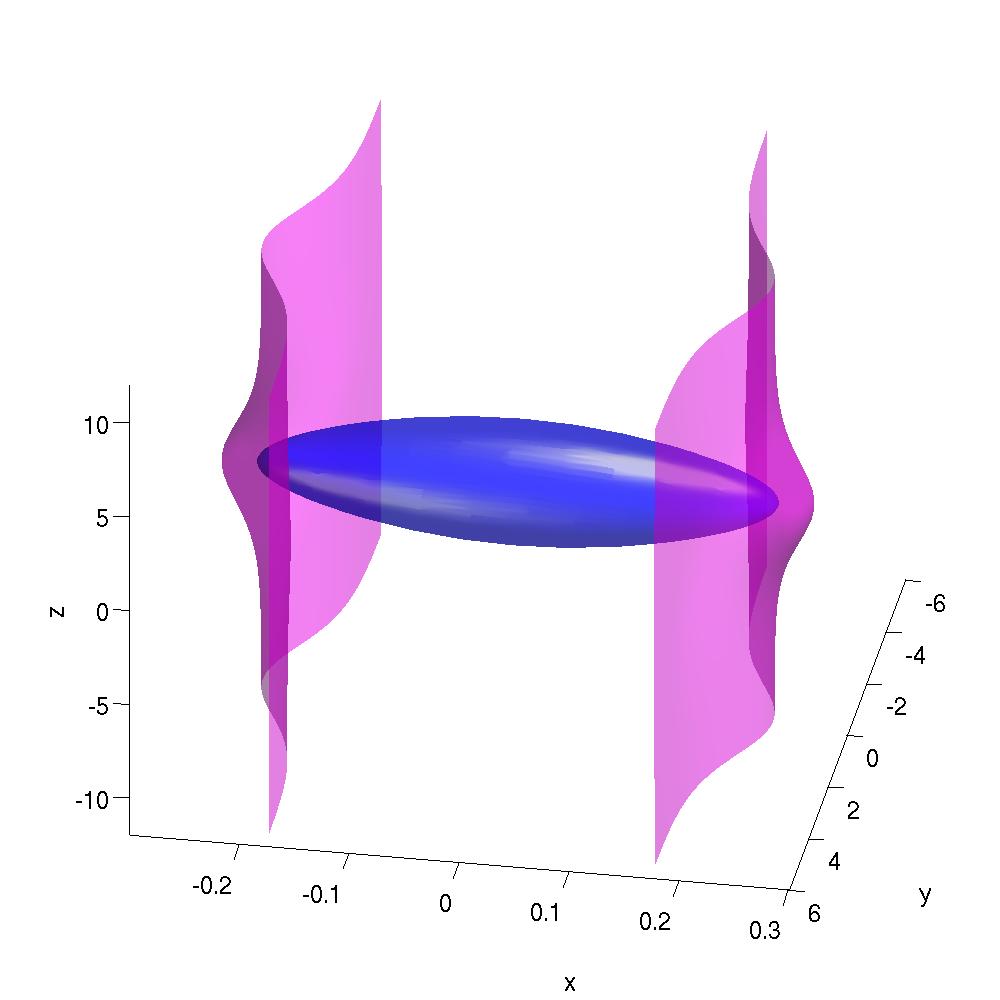}}
\subfigure[energy density]{\includegraphics[width=0.3\linewidth]{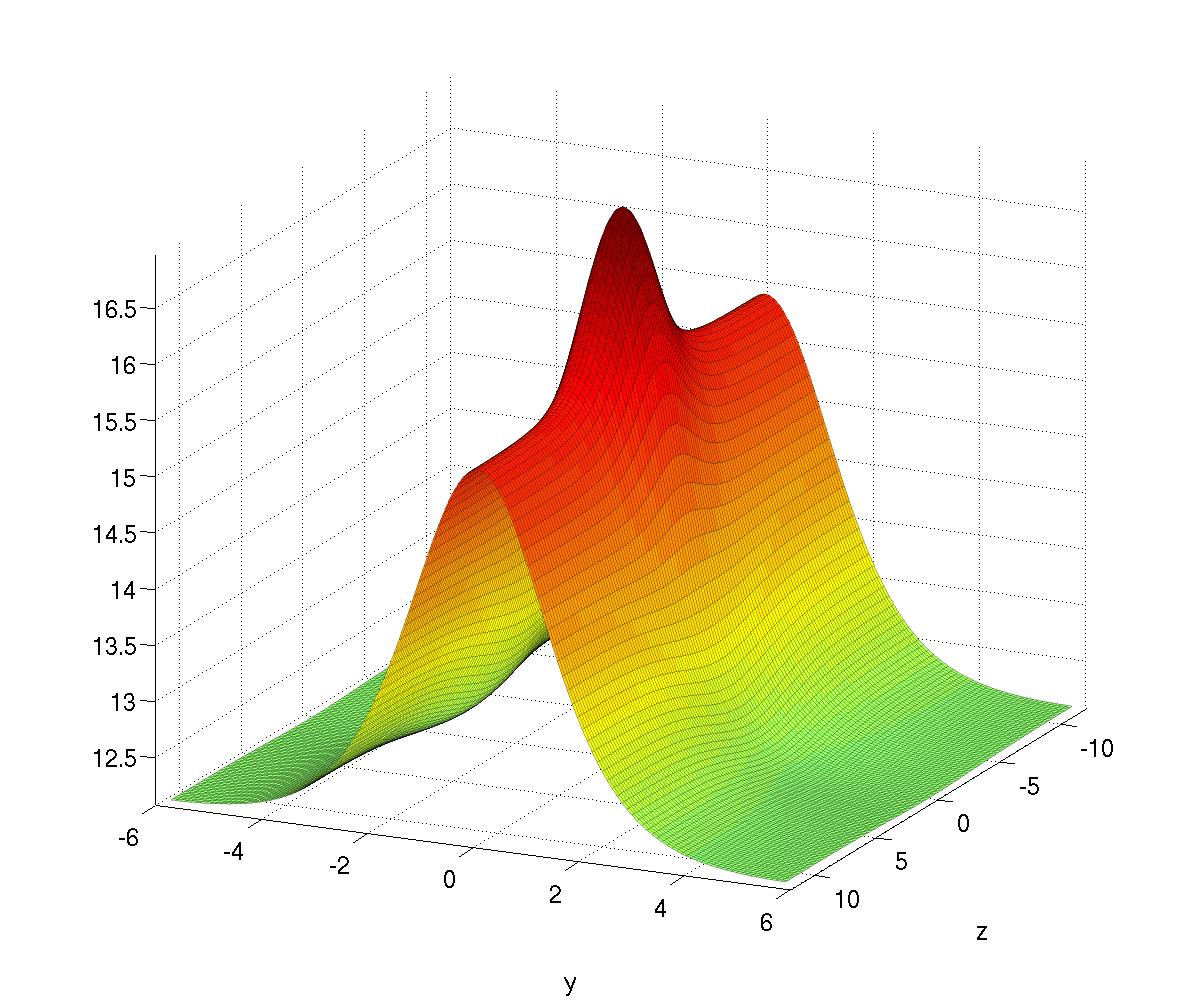}}
\subfigure[baryon charge density]{\includegraphics[width=0.3\linewidth]{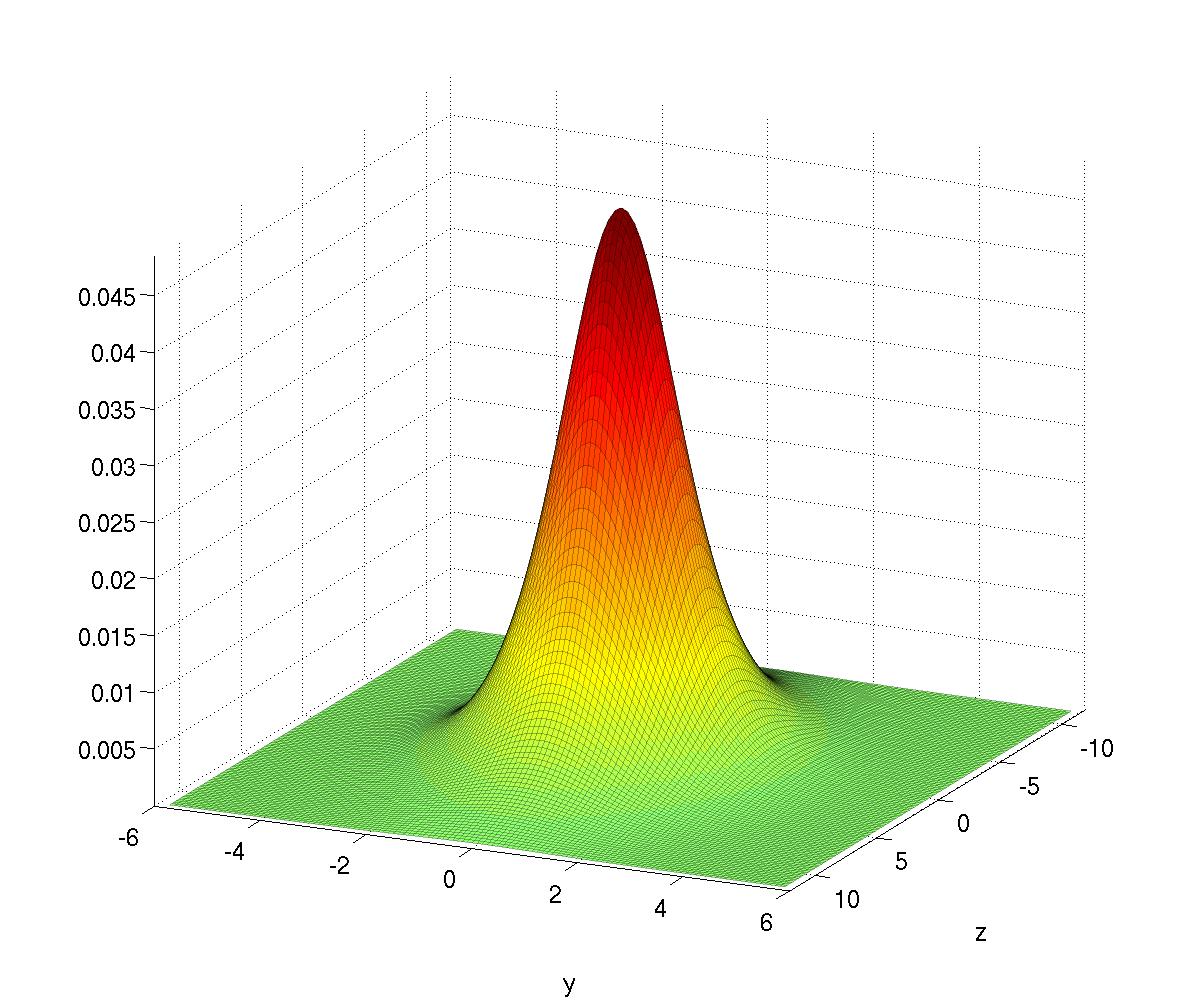}}
}
\caption{The domain wall with a domain line on which half a Skyrmion
  resides. For details see fig.~\ref{fig:DW_DL_S}. The calculation is
  done on a $129^3$ cubic lattice, $B^{\rm numerical}=0.492$ and the
  potential used is \eqref{eq:VhalfSkyrmion} with
  $m_4=4,m_3=2,m_2=1$. } 
\label{fig:DW_DL_HS}
\end{center}
\end{figure}

We now solve this system numerically and again the field used is
simply $\mathbf{n}(x,y,z)$ subject to Dirichlet boundary conditions
\eqref{eq:DBCx} and Neumann boundary conditions on the remaining sides
of the cube, yielding the configuration shown in
fig.~\ref{fig:DW_DL_HS}. 
Notice again that there is a small valley in the energy density near
the peak of the half-Skyrmion (with respect to that in
fig.~\ref{fig:DW_DL_S}), see fig.~\ref{fig:DW_DL_HS}b. 

\section{The walls with trapped baby-Skyrmion configurations\label{sec:trapped_babySkyrmions}}

\subsection{Trapped Skyrmions in a linear potential}

The next configuration we consider is a baby-Skyrmion living on the
domain wall in the $x$-direction, which due to the ``winding'' of the
wall enjoys a full Skyrmion charge. 
The baby-Skyrmion needs a mass term for compensating the pressure of
the 2-dimensional Skyrme term (as the kinetic term is marginal in
$2+1$ dimensions). 
For illustrative purposes, we again choose a simplifying Ansatz,
i.e.~we take to be 
\beq
\mathbf{n} = \left\{\sin f(x)\frac{y}{\rho}\sin g(\rho),\sin
f(x)\frac{z}{\rho}\sin g(\rho),\sin f(x)\cos g(\rho),\cos f(x)\right\}
\, , \label{eq:ansatz_baby_sk}
\eeq
where $\rho\equiv\sqrt{y^2+z^2}$.
Choosing the potential
\beq
V = -\frac{1}{2} m_3^2 n_3
+ \frac{1}{2} m_4^2 (1-n_4^2) \, , 
\label{eq:Vlinear_babysk}
\eeq
and plugging the above Ansatz and potential into the Lagrangian
\eqref{eq:LO4} we get
\begin{align}
-\mathcal{L} = \frac{1}{2} f_x^2 
+ \frac{1}{2} m_4^2 \sin^2 f
+ \frac{1}{2}\sin^2 f(1+f_x^2)\left[g_\rho^2 + \frac{1}{\rho^2}\sin^2
  g\right] 
+ \frac{1}{2\rho^2}\sin^4 f\sin^2(g) g_\rho^2
- \frac{1}{2}m_3^2\cos g\sin f \, .
\end{align}
It is easy to recognize the wall in $f$ spanned over the
$x$-direction. Notice however, that by setting $\sin f=1$ (which
corresponds to being in the middle of the domain wall), this effective
Lagrangian reduces to the baby-Skyrme model (for the baby-Skyrmion)
with the one exception that the standard kinetic term is enhanced by a
factor of $(1+f_x^2)$. The kinetic term for $f$ on the wall is in turn
proportional to $m_4^2$. 

\begin{figure}[!hpt]
\begin{center}
\mbox{\subfigure[isosurfaces]{\includegraphics[width=0.3\linewidth]{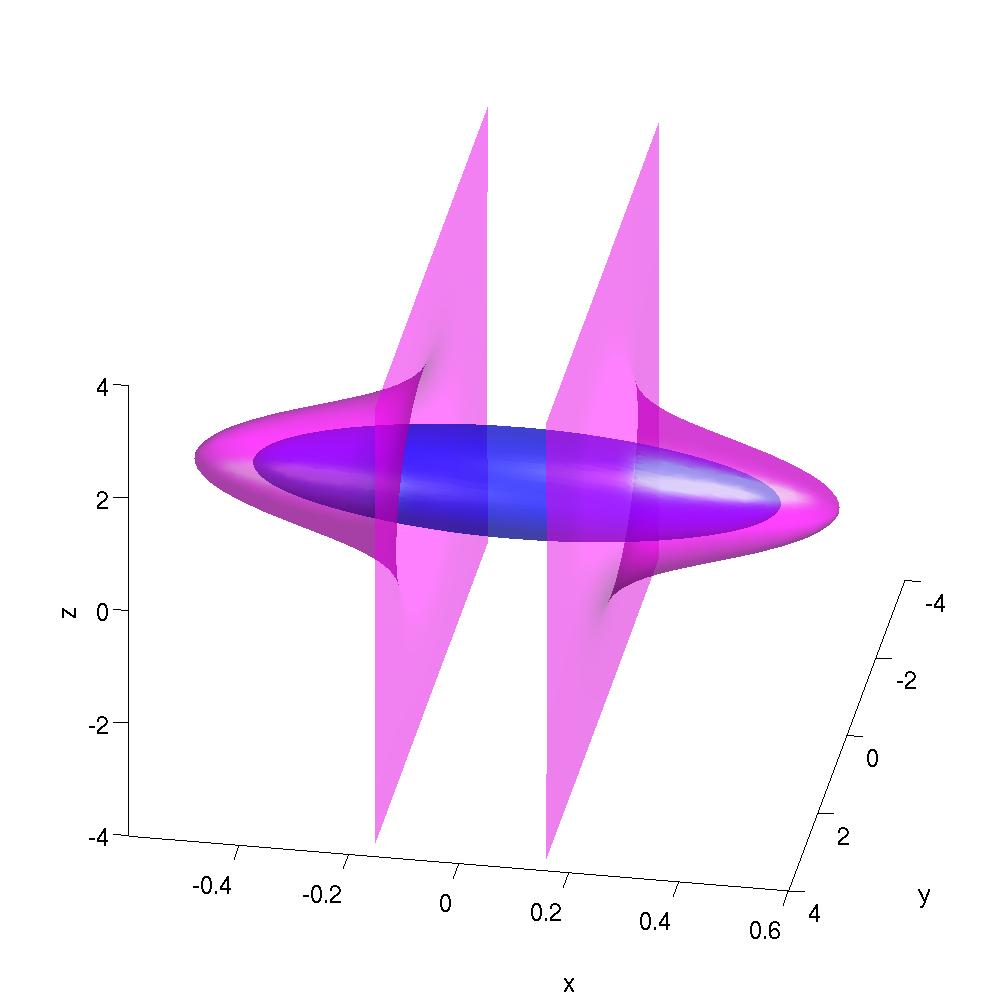}}
\subfigure[energy density]{\includegraphics[width=0.3\linewidth]{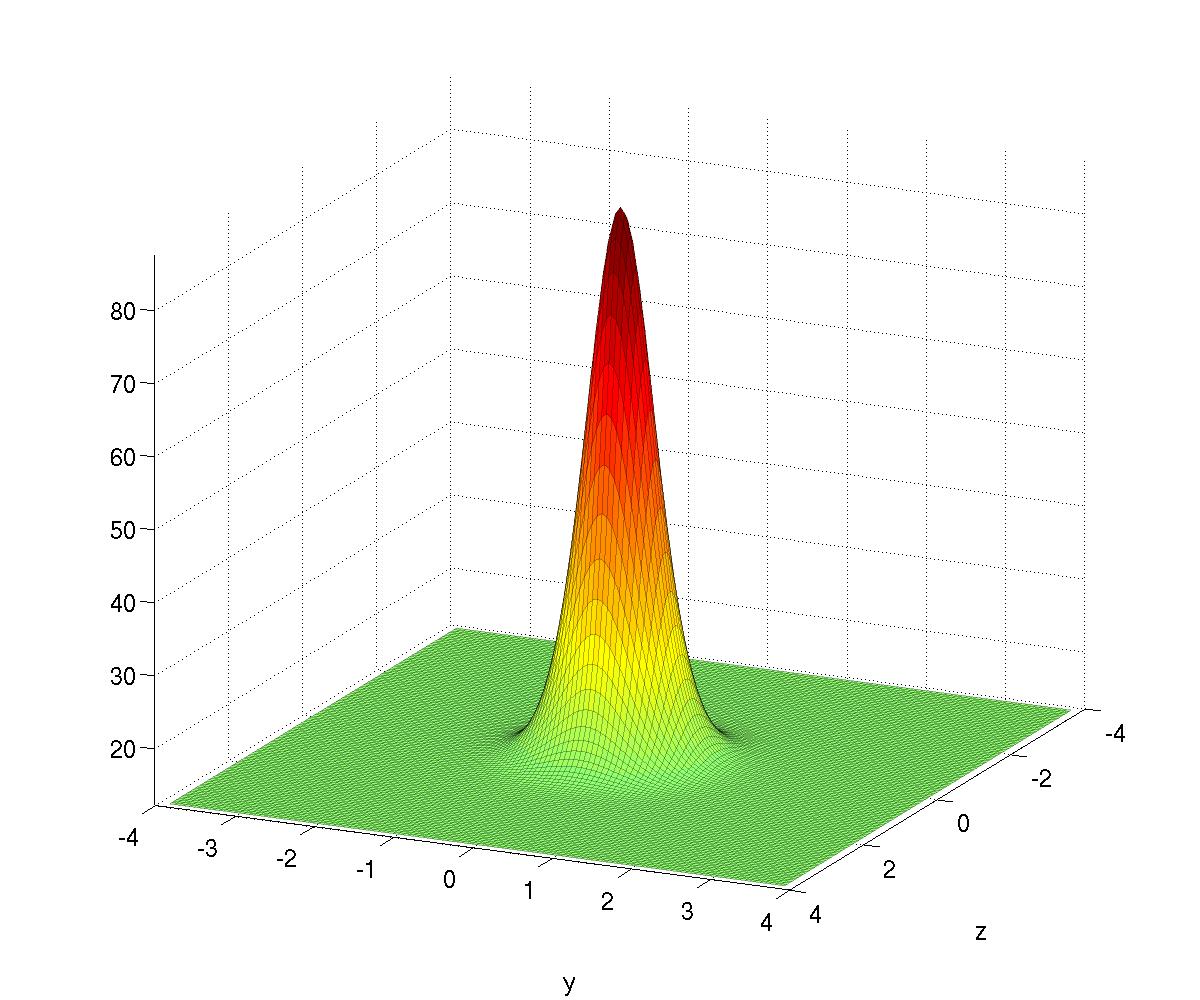}}
\subfigure[baryon charge density]{\includegraphics[width=0.3\linewidth]{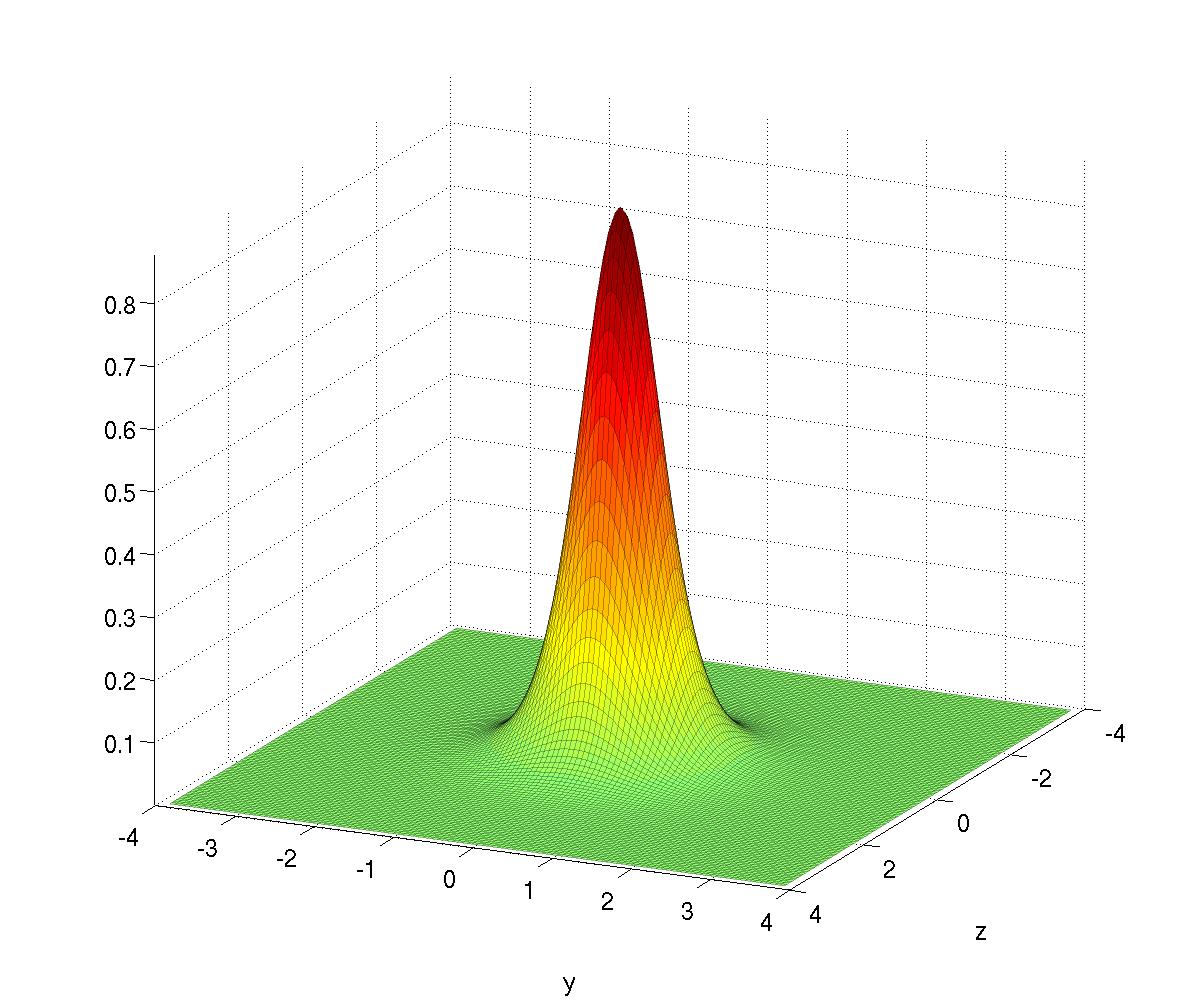}}
}
\caption{The domain wall with a trapped Skyrmion. For details see
  fig.~\ref{fig:DW_DL_S}. The calculation is done on a $129^3$ cubic
  lattice, $B^{\rm numerical}=0.998$ and the potential used is
  \eqref{eq:Vlinear_babysk} with $m_4=4,m_3=2$. }
\label{fig:DW_SL}
\end{center}
\end{figure}

Solving this system numerically with the generic field
$\mathbf{n}(x,y,z)$ subject to Dirichlet boundary conditions
\eqref{eq:DBCx} and Neumann boundary conditions on the remaining sides
of the cube, we obtain the configuration shown in
fig.~\ref{fig:DW_SL}.

\begin{figure}[!p]
\begin{flushleft}{\small $B^{\rm numerical}=1.997$:}\end{flushleft}
\begin{center}
\mbox{\subfigure{\includegraphics[width=0.3\linewidth]{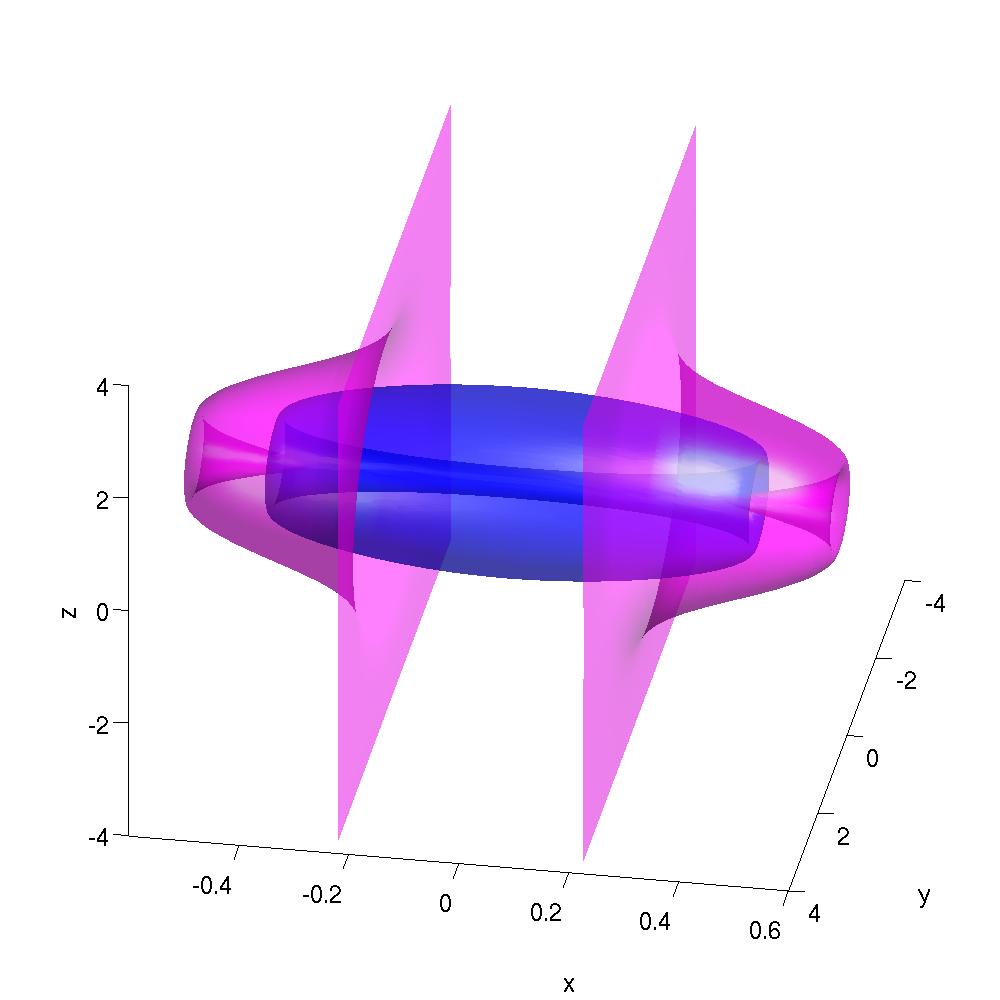}}
\subfigure{\includegraphics[width=0.3\linewidth]{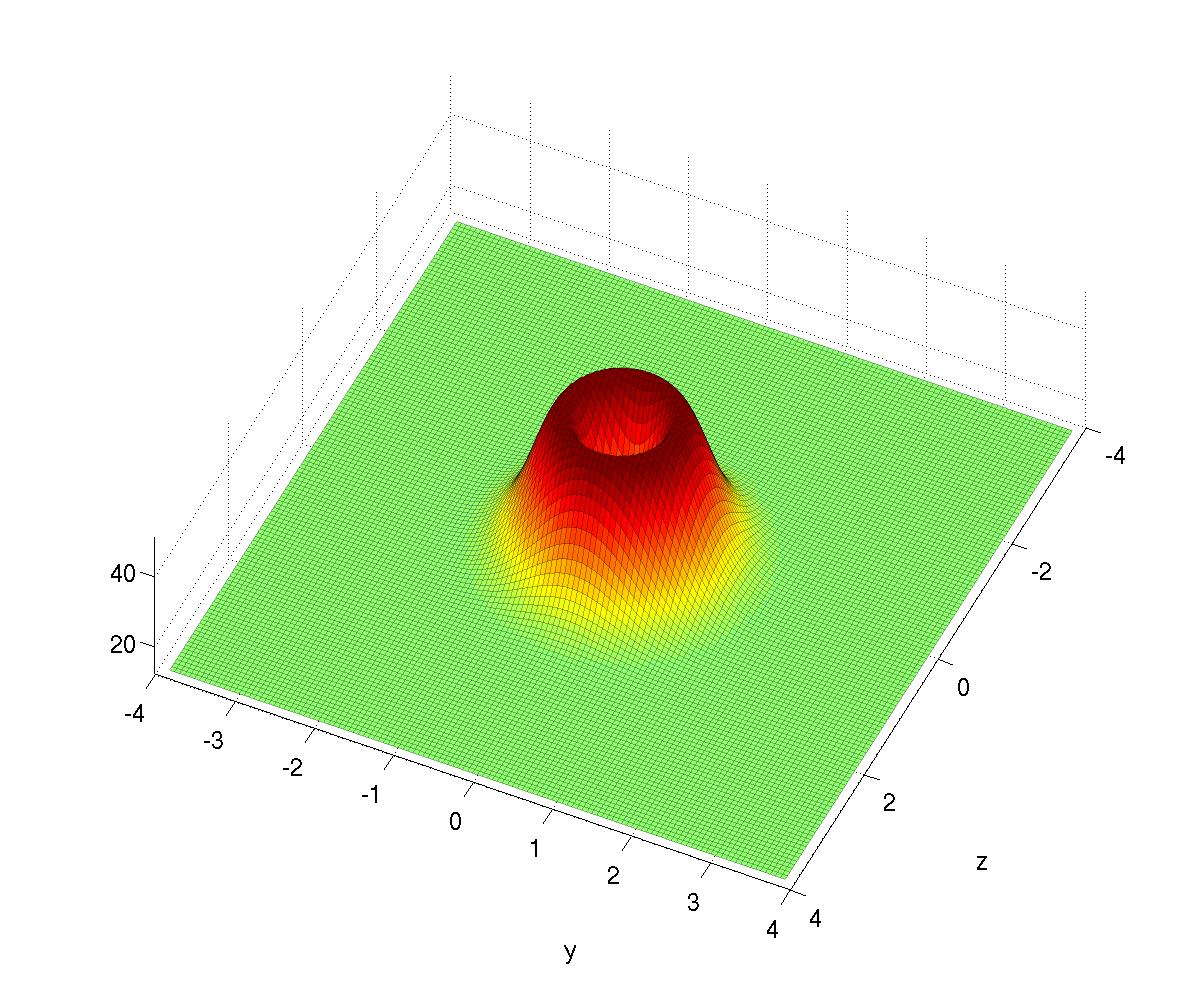}}
\subfigure{\includegraphics[width=0.3\linewidth]{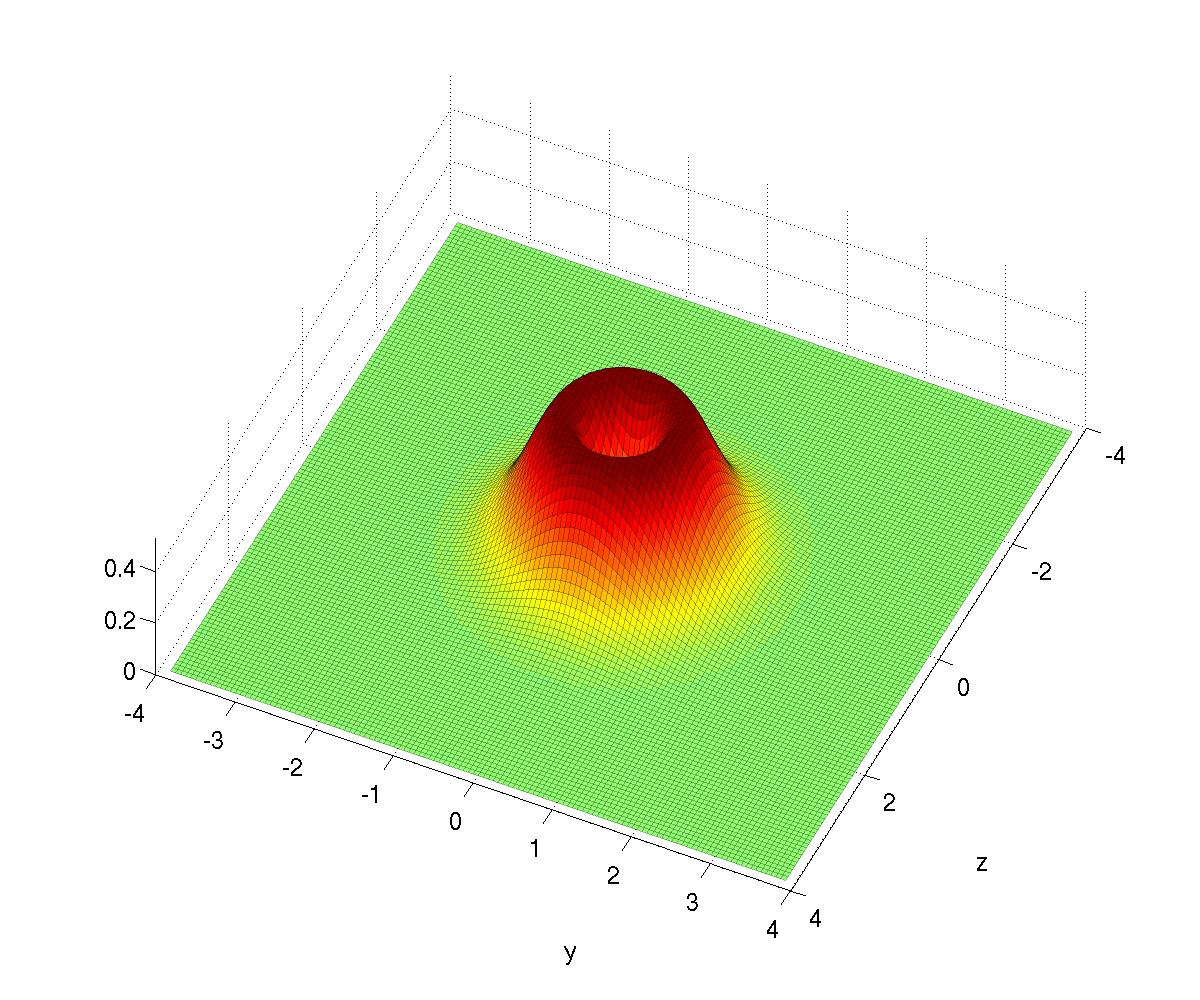}}
}
\end{center}
\begin{flushleft}{\small $B^{\rm numerical}=2.996$:}\end{flushleft}
\begin{center}
\mbox{\subfigure{\includegraphics[width=0.3\linewidth]{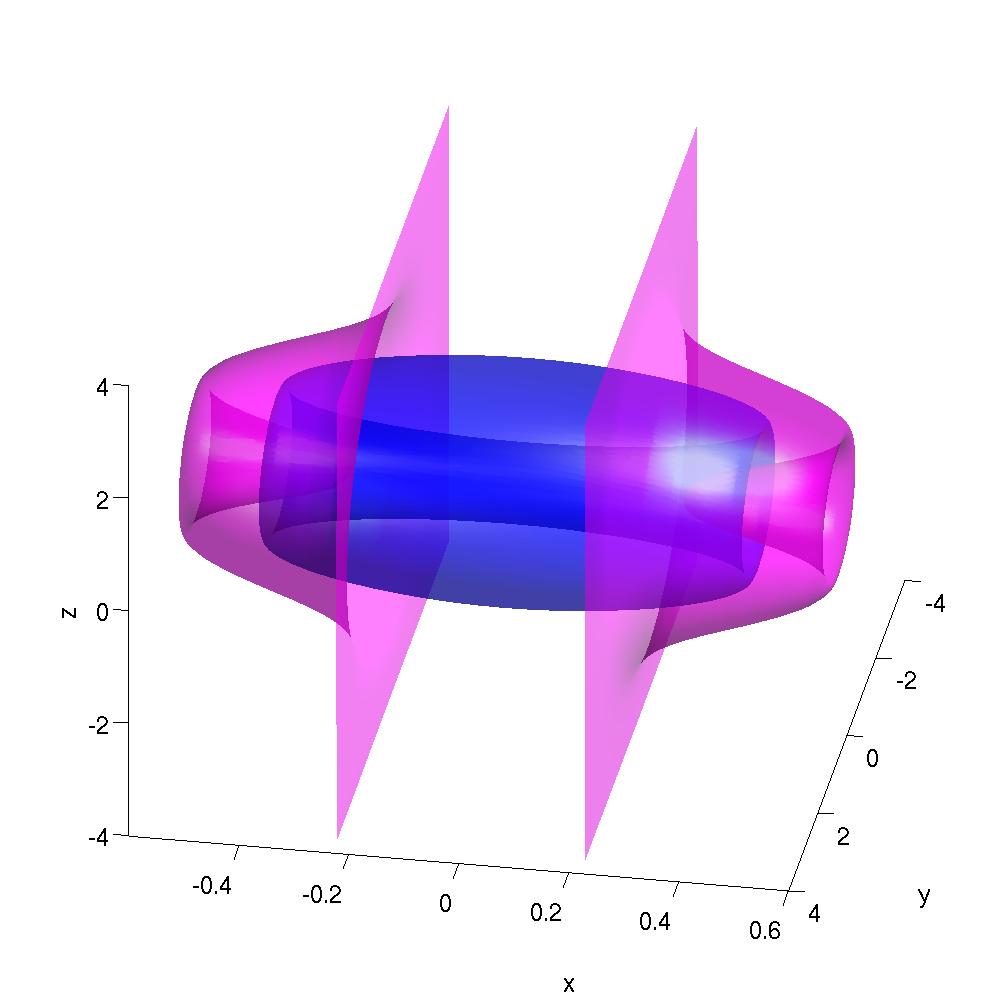}}
\subfigure{\includegraphics[width=0.3\linewidth]{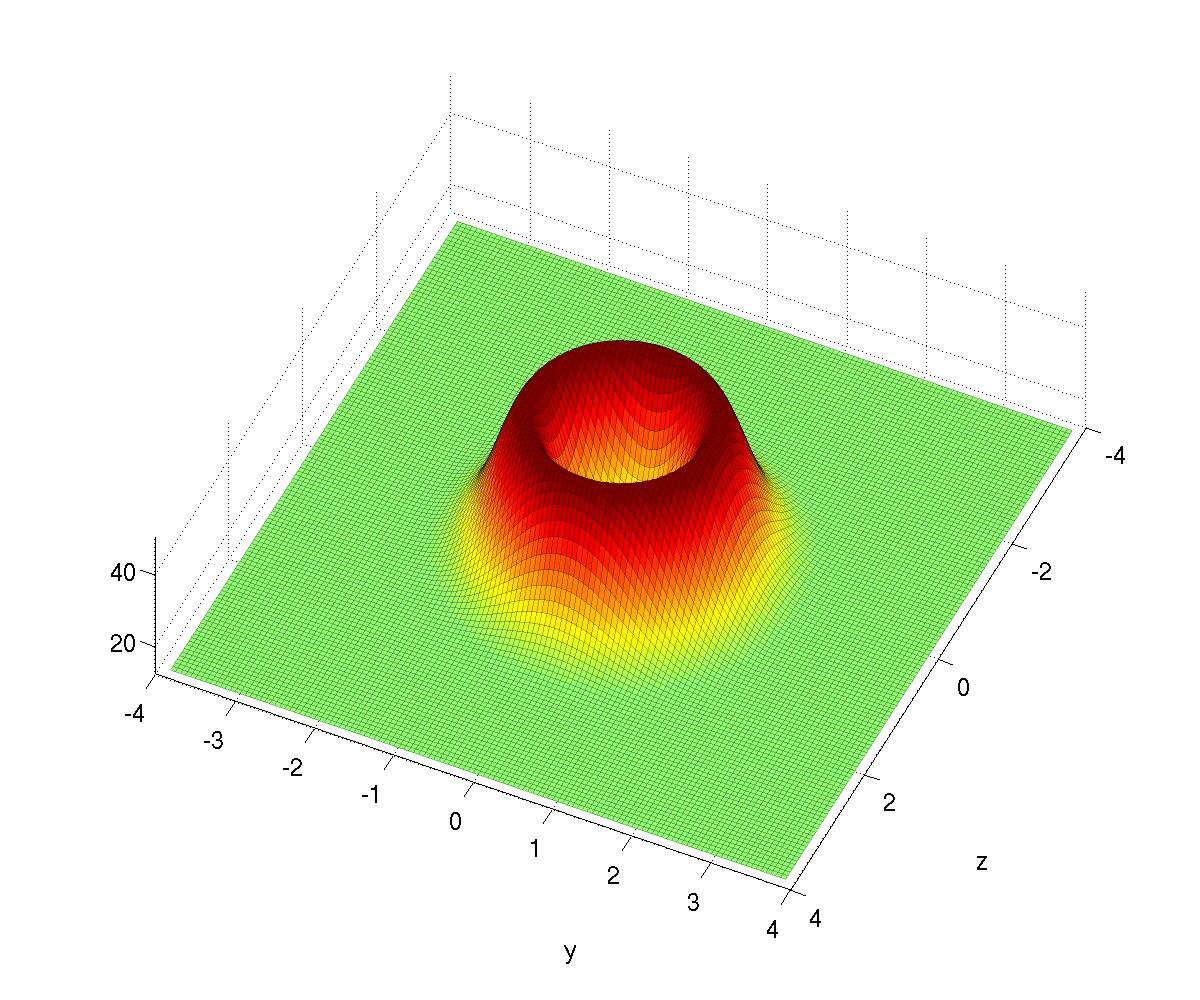}}
\subfigure{\includegraphics[width=0.3\linewidth]{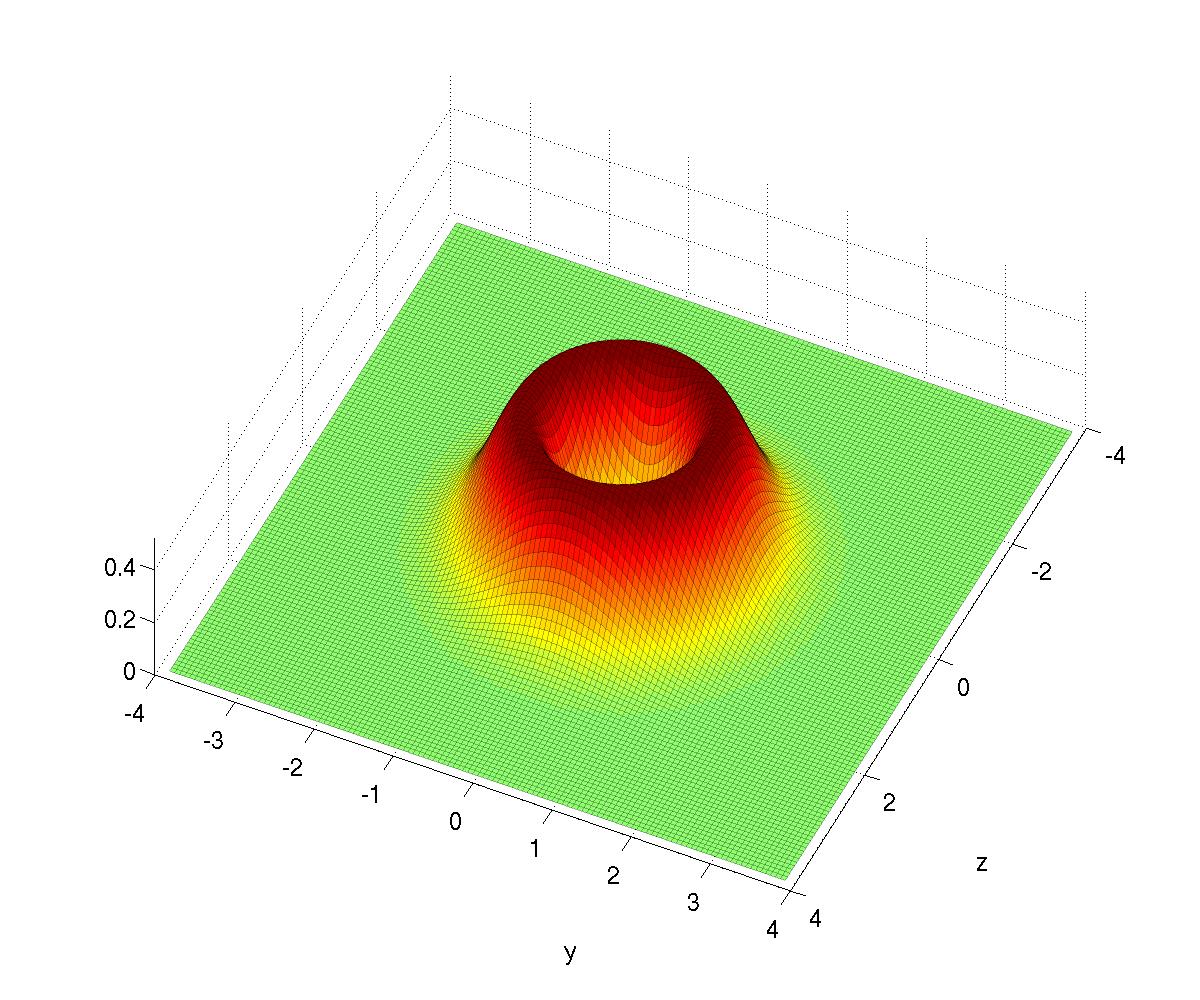}}
}
\end{center}
\begin{flushleft}{\small $B^{\rm numerical}=3.995$:}\end{flushleft}
\begin{center}
\mbox{\subfigure[isosurfaces]{\includegraphics[width=0.3\linewidth]{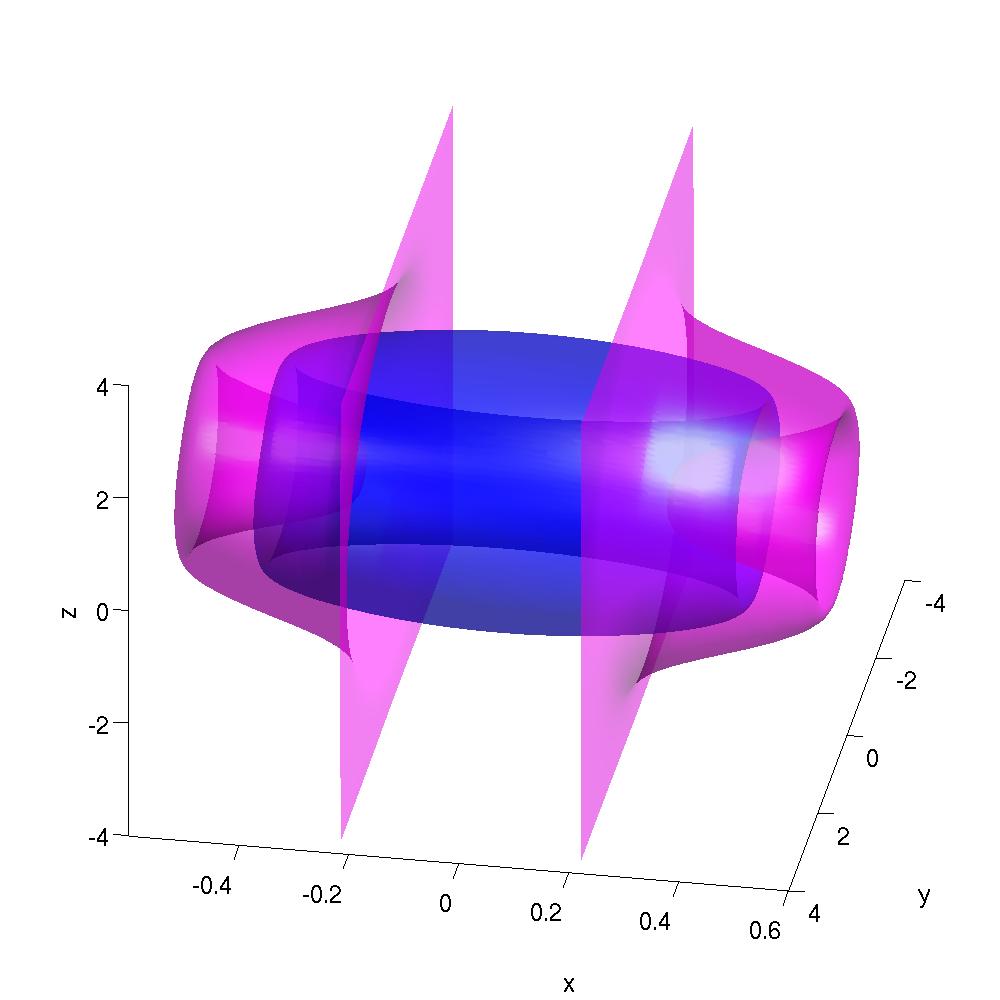}}
\subfigure[energy density]{\includegraphics[width=0.3\linewidth]{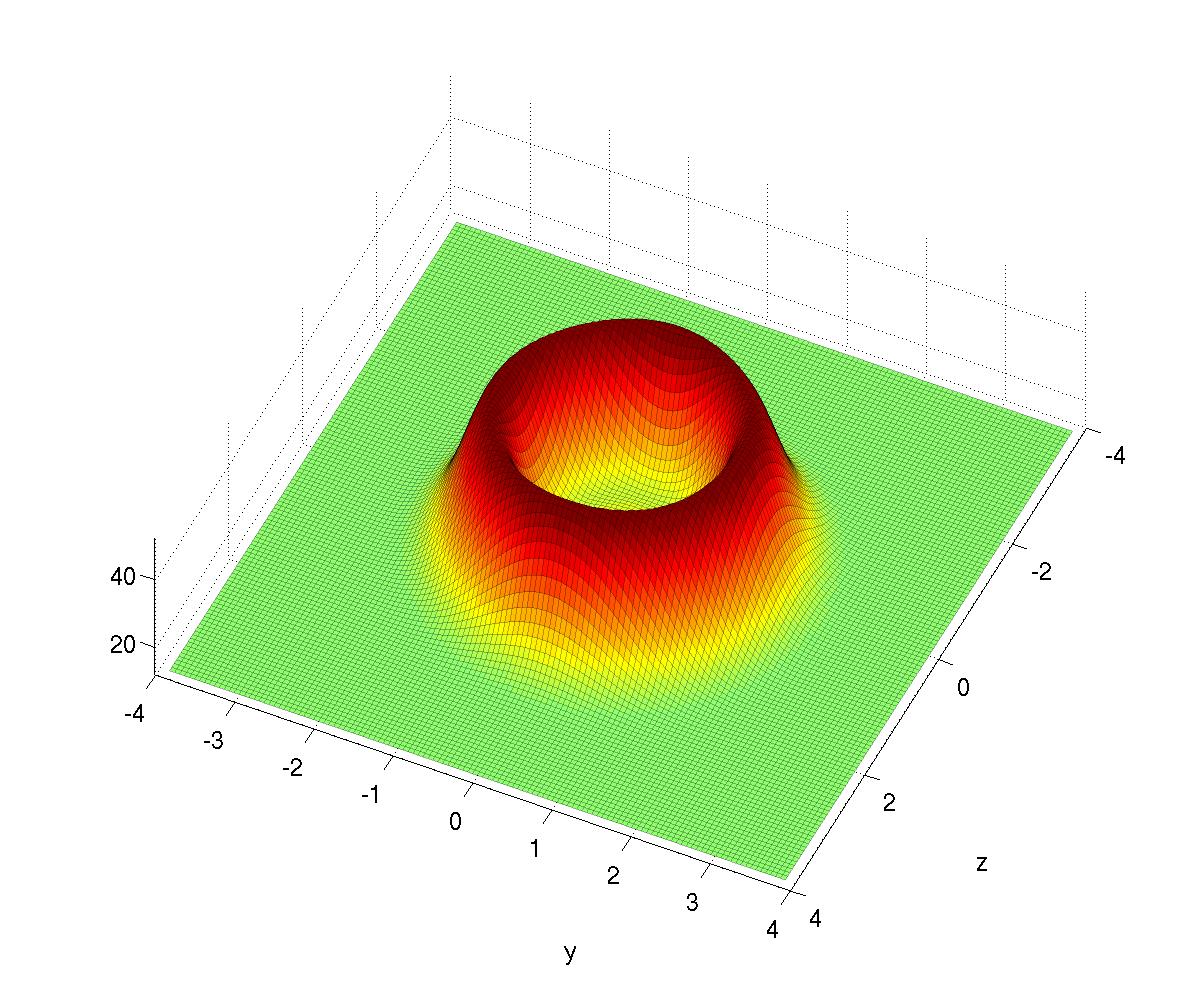}}
\subfigure[baryon charge density]{\includegraphics[width=0.3\linewidth]{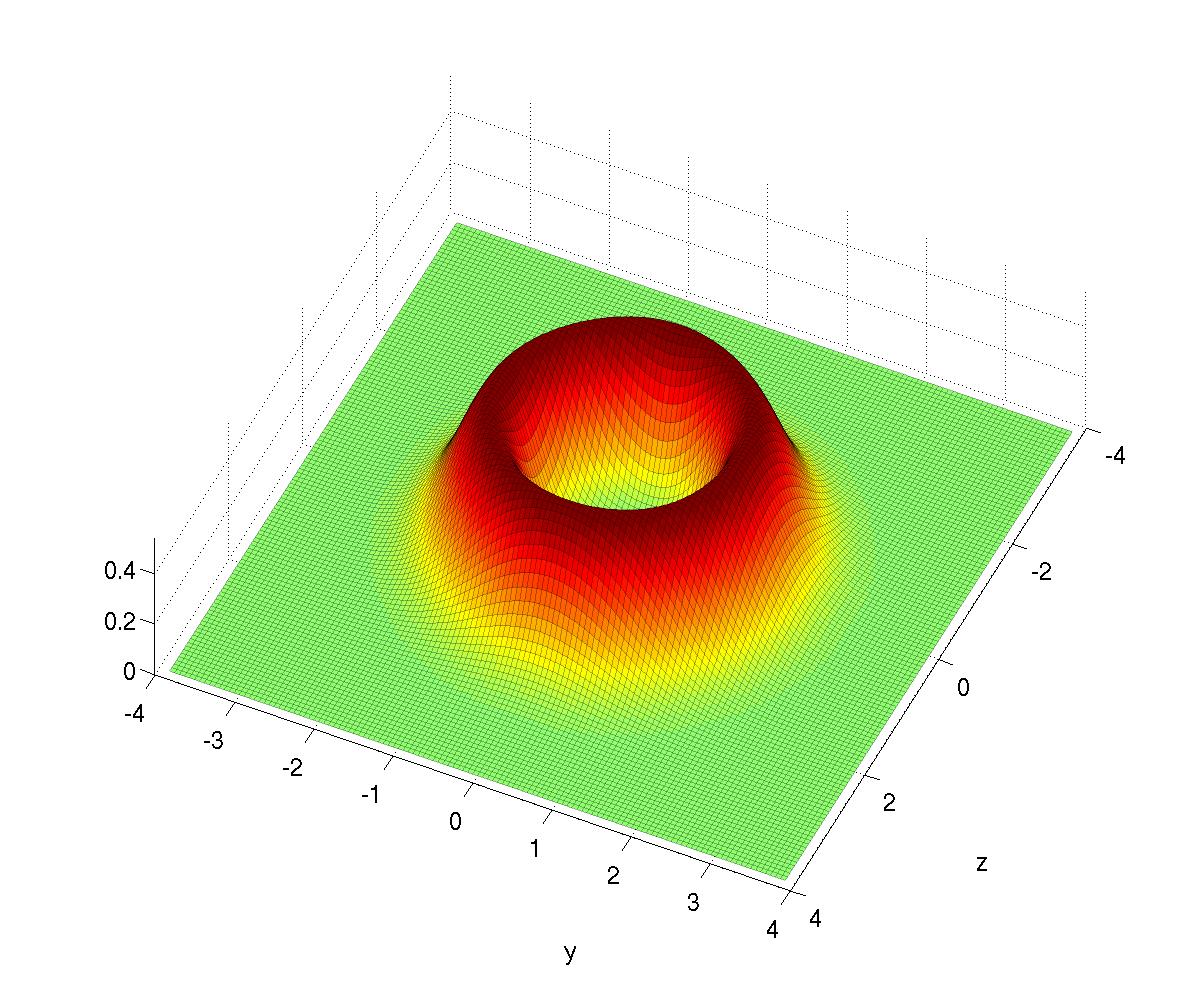}}
}
\caption{The domain wall with trapped $B=2,3,4$ Skyrmions. For
  details see fig.~\ref{fig:DW_DL_S}. The calculations are done on a
  $129^3$ cubic lattice and the potential used is
  \eqref{eq:Vlinear_babysk} with $m_4=4,m_3=2$.}
\label{fig:DW_2-4SL}
\end{center}
\end{figure}

We can trap more than one Skyrmion on the wall, by increasing the
winding of the baby-Skyrmion residing in the wall. For axially
symmetric solutions, we can consider the Ansatz
\beq
\mathbf{n} = \left\{\sin f(x)\sin k\phi\sin g(\rho),\sin
f(x)\cos k\phi\sin g(\rho),\sin f(x)\cos g(\rho),\cos f(x)\right\}
\, , \label{eq:ansatz_higher_baby_sk}
\eeq
where $\rho,\phi$ are polar coordinates on the wall. Keeping the
potential unchanged, we can write the Lagrangian
\begin{align}
-\mathcal{L} = \frac{1}{2} f_x^2 
+ \frac{1}{2} m_4^2 \sin^2 f
+ \frac{1}{2}\sin^2 f(1+f_x^2)\left[g_\rho^2 + \frac{k^2}{\rho^2}\sin^2
  g\right] 
+ \frac{k^2}{2\rho^2}\sin^4 f\sin^2(g) g_\rho^2
- \frac{1}{2}m_3^2\cos g\sin f \, .
\end{align}
The axially symmetric solutions are not stable in the planar
Faddeev-Skyrme model \cite{Piette:1994ug,Eslami:2000tj}; 
i.e.~the baby-Skyrmions with the potential
$\tfrac{1}{2}m_3^2(1-n_3)$ will form bound states resembling a baby-Skyrmion lattice \cite{baby-Skyrme-lattice}.  
Although the intuition of the baby-Skyrmions on $\mathbb{R}^2$ is
useful, our trapped baby-Skyrmions experience the curvature of the
wall which in turn induces terms and changes the coefficients of the 
Lagrangian, as we illustrated above and even more so without assuming
factorization in the variables. 
We find that the baby-Skyrmion rings are stable for baryon numbers 1,2
and 3 (see app.~\ref{app:relaxation_sequence} for a sequence of the
relaxation of an asymmetric configuration down to the 2-ring); in
fig.~\ref{fig:DW_2-4SL} are shown numerical 
solutions of higher-winding Skyrmions trapped on the wall with charge
$B=2,3,4$, respectively.

\begin{table}[!ht]
\begin{center}
\caption{Baryon charge and energy of the baby-Skyrmion in the linear
  potential (that of the domain wall is subtracted from the total
  energy). }
\label{tab:b/e_sk}
\begin{tabular}{l|llll}
$B$ & 1 & 2 & 3 & 4 \\
\hline
$B^{\rm numerical}$ & 0.998 & 1.997 & 2.996 & 3.995 \\
$E/B$ & $78.06\pm0.78$ & $73.43\pm0.12$ & $74.09\pm0.10$ & $75.36\pm0.08$
\end{tabular}
\end{center}
\end{table}

In table \ref{tab:b/e_sk} is shown the energy of the baby-Skyrmions, 
which is calculated by subtracting the wall contribution from the
total energy. It is observed that the three lowest baryon-charged
configurations are stable (minimum energy solutions 
for each $B=1,2,3$); the $B=2$ solution cannot decay into two 
$B=1$ solutions and the $B=3$ solution cannot
decay into a sum of the $B=1$ and $B=2$ solutions, 
whereas the $B=4$ solution is only meta-stable; 
although we could find the
solution, it can decay into two $B=2$ solutions. This can happen
either by quantum tunneling or at finite temperature or if some
kinematic perturbations impinge. 

\begin{figure}[!htp]
\begin{center}
\mbox{\subfigure[isosurfaces]{\includegraphics[width=0.3\linewidth]{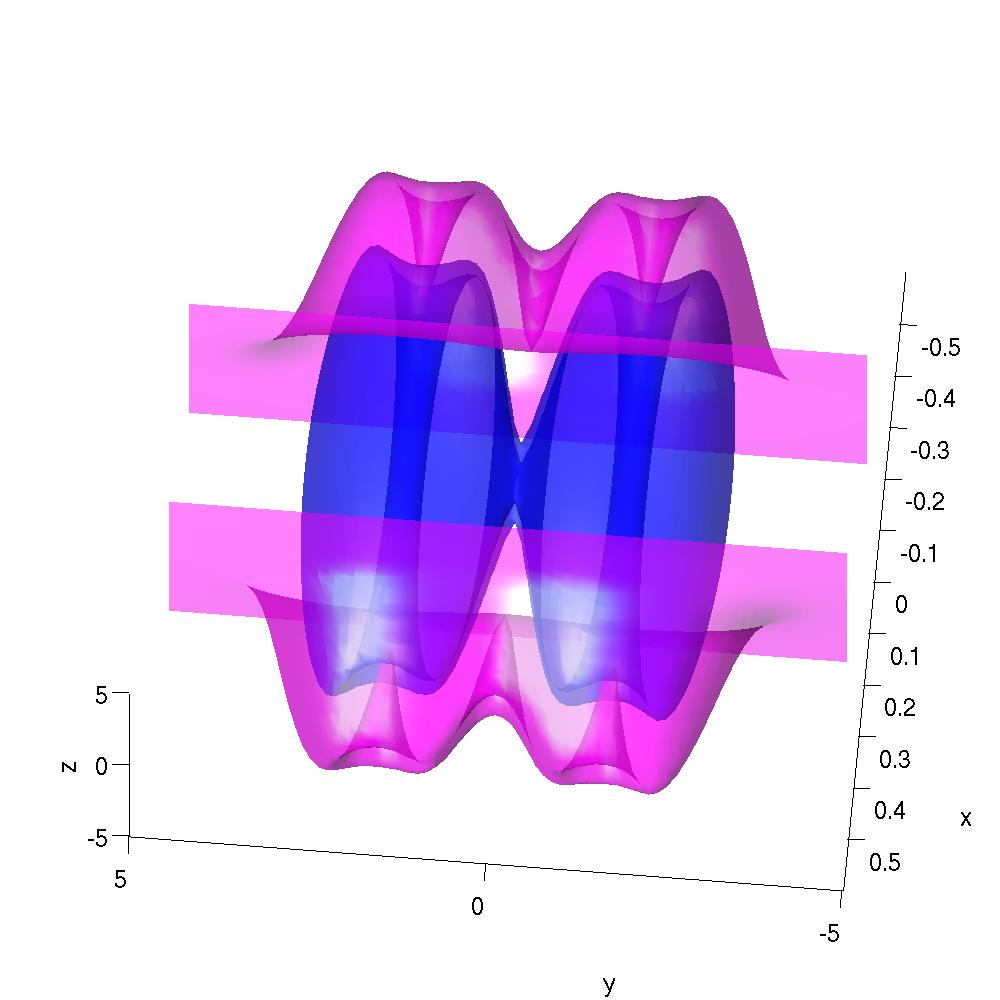}}
\subfigure[energy density]{\includegraphics[width=0.3\linewidth]{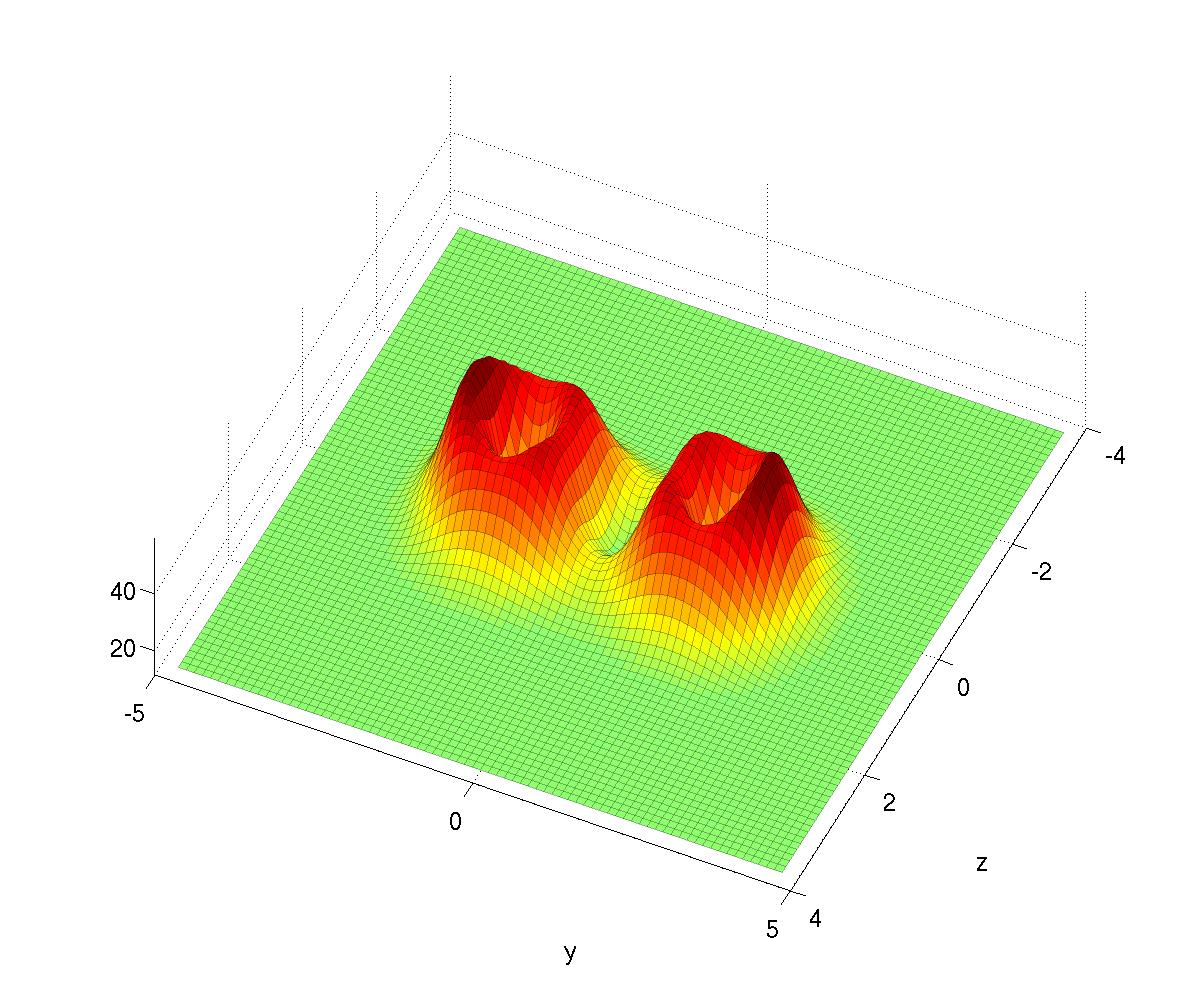}}
\subfigure[baryon charge density]{\includegraphics[width=0.3\linewidth]{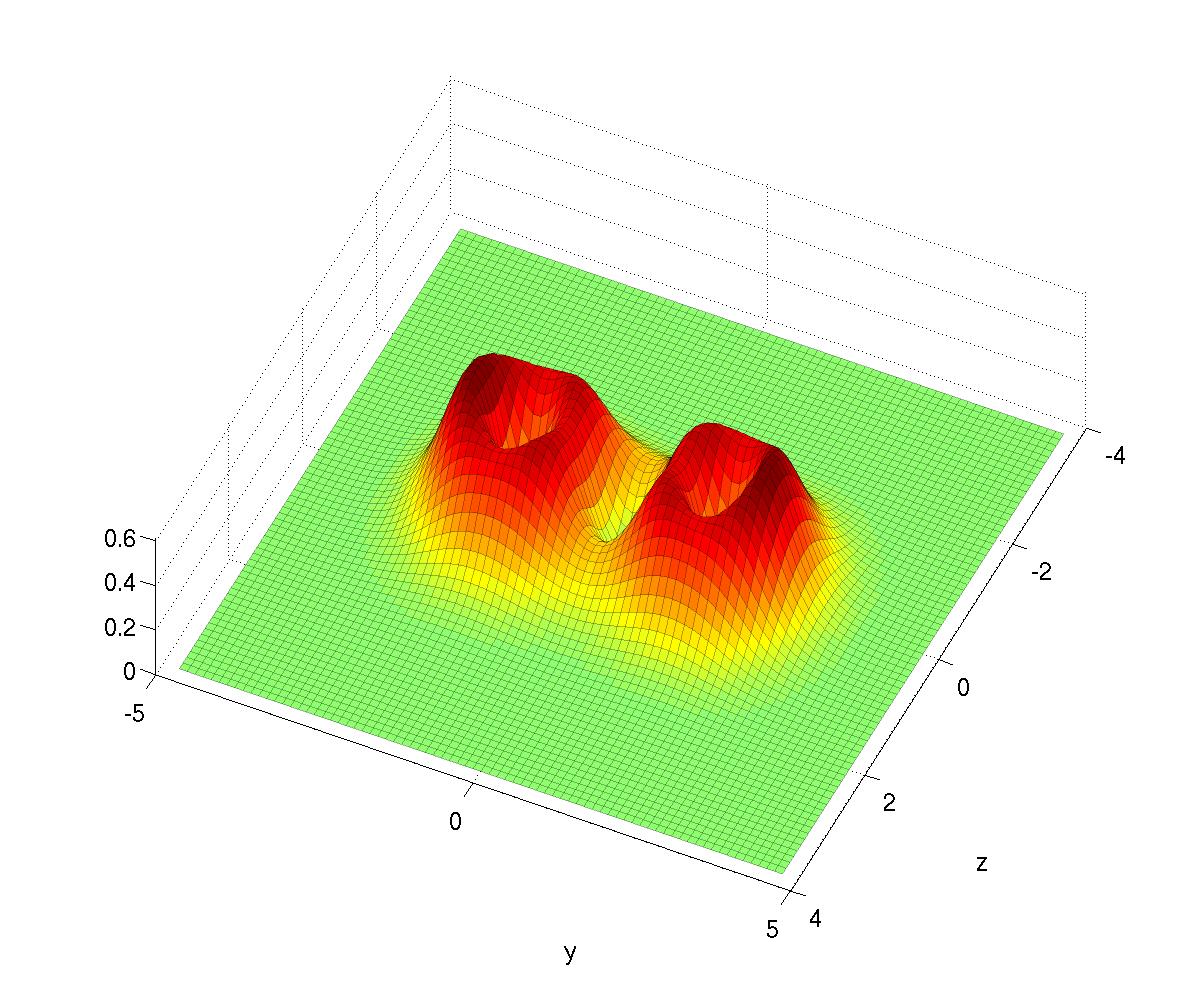}}
}
\caption{The domain wall with a minimal lattice structure of two
  trapped $B=2$ Skyrmions giving the total of a $B=4$ configuration. For
  details see fig.~\ref{fig:DW_DL_S}. The calculation is done on an
  $81^3$ cubic lattice, $B^{\rm numerical}=3.990$, $E=4\times 73.07$
  and the potential used is \eqref{eq:Vlinear_babysk} with
  $m_4=4,m_3=2$. }  
\label{fig:DW_2_2RINGS}
\end{center}
\end{figure}

As observed in table \ref{tab:b/e_sk}, the charge 4 baby-Skyrmion is
at best meta-stable. By starting with non-axially symmetric initial
conditions we can still obtain the 4-ring, but by skewing the setup
enough it falls into a lower-energy state, i.e.~composed by two
charge-2 rings. By starting with an asymmetric initial guess, we find
the configuration shown in fig.~\ref{fig:DW_2_2RINGS}. We notice that
the energy of this configuration is lower than that of the ring, whose 
energy is stated in table \ref{tab:b/e_sk}.

\begin{figure}[!htp]
\begin{center}
\mbox{\subfigure[isosurfaces]{\includegraphics[width=0.3\linewidth]{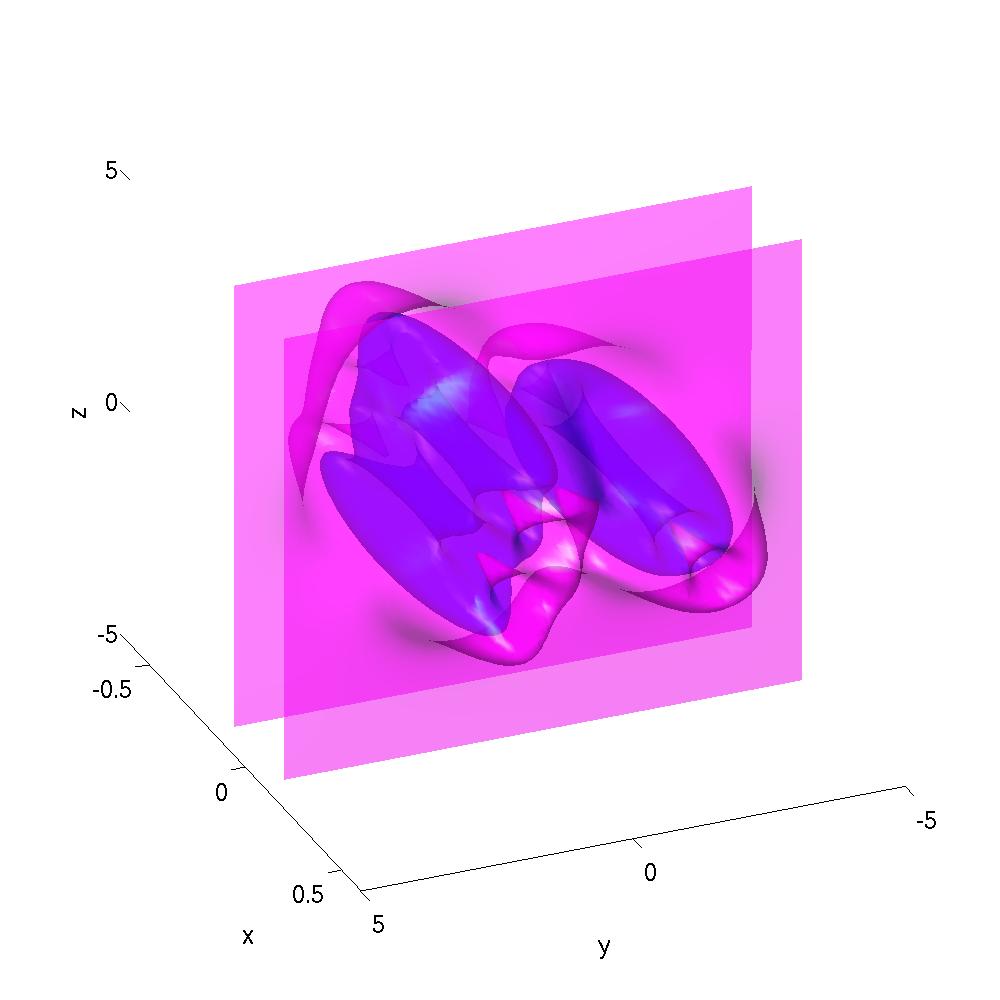}}
\subfigure[energy density]{\includegraphics[width=0.3\linewidth]{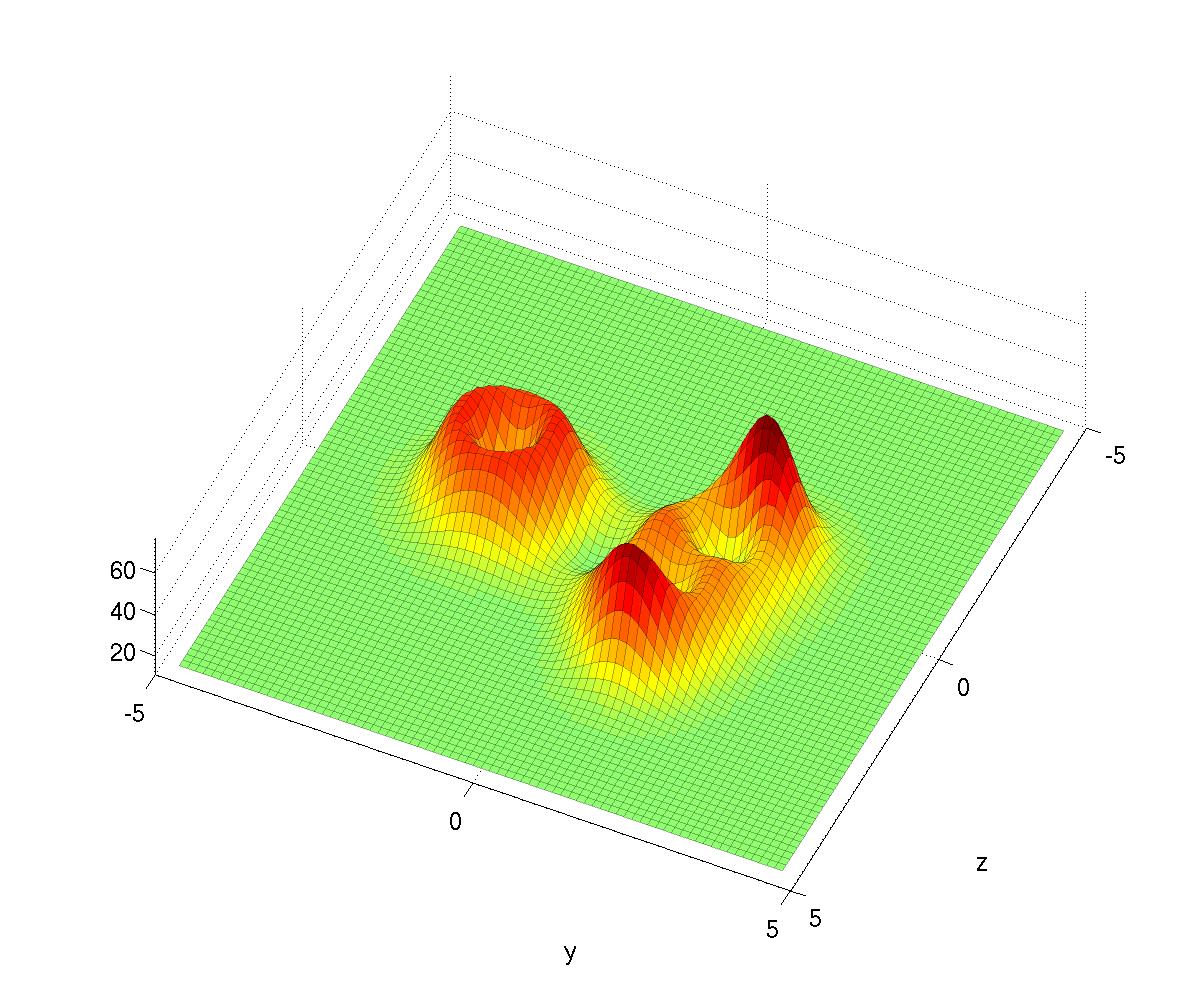}}
\subfigure[baryon charge density]{\includegraphics[width=0.3\linewidth]{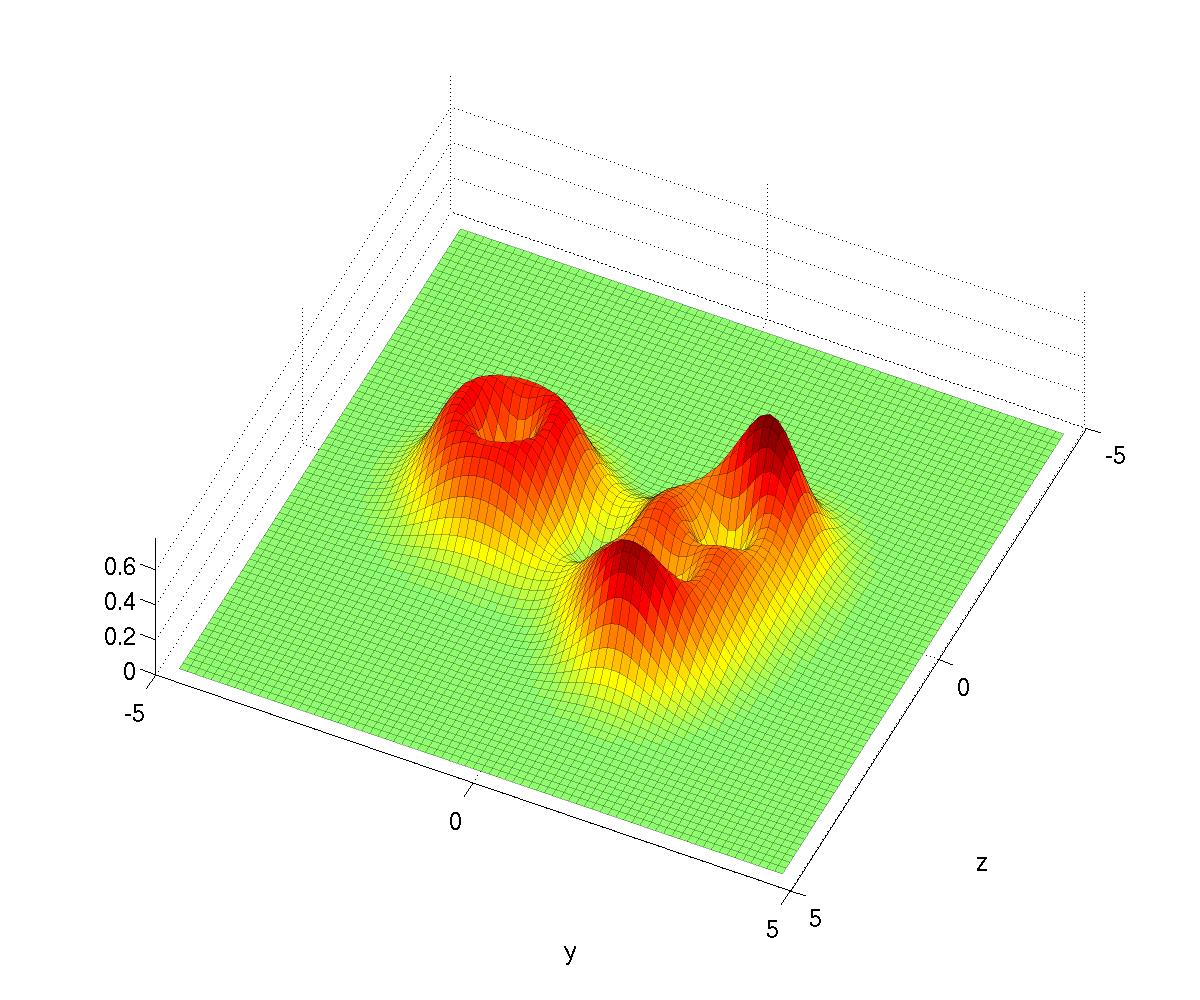}}
}
\caption{The domain wall with a bound state of a 2-ring and a deformed
  3-ring forming a $B=5$ configuration. For details see
  fig.~\ref{fig:DW_DL_S}. The calculation is done on an $81^3$ cubic
  lattice, $B^{\rm numerical}=4.988$, $E=5\times 73.61$ and the
  potential used is \eqref{eq:Vlinear_babysk} with $m_4=4,m_3=2$. }
\label{fig:DW_2.5_2RINGS}
\end{center}
\end{figure}

\begin{figure}[!htp]
\begin{center}
\mbox{\subfigure[isosurfaces]{\includegraphics[width=0.3\linewidth]{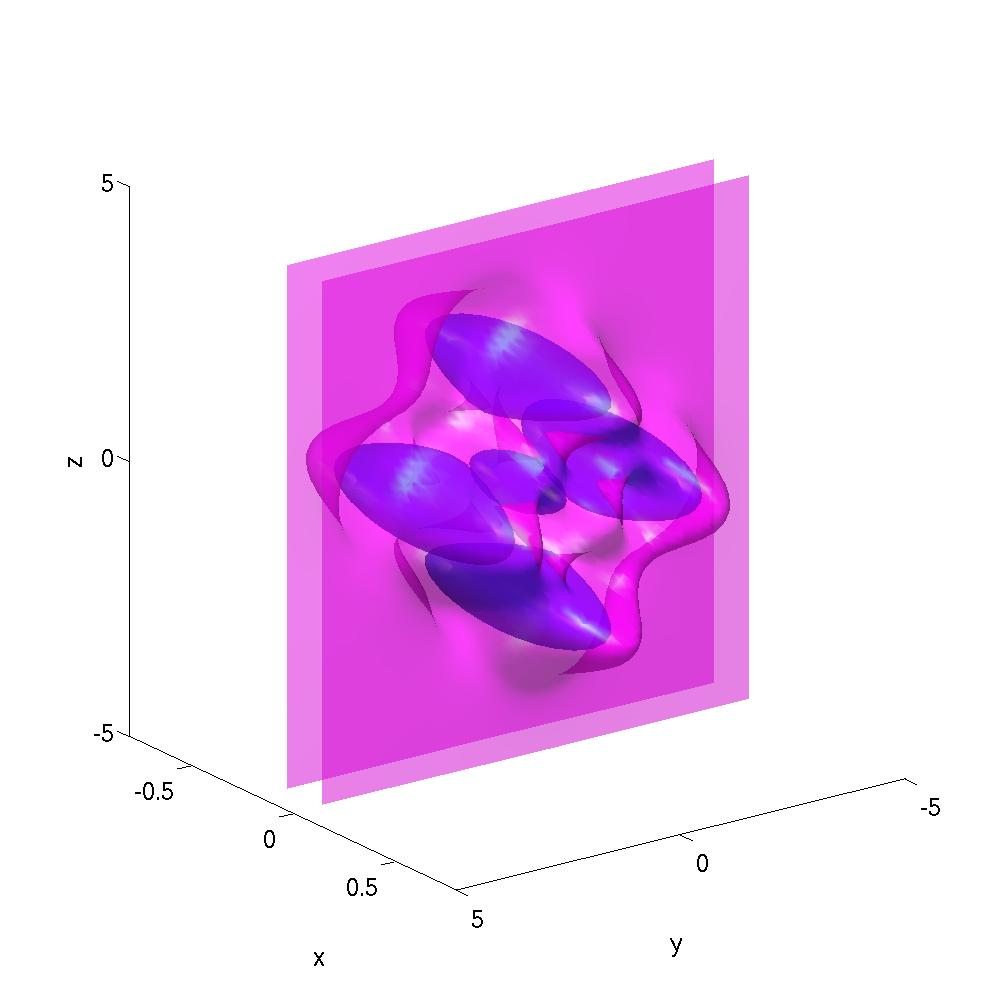}}
\subfigure[energy density]{\includegraphics[width=0.3\linewidth]{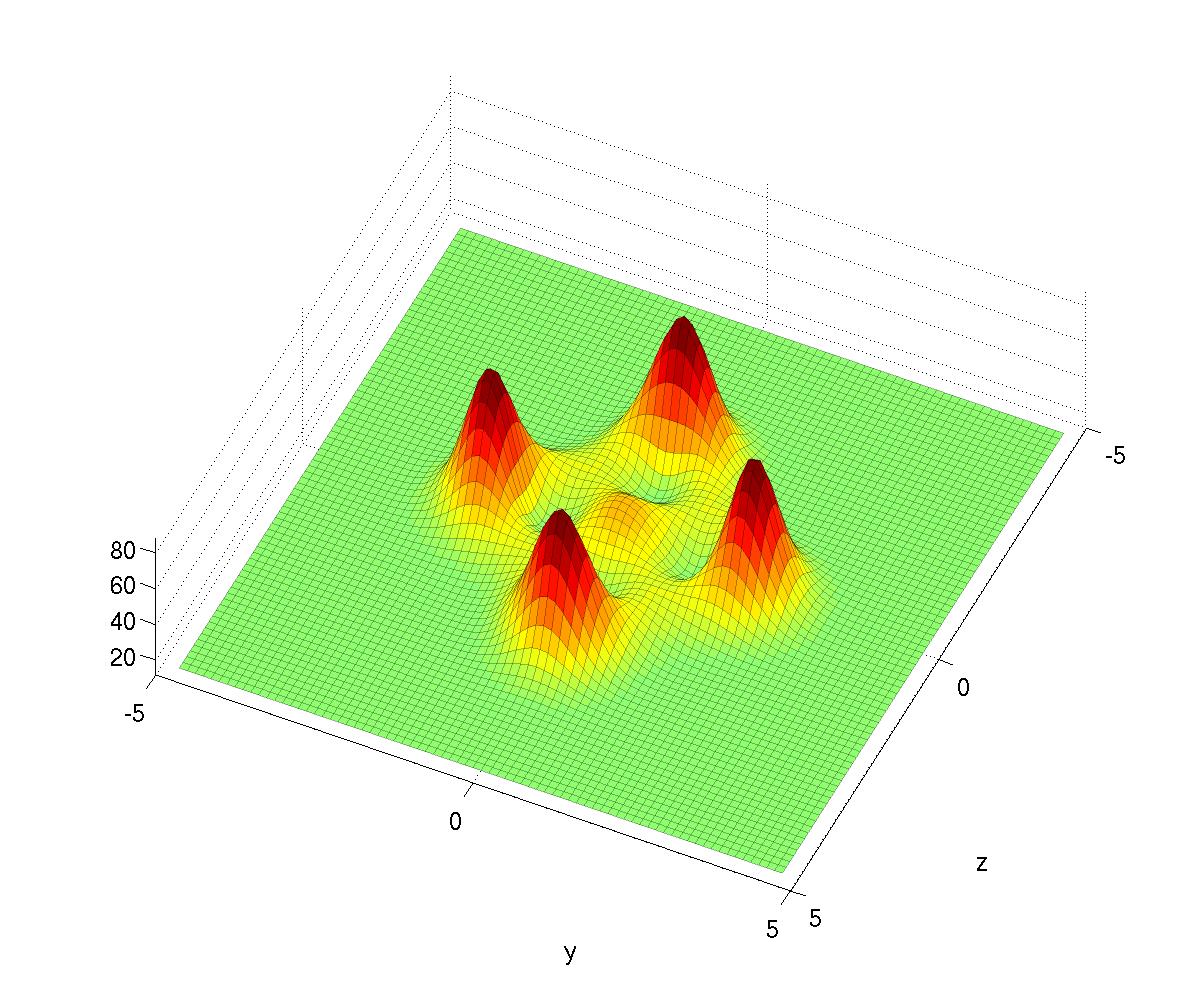}}
\subfigure[baryon charge density]{\includegraphics[width=0.3\linewidth]{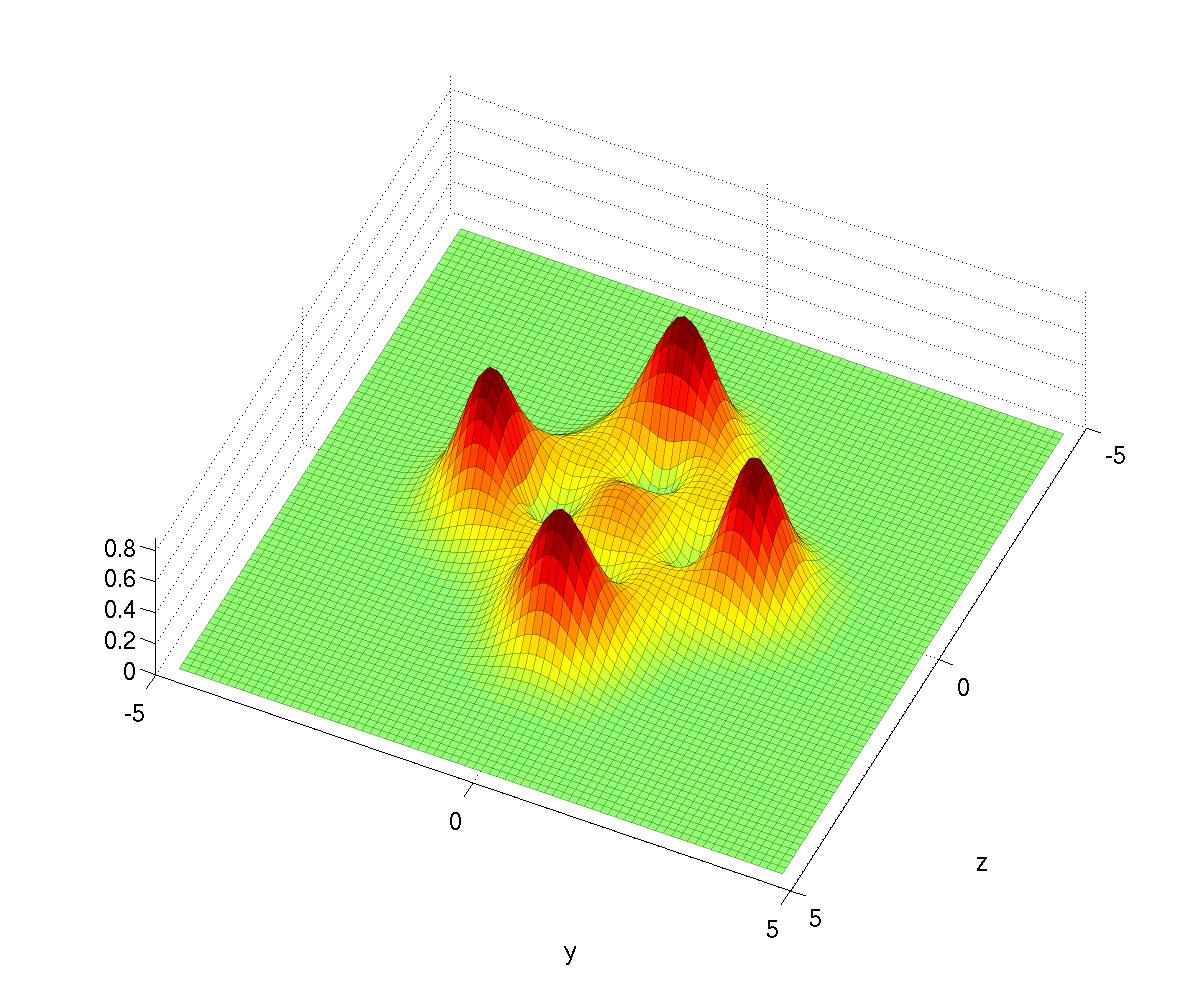}}
}
\caption{The domain wall with a symmetric initial condition for the
  5-ring which collapses to the shown bound state of single Skyrmions
  forming a meta-stable $B=5$ configuration. For details see
  fig.~\ref{fig:DW_DL_S}. The calculation is done on an $81^3$ cubic
  lattice, $B^{\rm numerical}=4.98$, $E=5\times 75.58$ and the
  potential used is \eqref{eq:Vlinear_babysk} with $m_4=4,m_3=2$. }
\label{fig:DW_5RING_collapsed}
\end{center}
\end{figure}

With this knowledge at hand, we study the $B=5$ solution with various
different initial guesses in order to find the lowest energy solution
(the globally stable one). An asymmetric initial guess upon numerical
relaxation gives rise to the solution shown in
fig.~\ref{fig:DW_2.5_2RINGS}, which is some sort of bound state of a
genuine 2-ring and another creature carrying charge 3. If on the other
hand we start with an axially symmetric initial guess, we find a
meta-stable solution which is shown in
fig.~\ref{fig:DW_5RING_collapsed}. The latter solution has a higher
energy per charge than the asymmetric solution, so eventually it will
decay to that by means of tunneling or perturbations etc. 
We observe that the lowest energy solution found for $B=5$ consists of
one 2-ring and the remaining charge. With table \ref{tab:b/e_sk}
in mind, we conjecture that for large $B$, the stable configurations
are formed of 2-rings situated in a lattice-like structure. 

\subsection{Trapped Skyrmions in a quadratic potential}

In this section we switch to baby Skyrmions trapped on the wall and
thus exhibiting full baryon charge in the modified potential
\cite{Kudryavtsev:1997nw,Weidig:1998ii}, namely the quadratic one
\beq
V = -\frac{1}{2}m_3^2 n_3^2 
+ \frac{1}{2}m_4^2 (1-n_4^2) \, .
\label{eq:Vquadratic_no_m2}
\eeq
Using the Ansatz \eqref{eq:ansatz_baby_sk} and the above given
potential, the Lagrangian \eqref{eq:LO4} yields
\begin{align}
-\mathcal{L} &= \frac{1}{2} f_x^2 
+ \frac{1}{2} (m_4^2-m_3^2) \sin^2 f
+ \frac{1}{2}\sin^2 f(1+f_x^2)\left[g_\rho^2 + \frac{1}{\rho^2}\sin^2
  g\right] 
+ \frac{1}{2\rho^2}\sin^4 f\sin^2(g) g_\rho^2 \non
&\phantom{=\ }
+ \frac{1}{2}m_3^2\sin^2 f\sin^2 g \, ,
\end{align}
from which we can identify the domain wall in $f$ with effective
mass-squared $(m_4^2-m_3^2)$ and a baby-Skyrmion with a quadratic mass
term as well as an enhanced factor of $(1+f_x^2)$ in front of its
kinetic term; $|f_x| \sim m_4$ on the domain wall. 

In principle the above effective Lagrangian should allow for a
ring-like baby-Skyrmion configuration; the problem however is
this. The kinetic term is a marginal term, whereas the Skyrme term
yields a pressure. With the enhanced prefactor of the kinetic term,
the generic configuration is a baby-Skyrmion with a peak in both the
energy distribution and the baryon charge distribution. Since the
enhancement comes about due to the steepness of the domain wall in
$f$, i.e.~it is proportional to $m_4^2$, we can lower this mass and in
turn increase the width of the domain wall. The condition however that
$m_3\ll m_4$ means that the mass term in the baby-Skyrmion effective
theory is becoming very small. This mass term is needed for the
baby-Skyrmion to exist on $\mathbb{R}^2$. 
Although we do not have a proof of non-existence for the ring-like
structure at present, we have not been able to find any such
configuration numerically, scanning over a range of parameters. 

\begin{figure}[!hpt]
\begin{center}
\mbox{\subfigure[isosurfaces]{\includegraphics[width=0.3\linewidth]{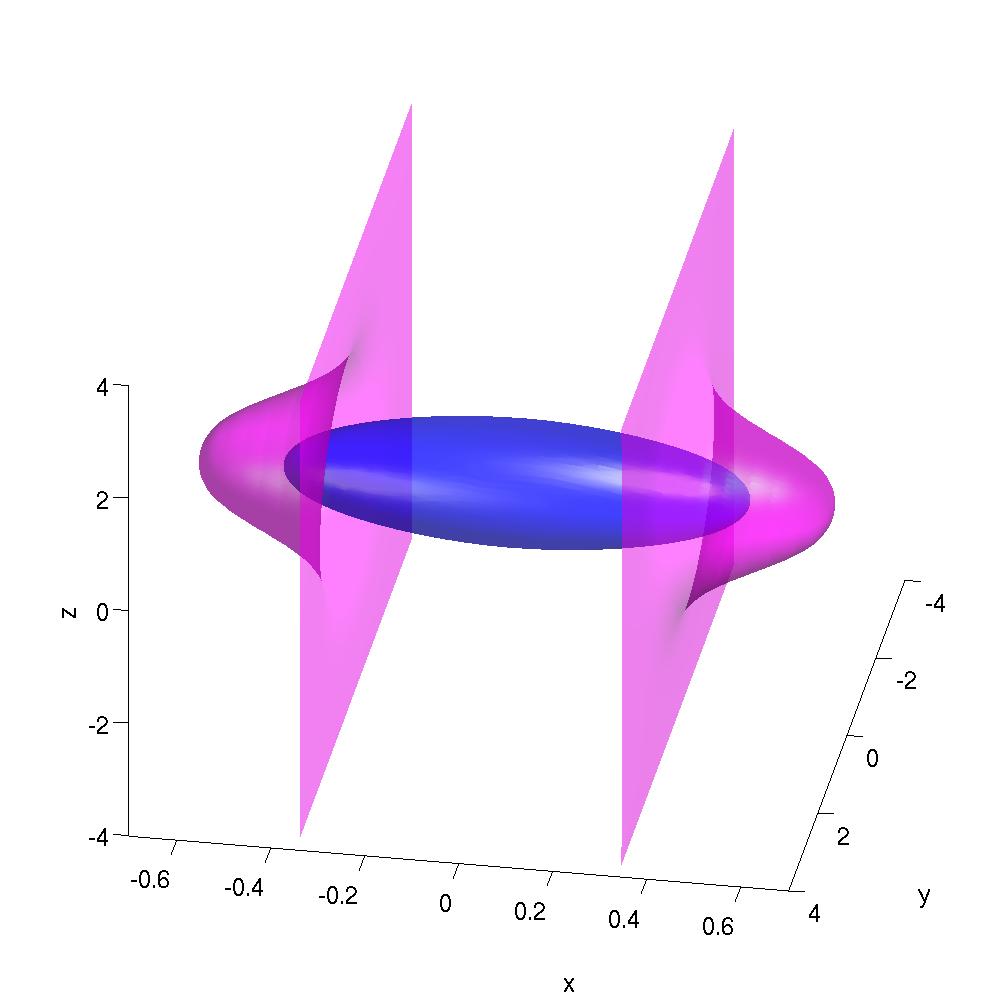}}
\subfigure[energy density]{\includegraphics[width=0.3\linewidth]{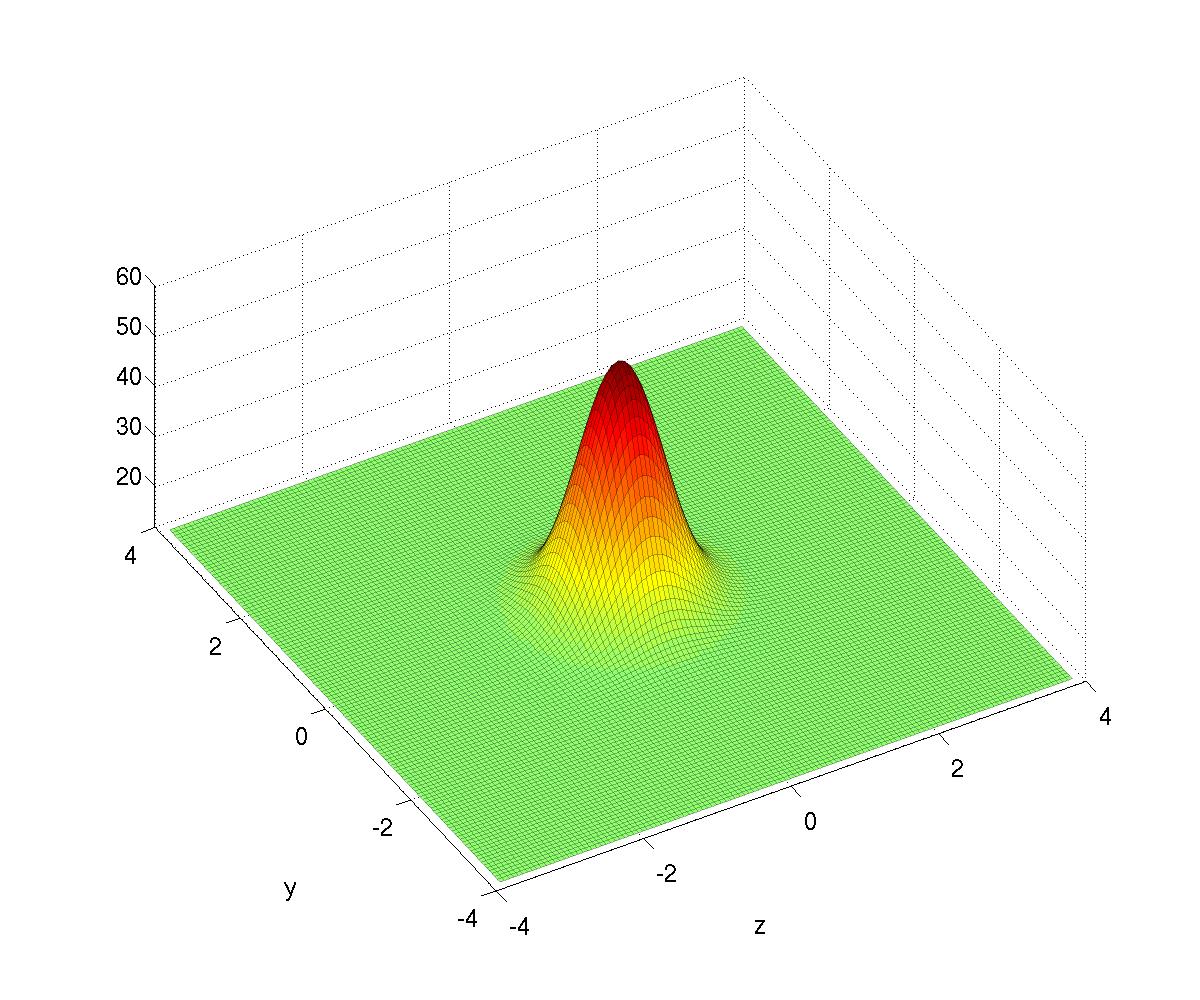}}
\subfigure[baryon charge density]{\includegraphics[width=0.3\linewidth]{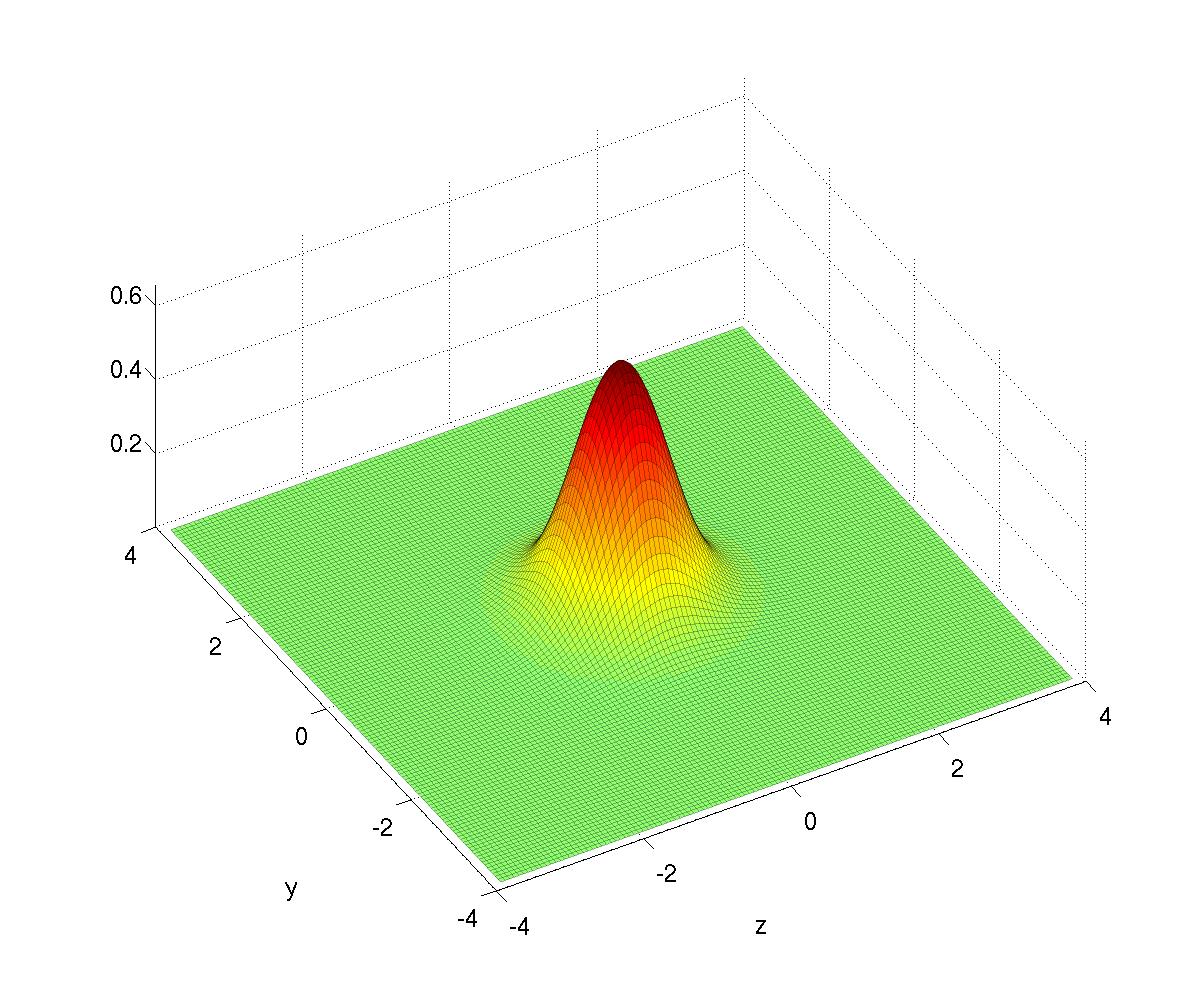}}
}
\caption{The domain wall with a trapped Skyrmion in a quadratic
  potential. Although having a flat top, the energy density is not a
  ring-like structure. For details see fig.~\ref{fig:DW_DL_S}. 
  The calculation is done on a $129^3$ cubic lattice, 
  $B^{\rm numerical}=0.998$ and the potential
  used is \eqref{eq:Vquadratic_no_m2} with $m_4=4,m_3=2$.} 
\label{fig:DW_SQ}
\end{center}
\end{figure}

For completeness we obtain numerical solutions for the single Skyrmion
trapped on the domain wall in the quadratic potential
\eqref{eq:Vquadratic_no_m2} and it is shown in
fig.~\ref{fig:DW_SQ}. It is observed that the top of the energy
density is more flattened out with respect to the Skyrmion in the
linear potential (see fig.~\ref{fig:DW_SL}). 

Since as explained above, the single charge Skyrmion configuration
does not differ much from the that with a linear potential, we turn to
the higher-winding ones. They are stable ring-like configurations
similar in nature to the ones in the linear potential. One twist,
however, is possible here as compared to the case with the linear
potential, namely that we can add a mass term for $n_2$, breaking
rotational symmetry on the wall ring. The potential is
\beq
V = -\frac{1}{2}m_2^2 n_2
- \frac{1}{2}m_3^2 n_3^2
+ \frac{1}{2}m_4^2 (1-n_4^2) \, , 
\label{eq:quadratic_kink}
\eeq
and if we plug this and the Ansatz \eqref{eq:ansatz_baby_sk} into the
Lagrangian \eqref{eq:LO4}, we get
\begin{align}
-\mathcal{L} &= \frac{1}{2} f_x^2  
+ \frac{1}{2} (m_4^2-m_3^2) \sin^2 f
+ \frac{1}{2}\sin^2 f(1+f_x^2)\left[g_\rho^2 + \frac{k^2}{\rho^2}\sin^2
  g\right] 
+ \frac{k^2}{2\rho^2}\sin^4 f\sin^2(g) g_\rho^2 \non
&\phantom{=\ }
+ \frac{1}{2}m_3^2\sin^2 f\sin^2 g
- \frac{1}{2}m_2^2\sin f\sin g
  \sum_{j=0}^k \begin{pmatrix} k\\j\end{pmatrix}
  \frac{y^j z^{k-j}}{\rho^k}
  \cos\left(\frac{\pi}{2}(k-j)\right) \, .
\end{align}
Identifying the pieces of the above Lagrangian one-by-one we have the
domain wall in $f$ with effective mass-squared $(m_4^2-m_3^2)$ and
a baby-Skyrmion with mass $m_3^2$ and an enhanced kinetic term by a
factor of $(1+f_x^2)$. 
Finally, there is an angle-dependent mass term (with the angle being
the usual one in polar coordinates on the wall) inducing $k$
sine-Gordon kinks on top of the ring-like baby-Skyrmion
structure. This type of object has been already realized in the planar 
case (i.e.~in $2+1$ dimensions) in the literature, see
e.g.~\cite{Kobayashi:2013ju}.

\begin{figure}[!p]
\begin{flushleft}$B^{\rm numerical}=1.996$:\end{flushleft}
\begin{center}
\mbox{\subfigure{\includegraphics[width=0.3\linewidth]{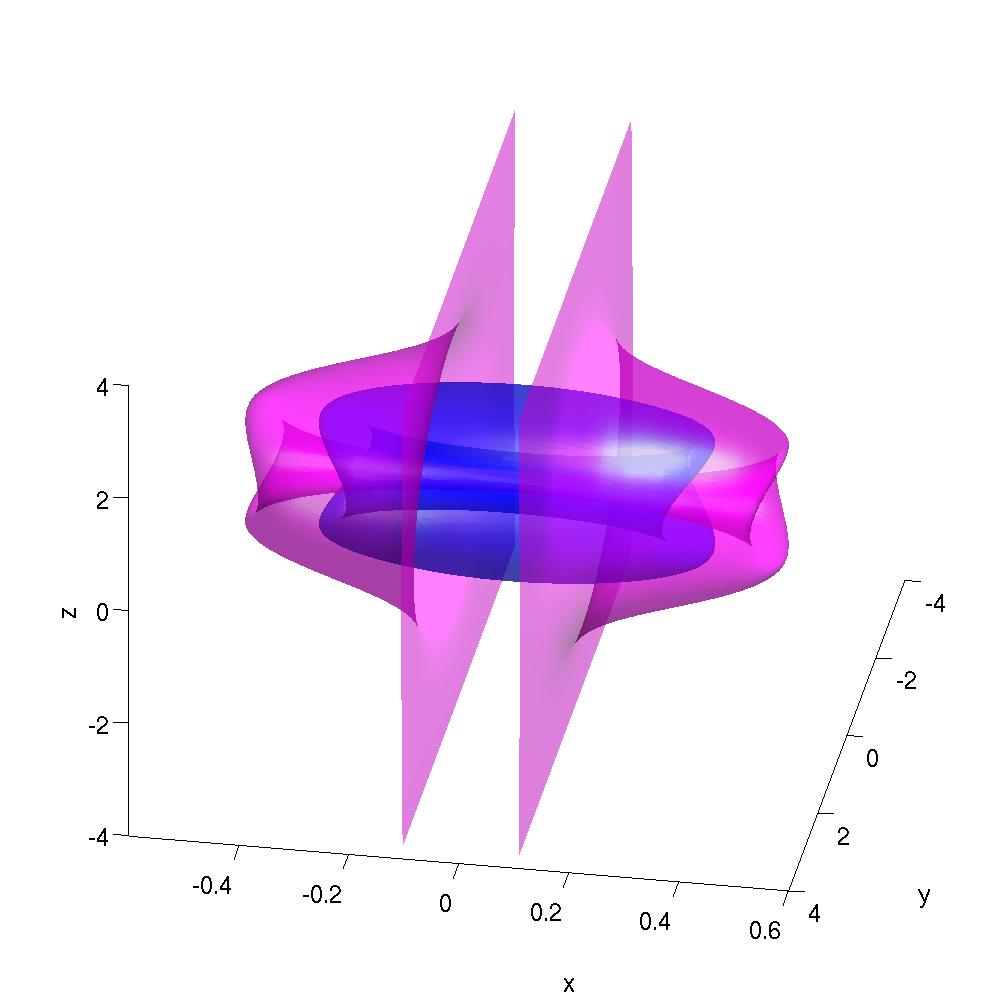}}
\subfigure{\includegraphics[width=0.3\linewidth]{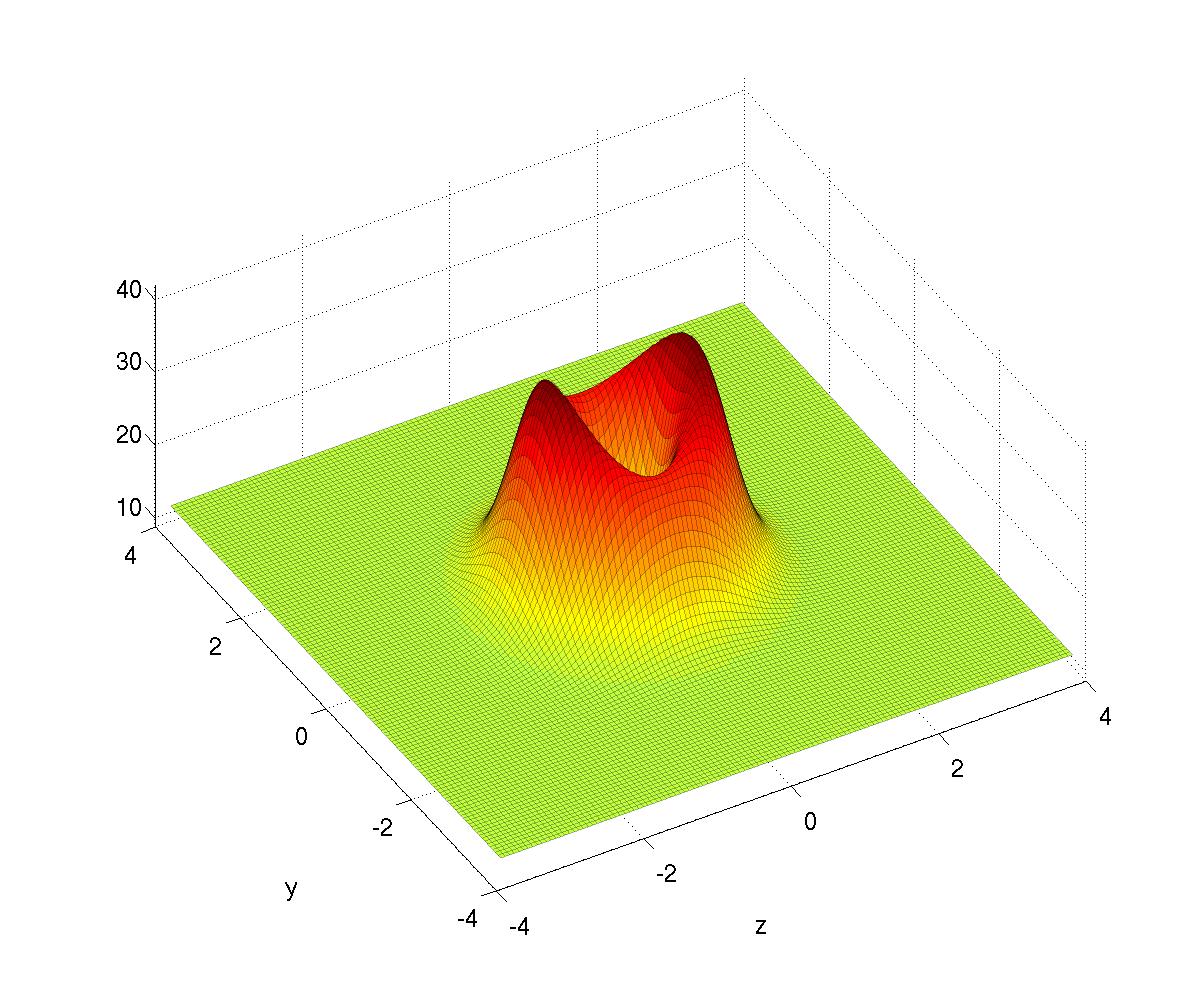}}
\subfigure{\includegraphics[width=0.3\linewidth]{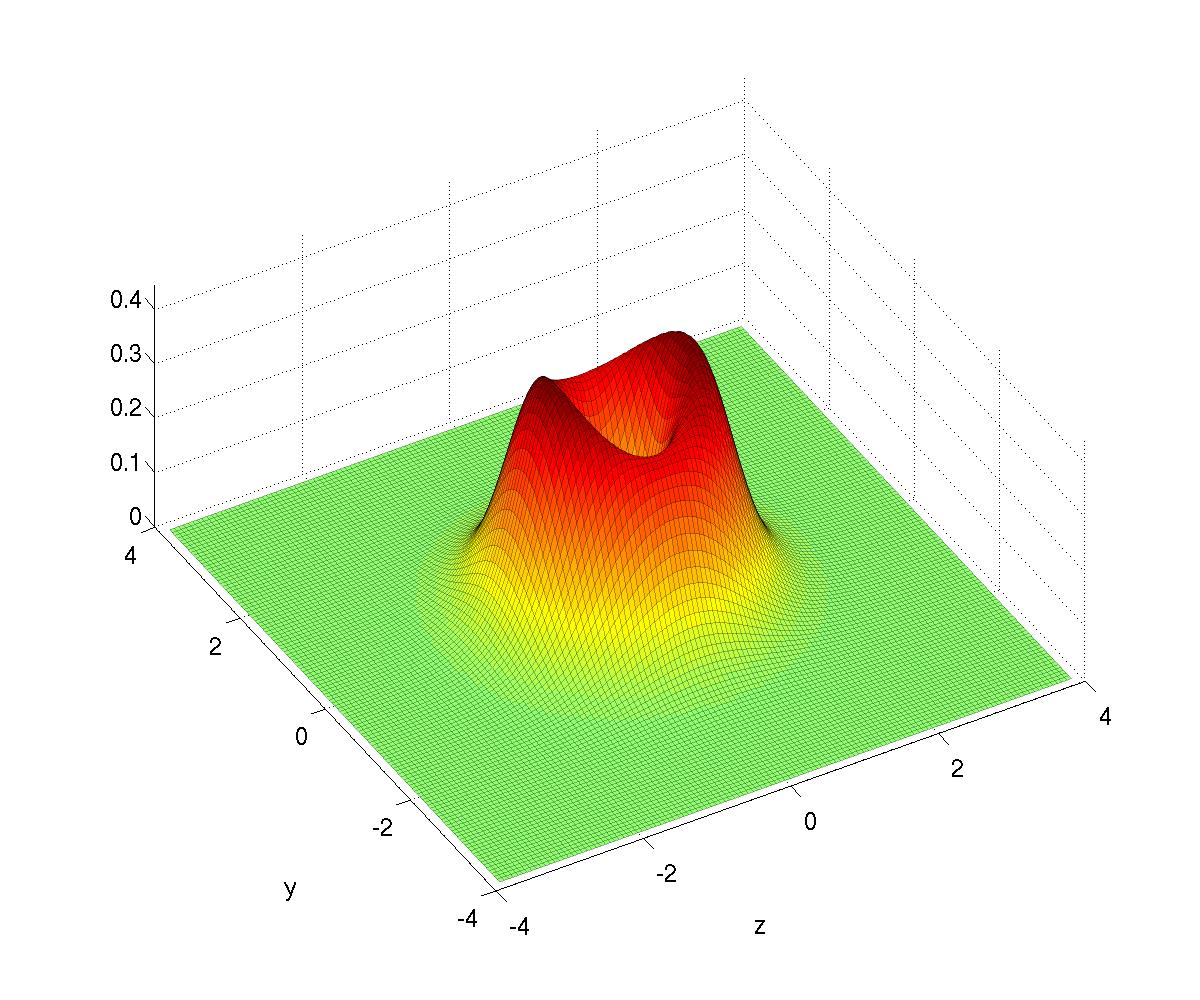}}
}
\end{center}
\begin{flushleft}$B^{\rm numerical}=2.991$:\end{flushleft}
\begin{center}
\mbox{\subfigure{\includegraphics[width=0.3\linewidth]{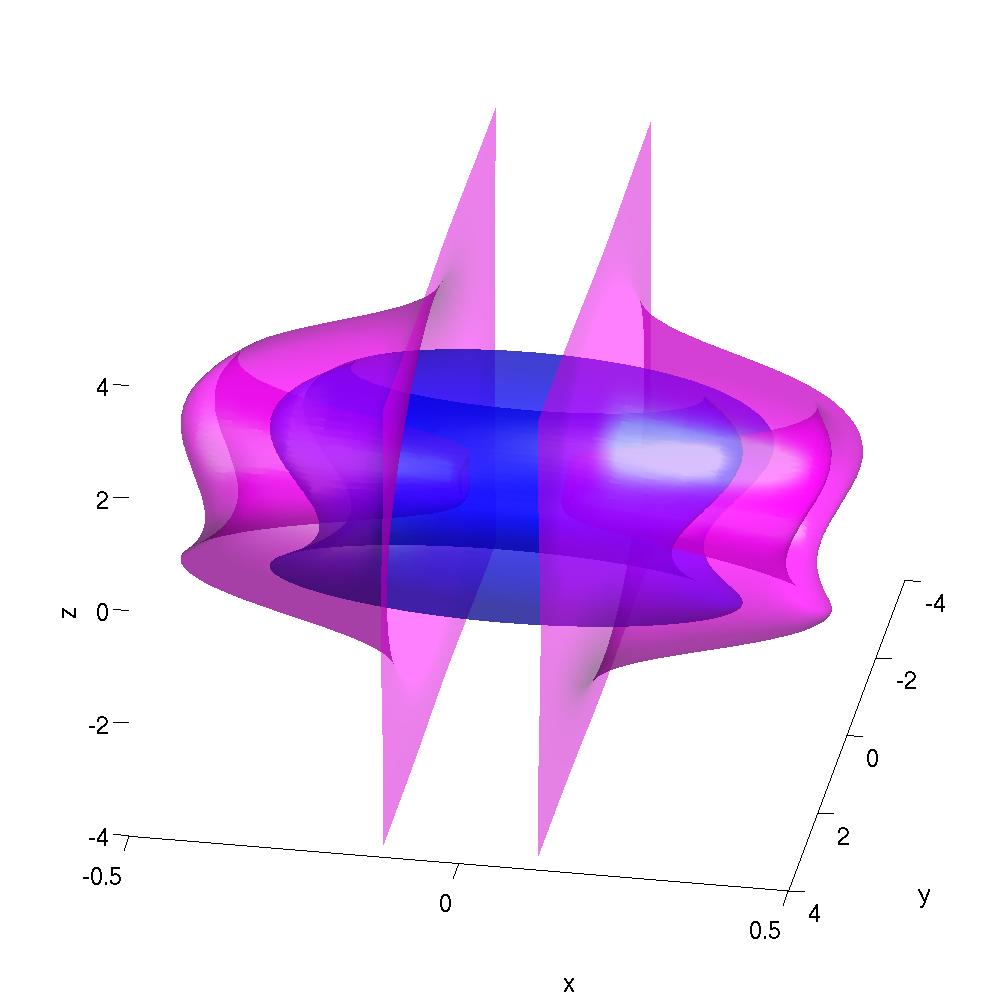}}
\subfigure{\includegraphics[width=0.3\linewidth]{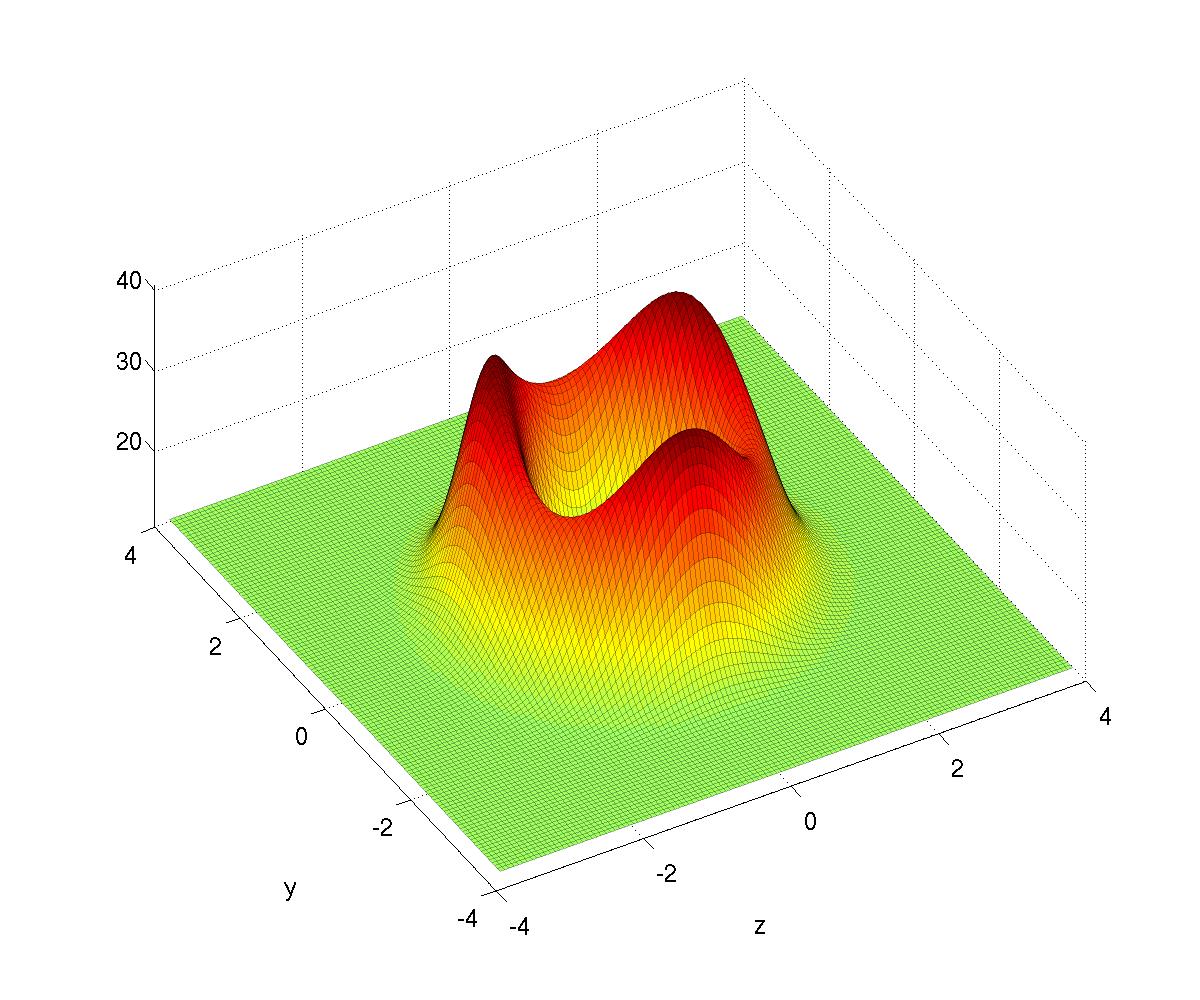}}
\subfigure{\includegraphics[width=0.3\linewidth]{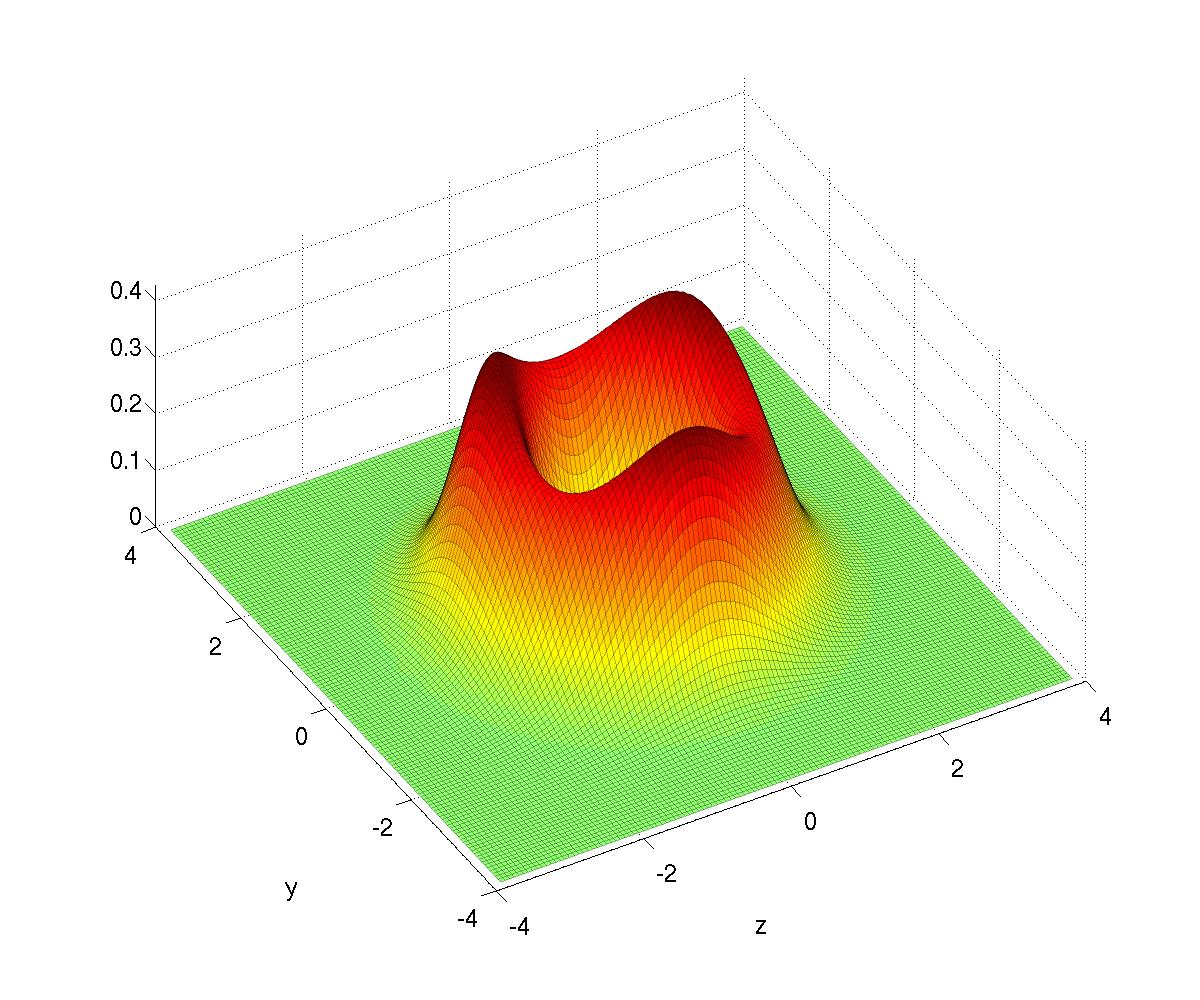}}
}
\end{center}
\begin{flushleft}$B^{\rm numerical}=3.992$:\end{flushleft}
\begin{center}
\mbox{\subfigure[isosurfaces]{\includegraphics[width=0.3\linewidth]{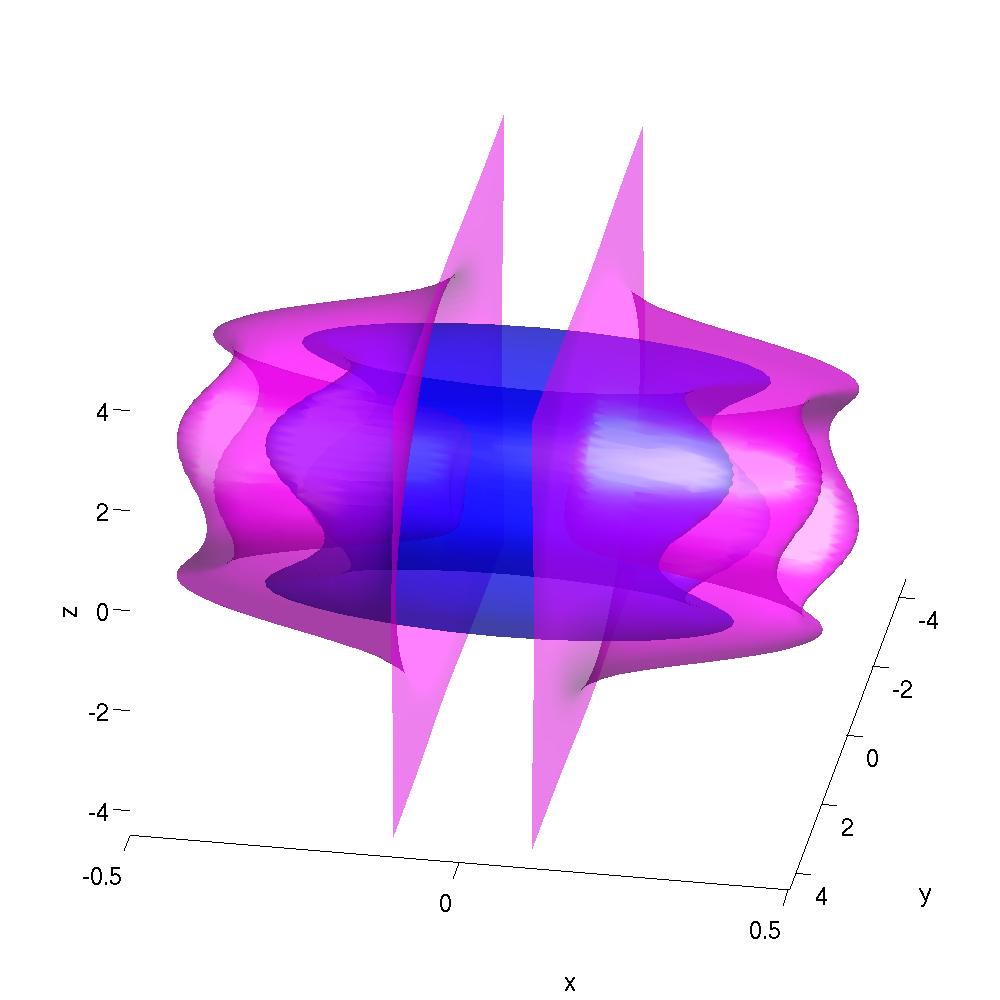}}
\subfigure[energy density]{\includegraphics[width=0.3\linewidth]{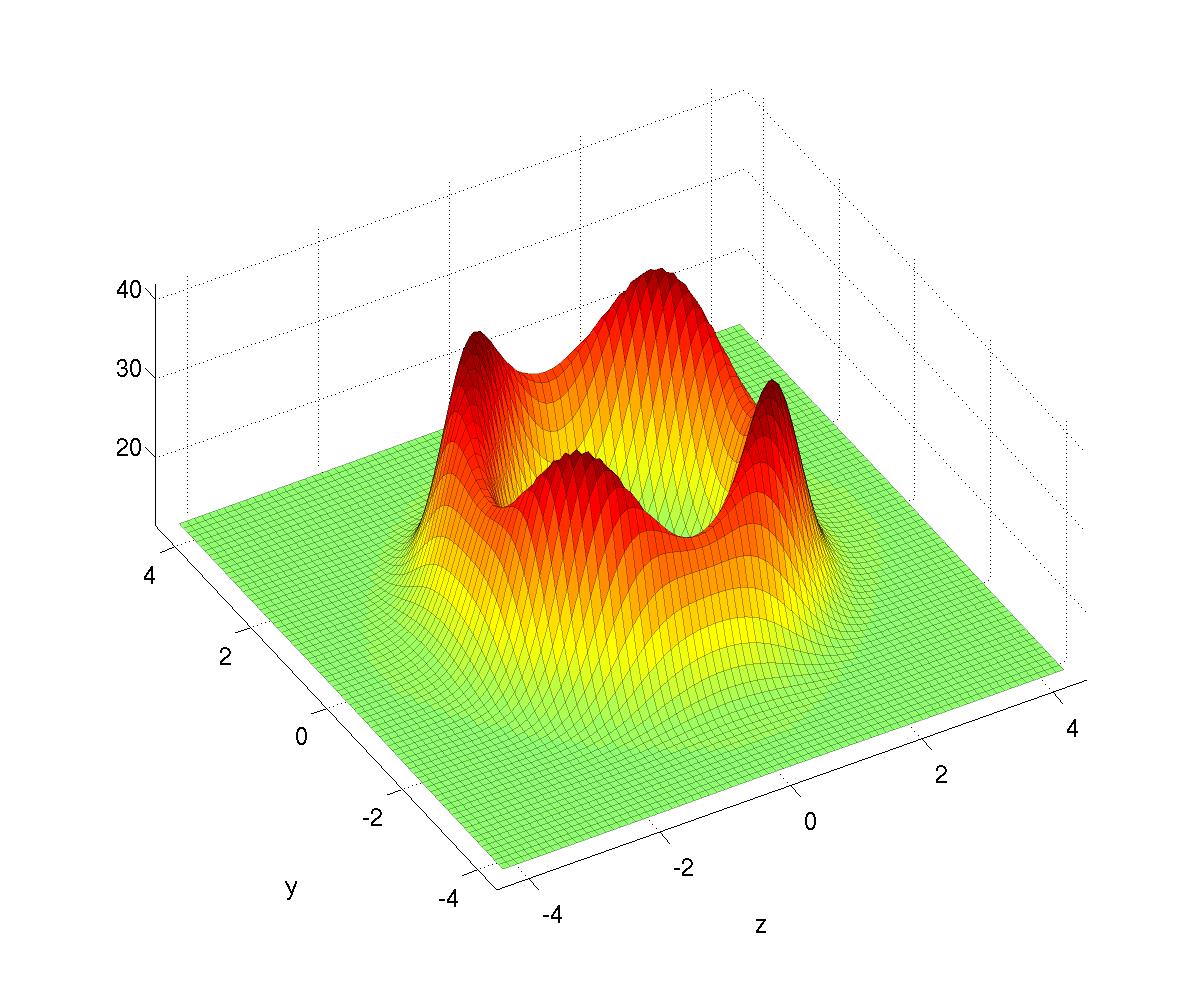}}
\subfigure[baryon charge density]{\includegraphics[width=0.3\linewidth]{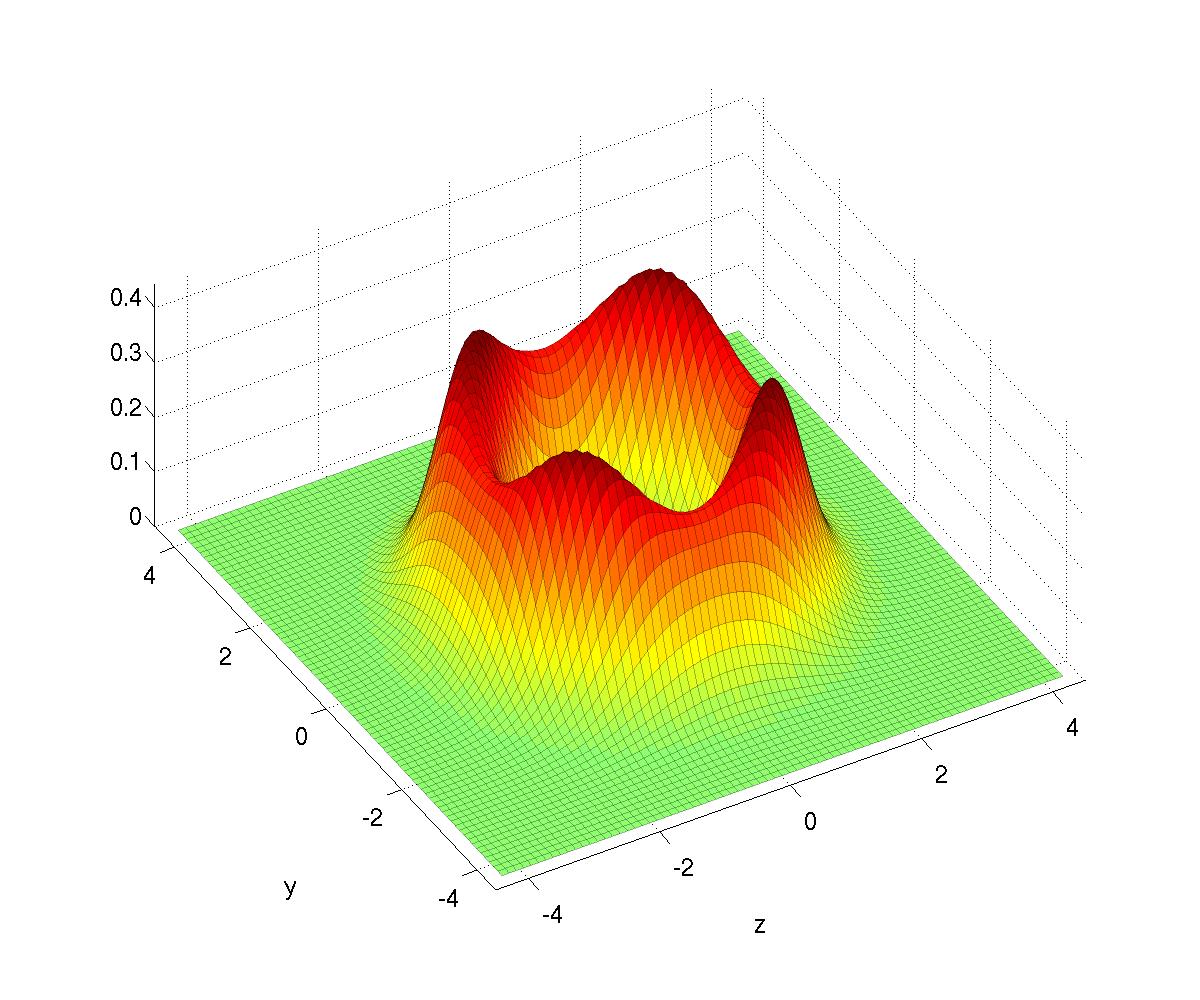}}
}
\caption{The domain wall with $2,3,4$ sine-Gordon kinks living on trapped
  $B=2,3,4$ Skyrmions. For details see fig.~\ref{fig:DW_DL_S}. The
  calculations are done on $129^3,129^3,81^2$ cubic lattices and the
  potential used is \eqref{eq:quadratic_kink} with $m_4=4,m_3=2,m_2=1$.}
\label{fig:DW_2-4SQ}
\end{center}
\end{figure}

\begin{figure}[!htp]
\begin{flushleft}$B^{\rm numerical}=4.992$:\end{flushleft}
\begin{center}
\mbox{
\subfigure[isosurfaces]{\includegraphics[width=0.3\linewidth]{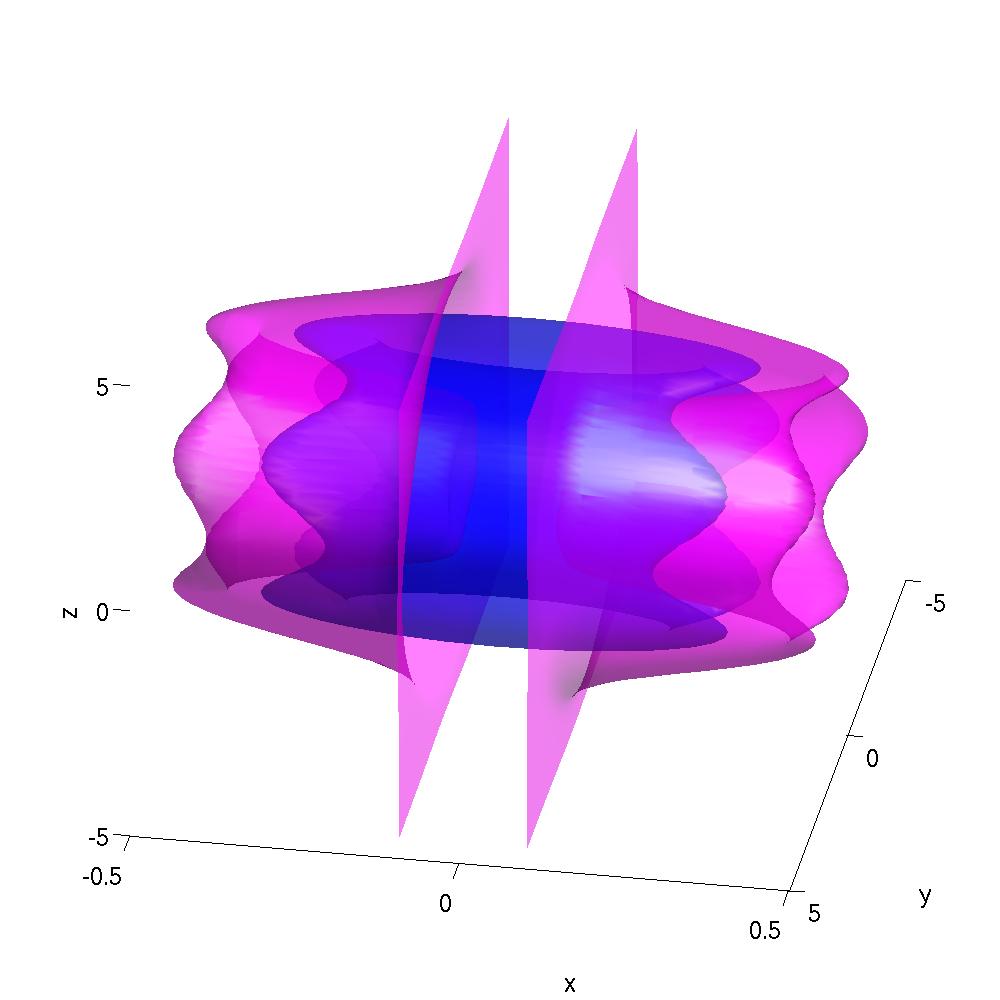}}
\subfigure[energy
    density]{\includegraphics[width=0.3\linewidth]{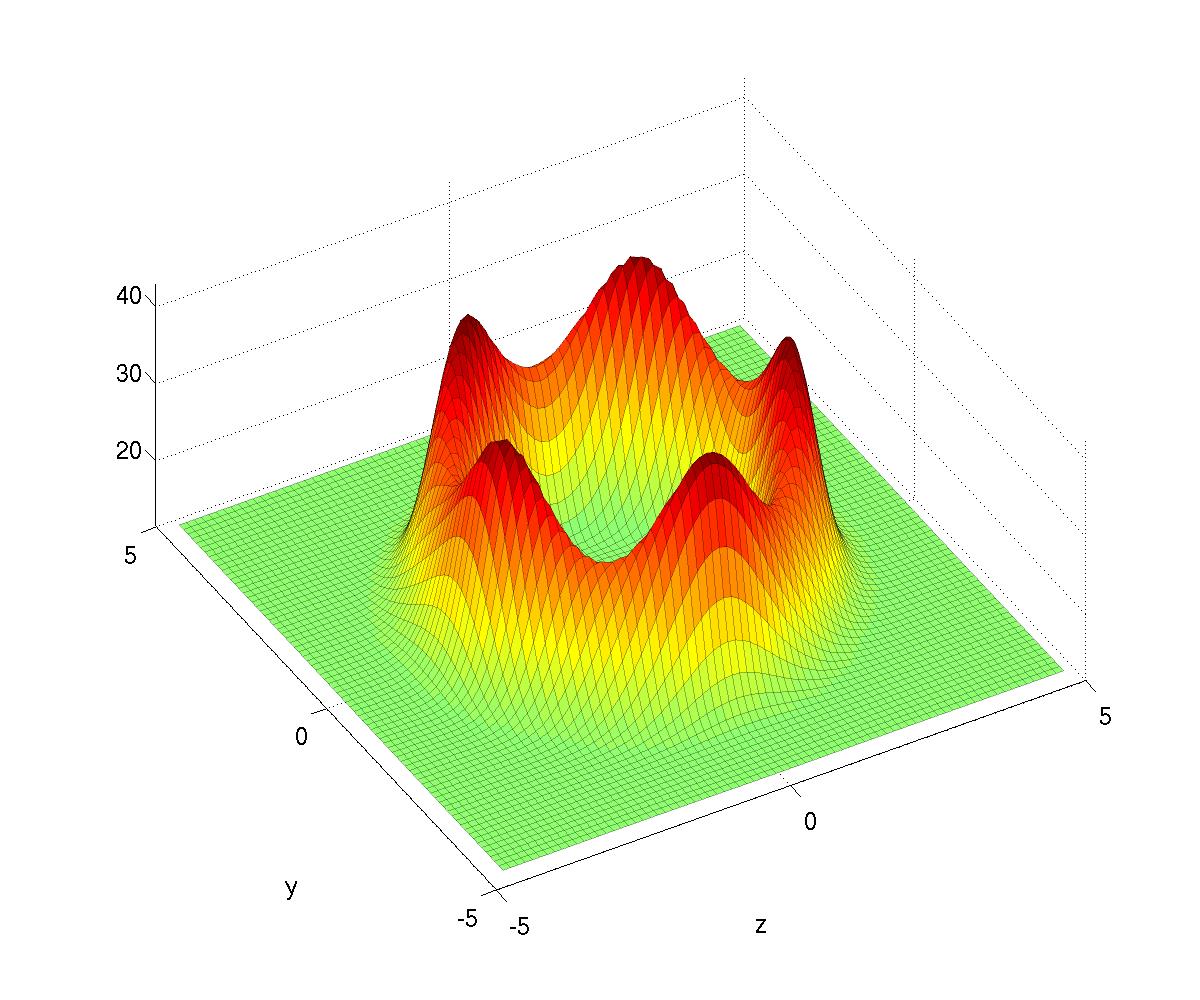}}
\subfigure[baryon charge
  density]{\includegraphics[width=0.3\linewidth]{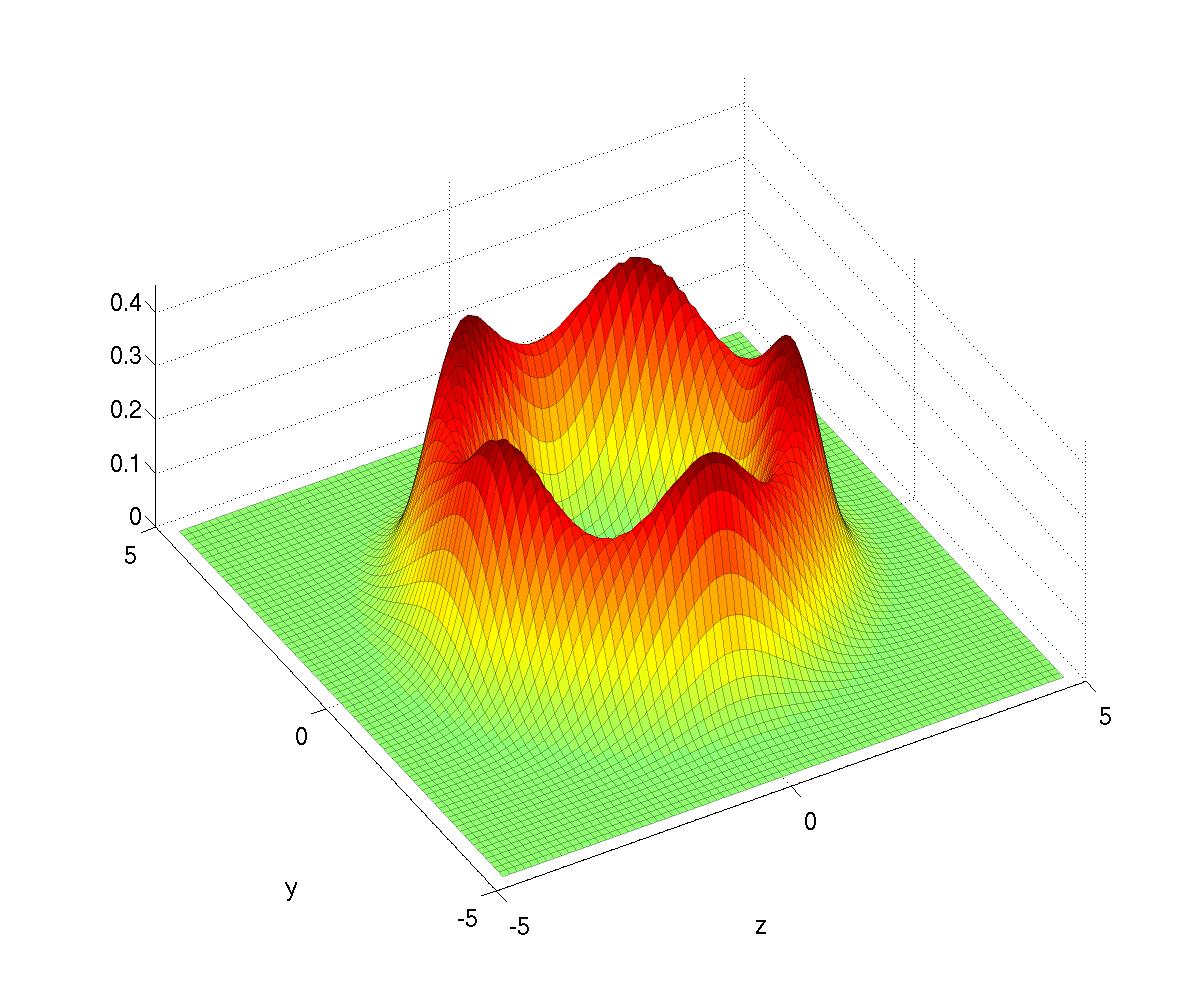}}}
\caption{The domain wall with $5$ sine-Gordon kinks living on a trapped
  $B=5$ Skyrmion. For details see fig.~\ref{fig:DW_DL_S}. The
  calculation is done on an $81^3$ cubic lattice and the potential used
  is \eqref{eq:quadratic_kink} with $m_4=4,m_3=2,m_2=1$.}  
\label{fig:DW_5SQ}
\end{center}
\end{figure}

After being in the right mind-set, we can obtain the numerical
solutions, which we again find using the generic field 
$\mathbf{n}(x,y,z)$, subject to the boundary condition \eqref{eq:DBCx}
and Neumann conditions in the other two directions. The solutions are
shown in figs.~\ref{fig:DW_2-4SQ} and \ref{fig:DW_5SQ}.

\begin{figure}[!htp]
\begin{center}
\mbox{
\subfigure[isosurfaces]{\includegraphics[width=0.3\linewidth]{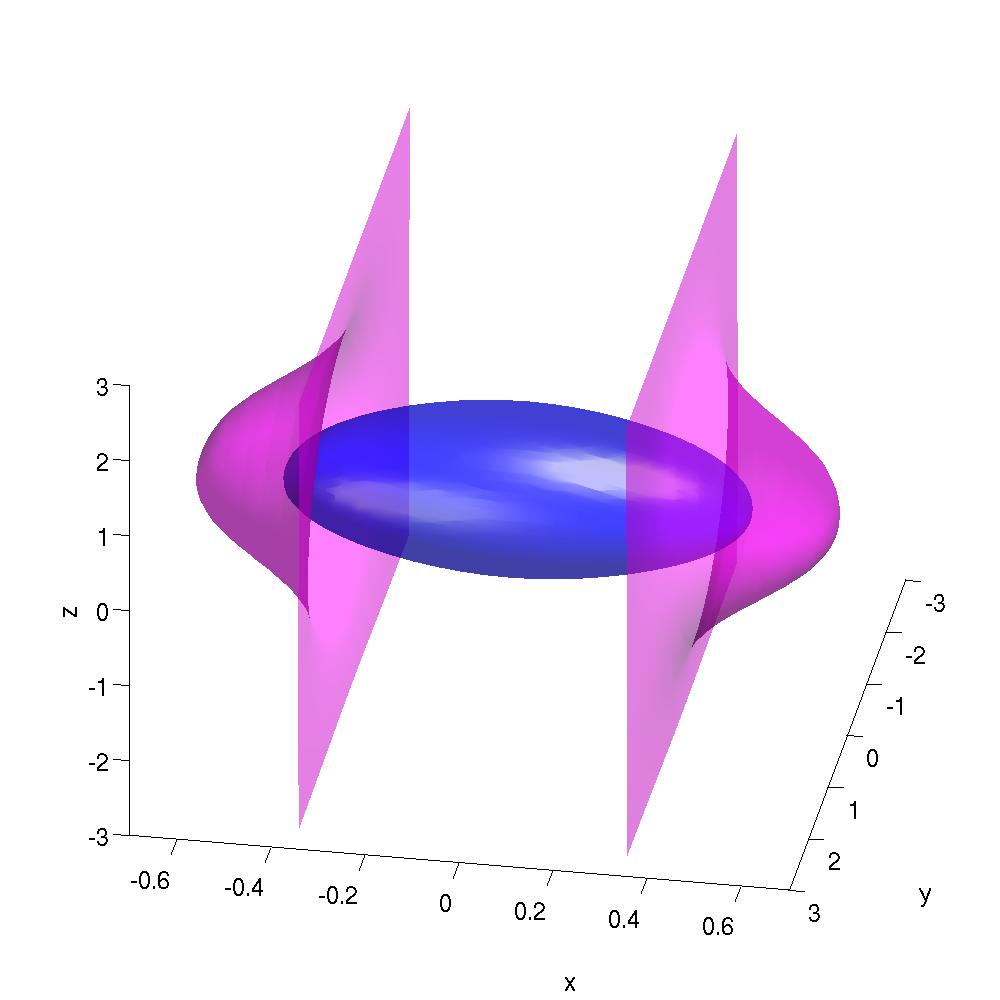}}
\subfigure[energy
    density]{\includegraphics[width=0.3\linewidth]{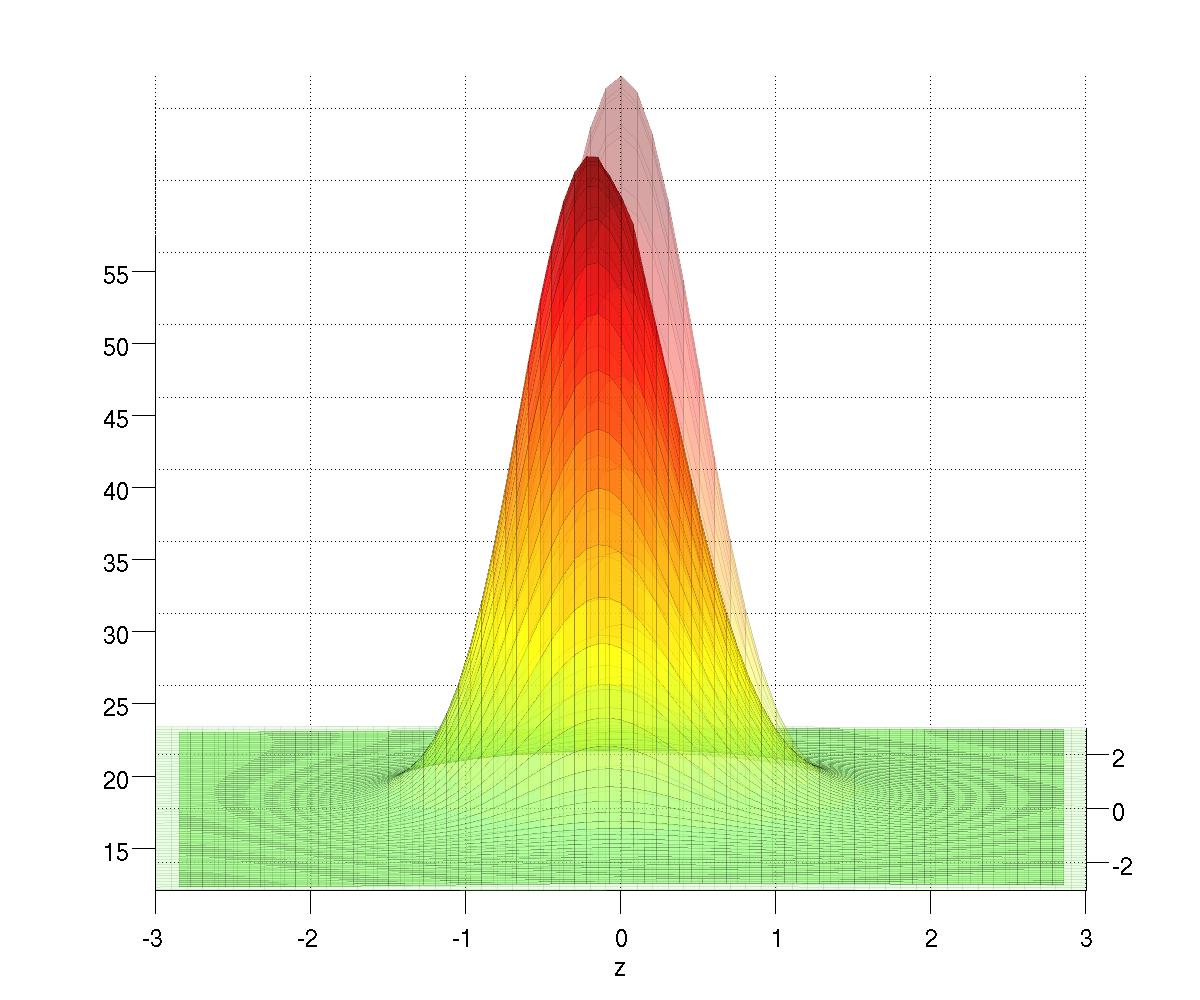}}
\subfigure[baryon charge
  density]{\includegraphics[width=0.3\linewidth]{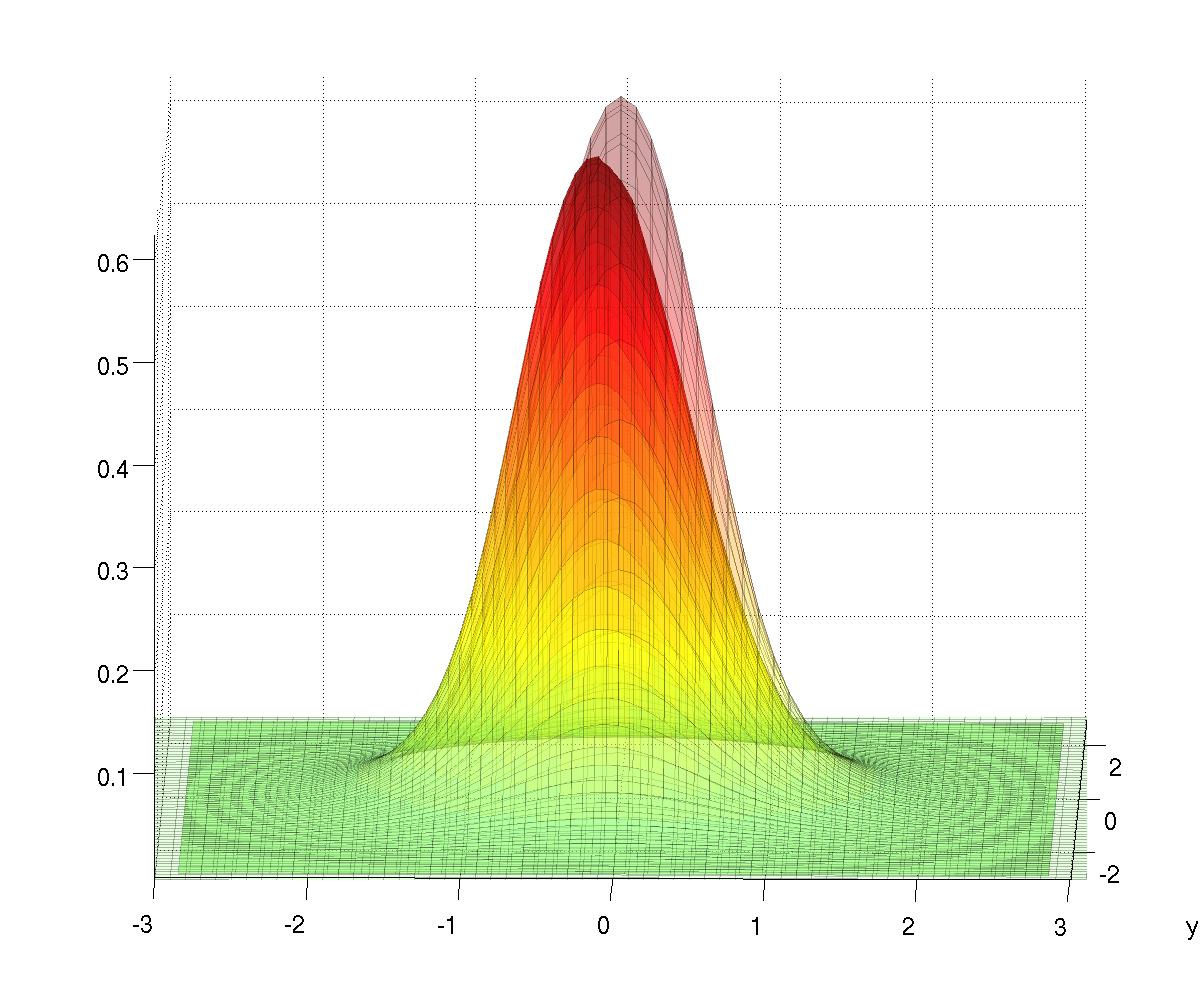}}}
\caption{The domain wall with a trapped Skyrmion in a quadratic
  potential (with a kink mass). 
  The energy density and baryon charge density on the middle of the
  wall ($x=0$) the Skyrmion in the potential \eqref{eq:quadratic_kink}
  are shown with those of fig.~\ref{fig:DW_SQ} as a reference in a
  transparent shade. The calculation is done on an $81^3$ cubic
  lattice and $m_4=4,m_3=2,m_2=1$.} 
\label{fig:skring_comparison}
\end{center}
\end{figure}

For completeness we obtain also the single Skyrmion in the potential
\eqref{eq:quadratic_kink}. The linear mass term for $n_2$ skews the
Skyrmion configuration. For comparison we show the energy density with
that of fig.~\ref{fig:DW_SQ} being a transparent shade in the
background, see fig.~\ref{fig:skring_comparison}.

\begin{table}[!ht]
\begin{center}
\caption{Baryon charge and energy of the baby-Skyrmion in the
  quadratic potential including the energy of the kinks living on the
  Skyrmion-ring (the energy of the domain wall is subtracted from the 
  total energy). }
\label{tab:b/e_skq}
\begin{tabular}{l|lllll}
\multicolumn{5}{c}{$m_2=1$ (with kinks)}\\
\hline\hline
$B$ & 1 & 2 & 3 & 4 & 5\\
\hline
$B^{\rm numerical}$ & 0.985 & 1.996 & 2.991 & 3.992 & 4.992 \\
$E/B$ & 74.11 & $69.32\pm0.06$ & $68.05\pm0.03$ & 67.79 & 67.66 \\[10pt]
\multicolumn{5}{c}{$m_2=0$ (without kinks)}\\
\hline\hline
$B$ & 1 & 2 & 3 & 4 & 5 \\
\hline
$B^{\rm numerical}$ & 0.998 & 1.995 & 2.990 & 3.992 & 4.992 \\
$E/B$ & $76.18\pm0.12$ & 69.28 & 67.93 & 67.61 & 67.47
\end{tabular}
\end{center}
\end{table}

In table \ref{tab:b/e_skq} is shown the energy of the baby-Skyrmion, 
which is calculated by subtracting the wall contribution from the
total energy. All the ring-like solutions we have found are stable
(i.e.~minimum energy configurations).

\section{Summary and Discussion\label{sec:discussion}}

In this paper we have studied domain walls with entrapped Skyrmions in
various guises. The first ones are full and half-charged Skyrmions
manifested as kinks on domain lines carrying 3-dimensional (full and
half) Skyrmion charge. 
This line of systems is interesting because it is not dependent on
having a Skyrme term and thus might find applications in
condensed-matter systems. 
The next kind of configurations is a domain wall with endowed
baby-Skyrmions carrying again full Skyrmion (baryon) charge. This type
of configuration needs both the Skyrme term as well as a mass
term. The mass term can be of any type, which from the point of view
of the pion is unimportant. From a topological point of view, however,
it matters. The two potentials under study in this paper are the
traditional linear potential and the quadratic potential which allows
for a domain wall. In the planar baby-Skyrmion model it is known that 
the linear potential does not allow for stable $B>1$ configurations
other than the lattice whilst the quadratic potential equips the model
with stable ring-like configurations \cite{Eslami:2000tj}. 

For the quadratic potential we find as in the planar case, that the
higher-winding baby-Skyrmions resident on the domain wall are indeed
stable for the baryon numbers we have investigated, i.e.~$B=1-5$. 
However, for the linear potential we find ourselves with a surprise;
the stable multi-charged configuration is made of charge 2 and 3
ring-like configurations. We conjecture that the large charge
configuration eventually becomes a baby-Skyrmion lattice composed of
charge-2 ring-like object as depicted in fig.~\ref{fig:DW_2_2RINGS}. 

Another comment in store is due the backreaction of the soliton living
on the domain wall. In the case of the Skyrmion and half-Skyrmion
trapped on the domain line living on the domain wall, we observe a
drop in the energy density on both sides of the peak. We interpret
this as a binding energy of the Skyrmion. 

We have seen in this paper that stable baby-Skyrmion configurations 
in baby-Skyrme theories induced on a domain wall are quite different 
for the linear potential $-n_3$ compared to those of the corresponding
planar theory, whereas for the quadratic potential $-n_3^2$, the
induced theories and their corresponding planar theories are
qualitatively alike.
Structures and dynamics of baby-Skyrmions in the planar
$(2+1)$-dimensional case with other choices of potentials have been
studied thus far in
e.g.~refs.~\cite{Jaykka:2010bq,Kobayashi:2013aja,Jaykka:2011ic}. 
Studying stable configurations in these models induced on a domain
wall will be interesting to explore. 
In particular, the potential $-n_3^2$ will be the most interesting one 
because it is quite common in condensed-matter systems and admits
molecule-type structures \cite{Jaykka:2010bq,Kobayashi:2013aja}.

Another interesting possibility which we have left for future work is
to consider a compactification of the domain wall to a sphere along
the lines of ref.~\cite{Gudnason:2013qba}, which is possible by
changing the sign of the Skyrme term and augment the theory by a
sextic derivative term; in this case the domain line with sine-Gordon
kinks can still exist.

\section*{Acknowledgments}

The work of MN is supported in part by Grant-in-Aid for Scientific
Research (No. 25400268) and by the ``Topological Quantum Phenomena'' 
Grant-in-Aid for Scientific Research on Innovative Areas (No. 25103720)  
from the Ministry of Education, Culture, Sports, Science and Technology 
(MEXT) of Japan. 
MN thanks for the warm hospitality at Nordita during his stay where
part of the work was carried out. 

\appendix
\section{Numerical relaxation}\label{app:lattice}

We use a rather simple relaxation on a cubic lattice with an imaginary
time that cools the configuration down to a state where the equations
of motion are satisfied and the topological charge contained is
measured as a check. The configurations are calculated on $129^3$ (or
$81^3$) 
spatial lattices with a second order finite difference method and
cooled in the imaginary time direction with a forward time step (FTS)
method and time step of size $\lesssim (0.3-0.5){\rm
  min}(h_x^4,h_y^4,h_z^4)$. This choice was made for the simplicity in
the implementation. Our code is implemented in C++.

\section{Skyrmion and half-Skyrmion on a domain line on the domain
  wall without the Skyrme term\label{app:normal_kinetic}}

\begin{figure}[!hpt]
\begin{center}
\mbox{\subfigure[isosurfaces]{\includegraphics[width=0.3\linewidth]{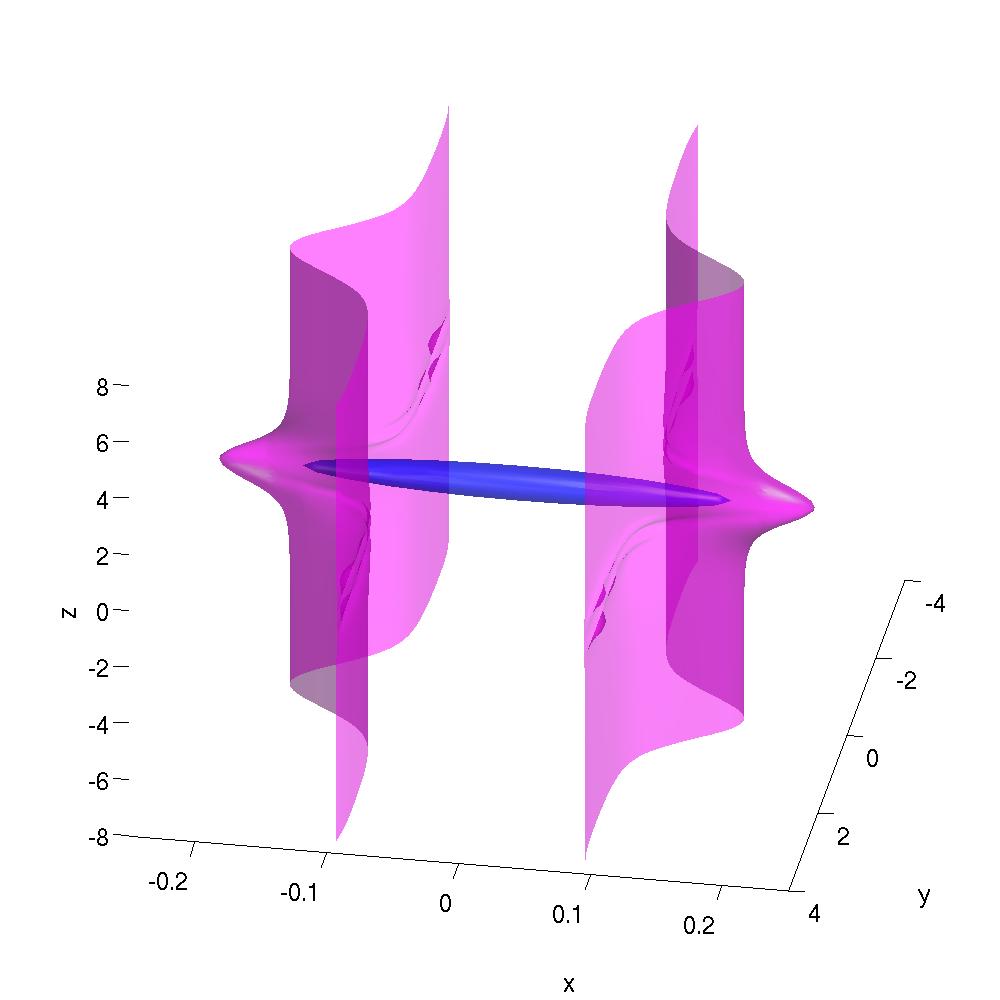}}
\subfigure[energy density]{\includegraphics[width=0.3\linewidth]{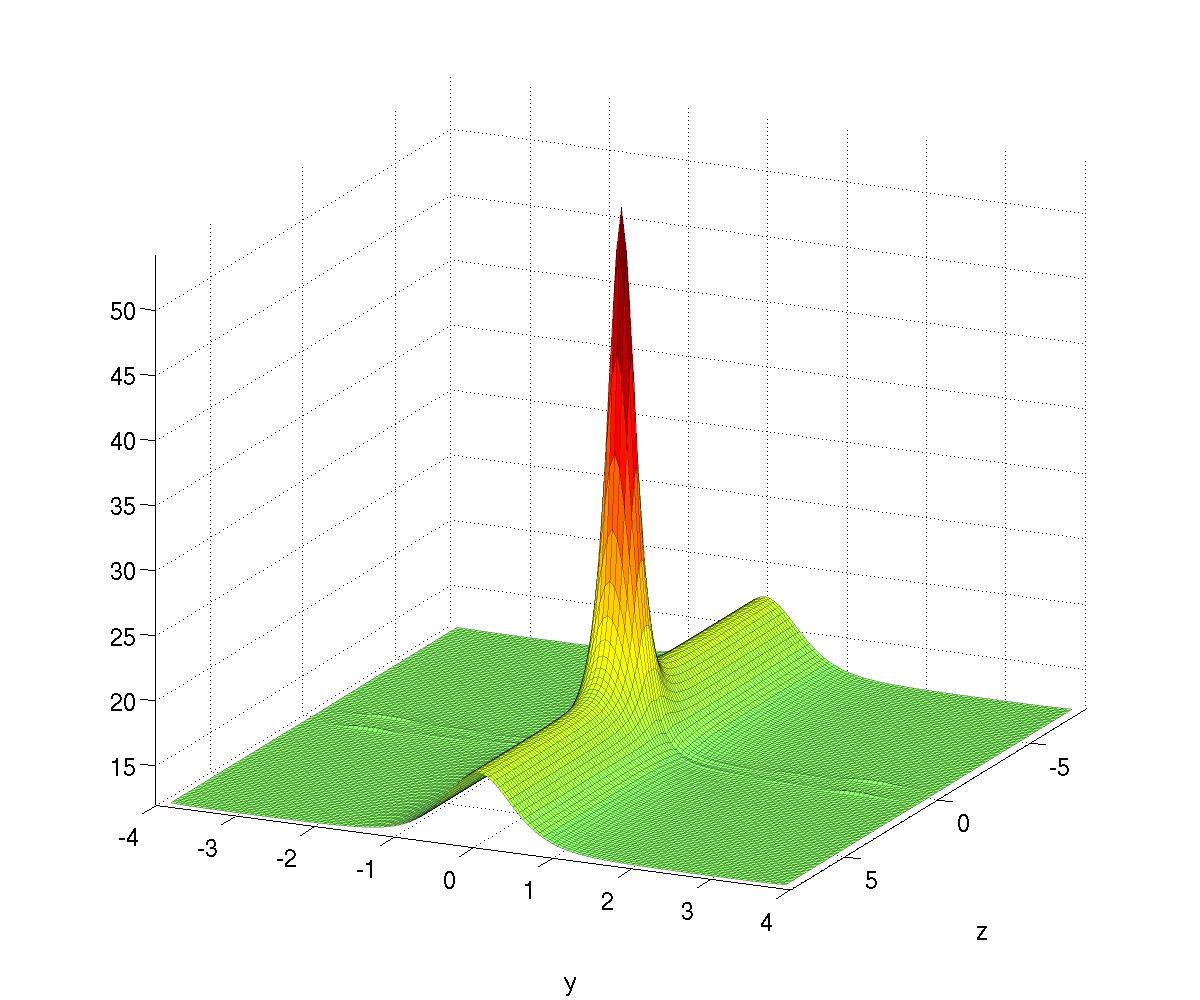}}
\subfigure[baryon charge density]{\includegraphics[width=0.3\linewidth]{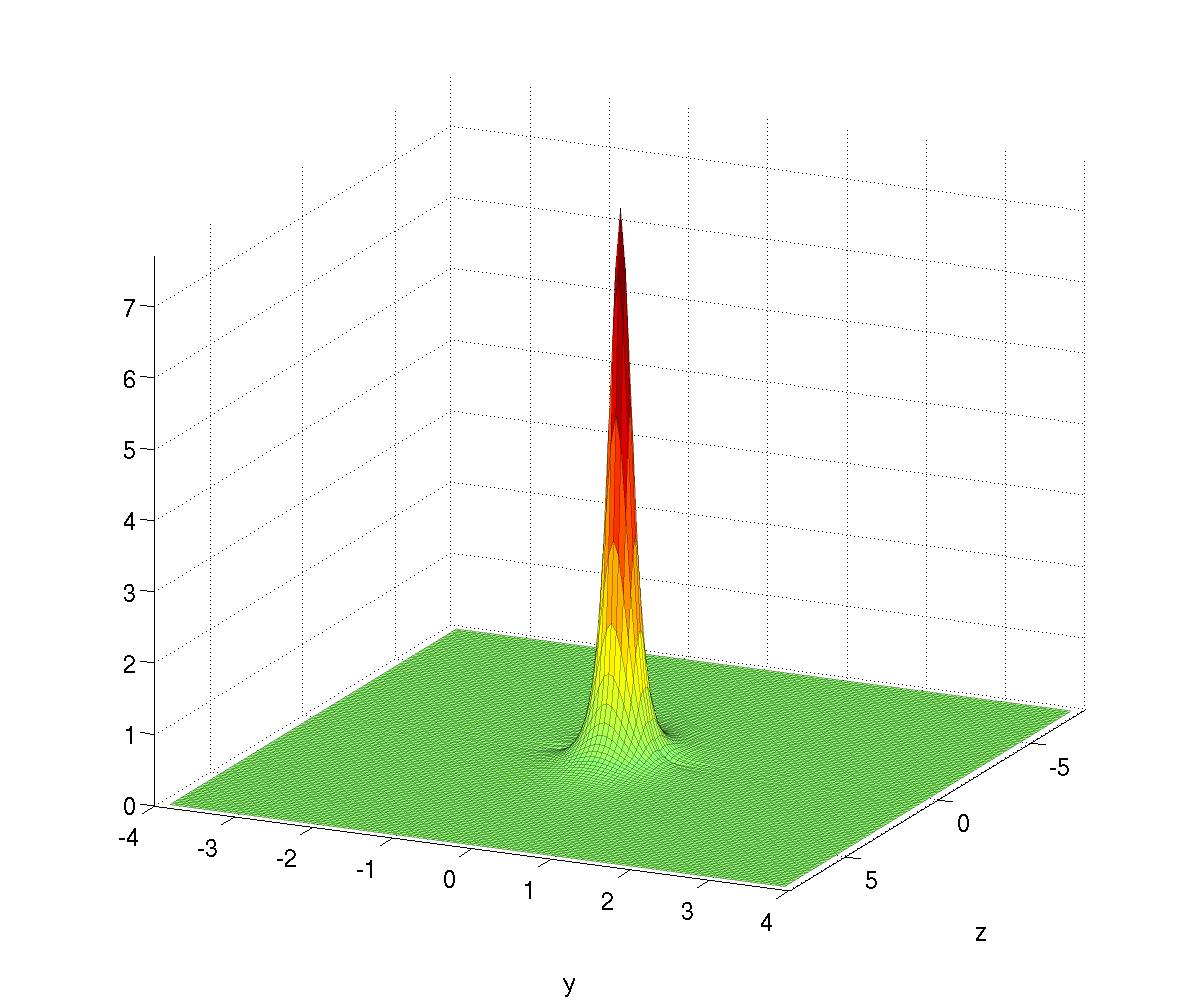}}
}
\caption{The domain wall with a domain line on which a Skyrmion
  resides in the theory without the Skyrme term. The calculation is
  done on a $129^3$ cubic lattice, $B^{\rm numerical}=0.9995$ and the
  potential used is \eqref{eq:VfullSkyrmion} with
  $m_4=4,m_3=2,m_2=1$. }
\label{fig:DW_DL_S_NORMAL}
\end{center}
\end{figure}

\begin{figure}[!hpt]
\begin{center}
\mbox{\subfigure[isosurfaces]{\includegraphics[width=0.3\linewidth]{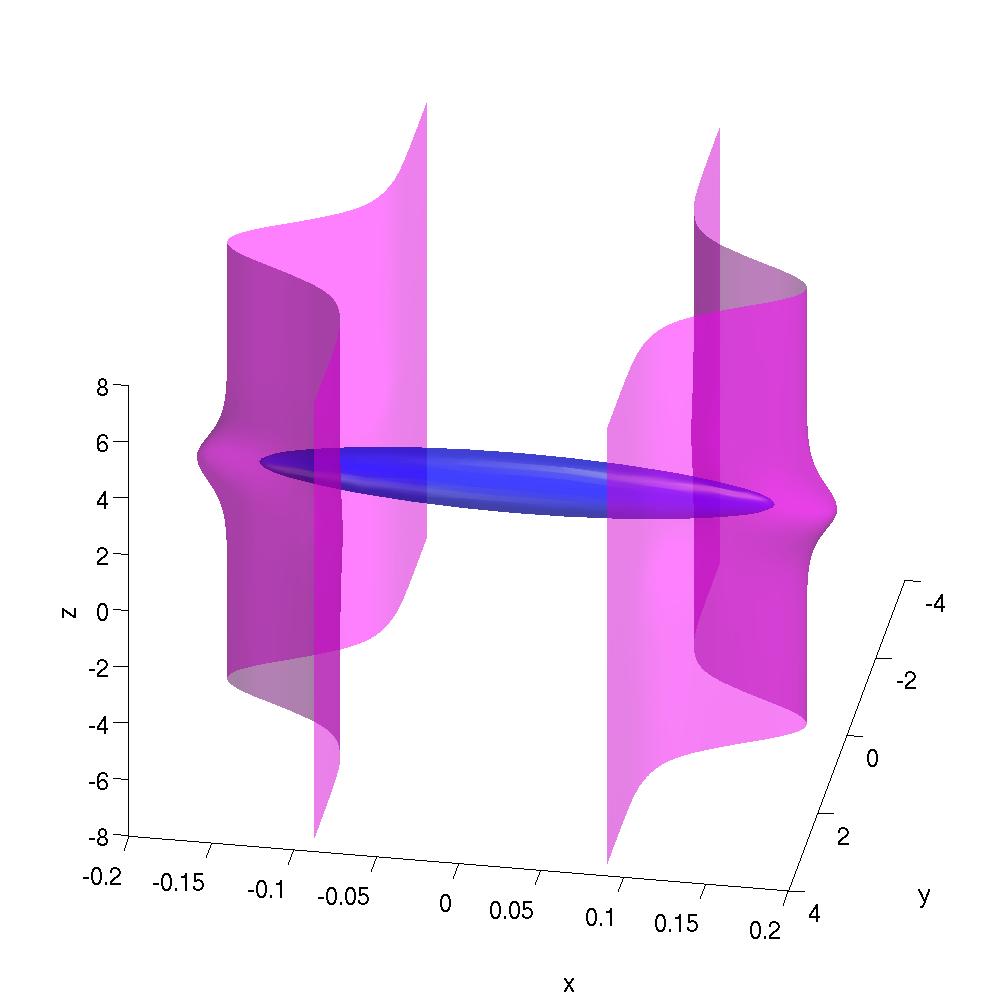}}
\subfigure[energy density]{\includegraphics[width=0.3\linewidth]{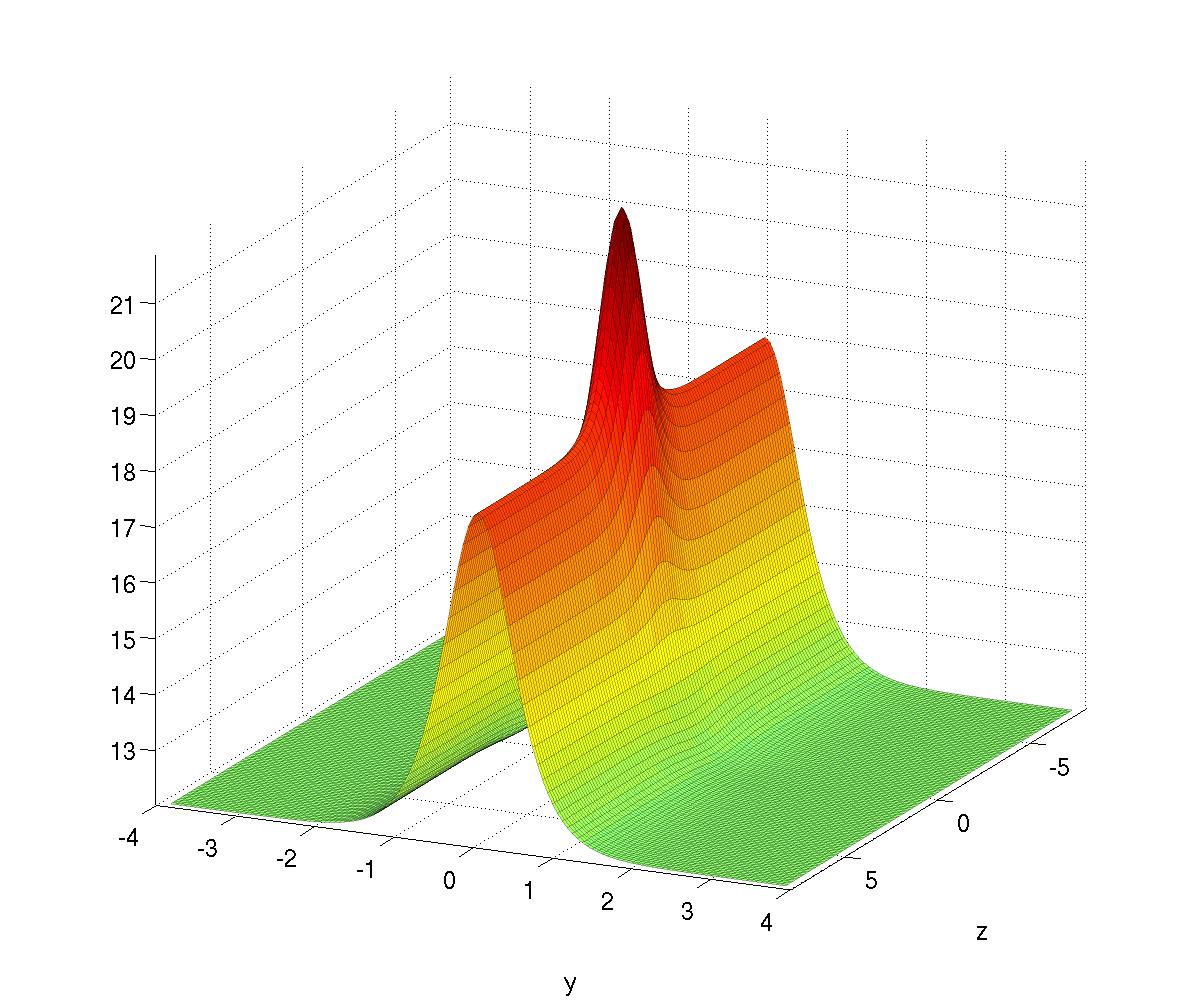}}
\subfigure[baryon charge density]{\includegraphics[width=0.3\linewidth]{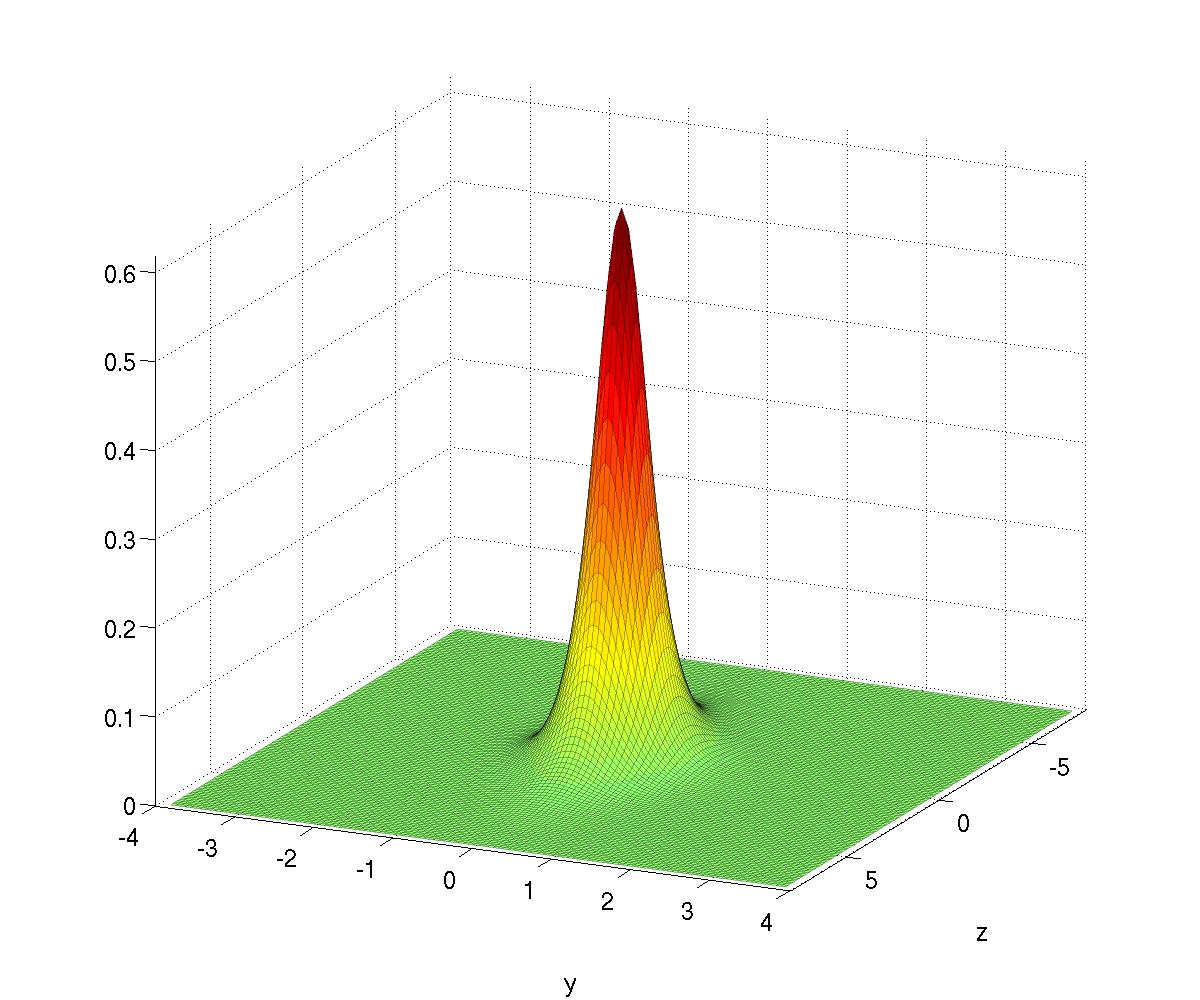}}
}
\caption{The domain wall with a domain line on which half a Skyrmion
  resides. The calculation is done on a $129^3$ cubic lattice, 
  $B^{\rm numerical}=0.49993$ and the potential used is
  \eqref{eq:VhalfSkyrmion} with $m_4=4,m_3=2,m_2=1$. }
\label{fig:DW_DL_HS_NORMAL}
\end{center}
\end{figure}

In figs.~\ref{fig:DW_DL_S_NORMAL} and \ref{fig:DW_DL_HS_NORMAL} are
shown the configurations containing the domain wall with a domain line
endowing a full Skyrmion or a half-Skyrmion, respectively. The figures
are shown for the purpose of possible interest in the configurations
not enjoying the Skyrme term, for instance in condensed-matter
physics. We observe a pair of ``lines'' coming out from the Skyrmion
in fig.~\ref{fig:DW_DL_S_NORMAL}, but we do not have an explanation
for their presence. 

\section{Asymmetric initial condition for a double-winding Skyrmion in
  a linear potential\label{app:relaxation_sequence}}

\begin{figure}[!htp]
\begin{center}
\mbox{\subfigure{\includegraphics[width=0.22\linewidth]{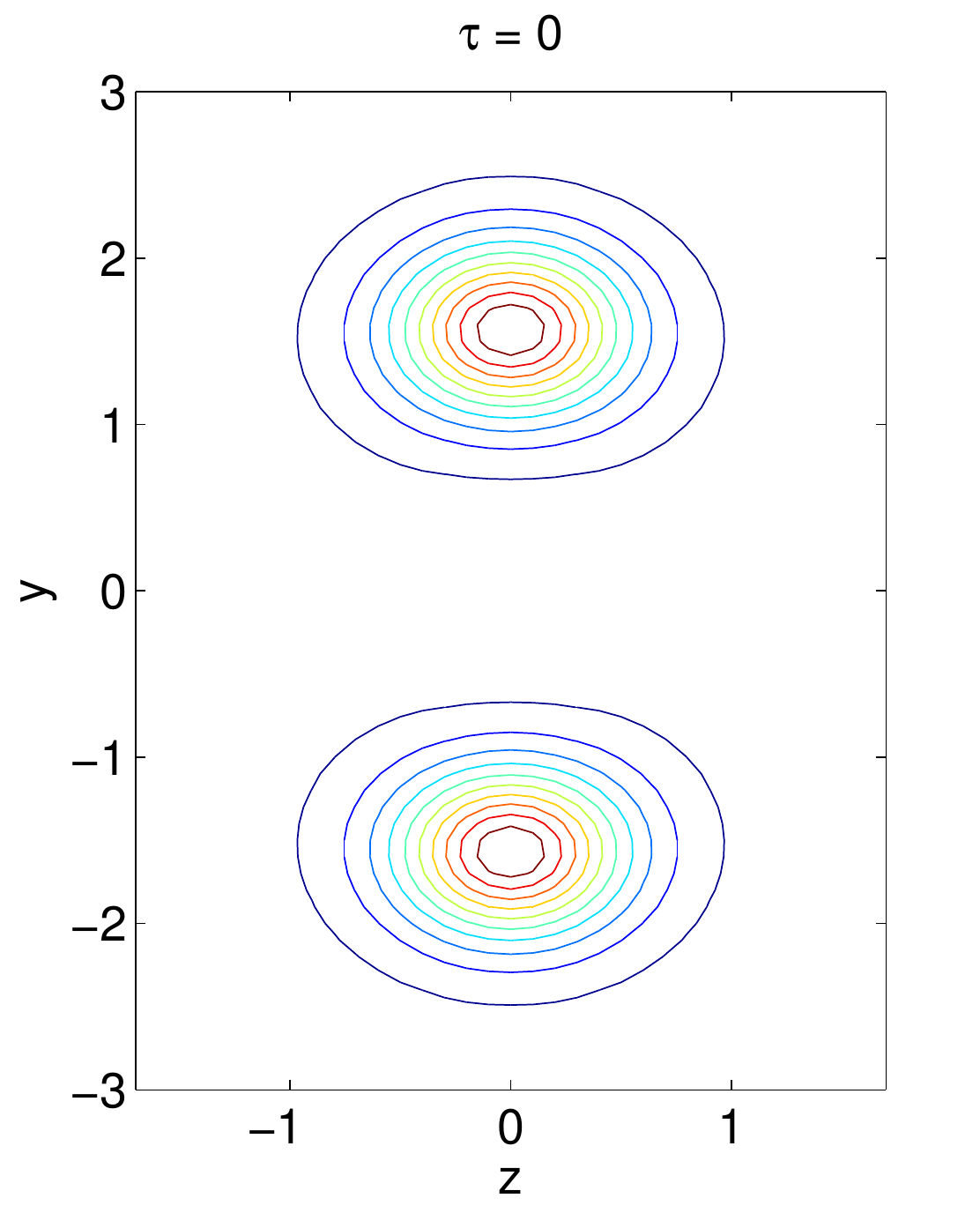}}
\subfigure{\includegraphics[width=0.22\linewidth]{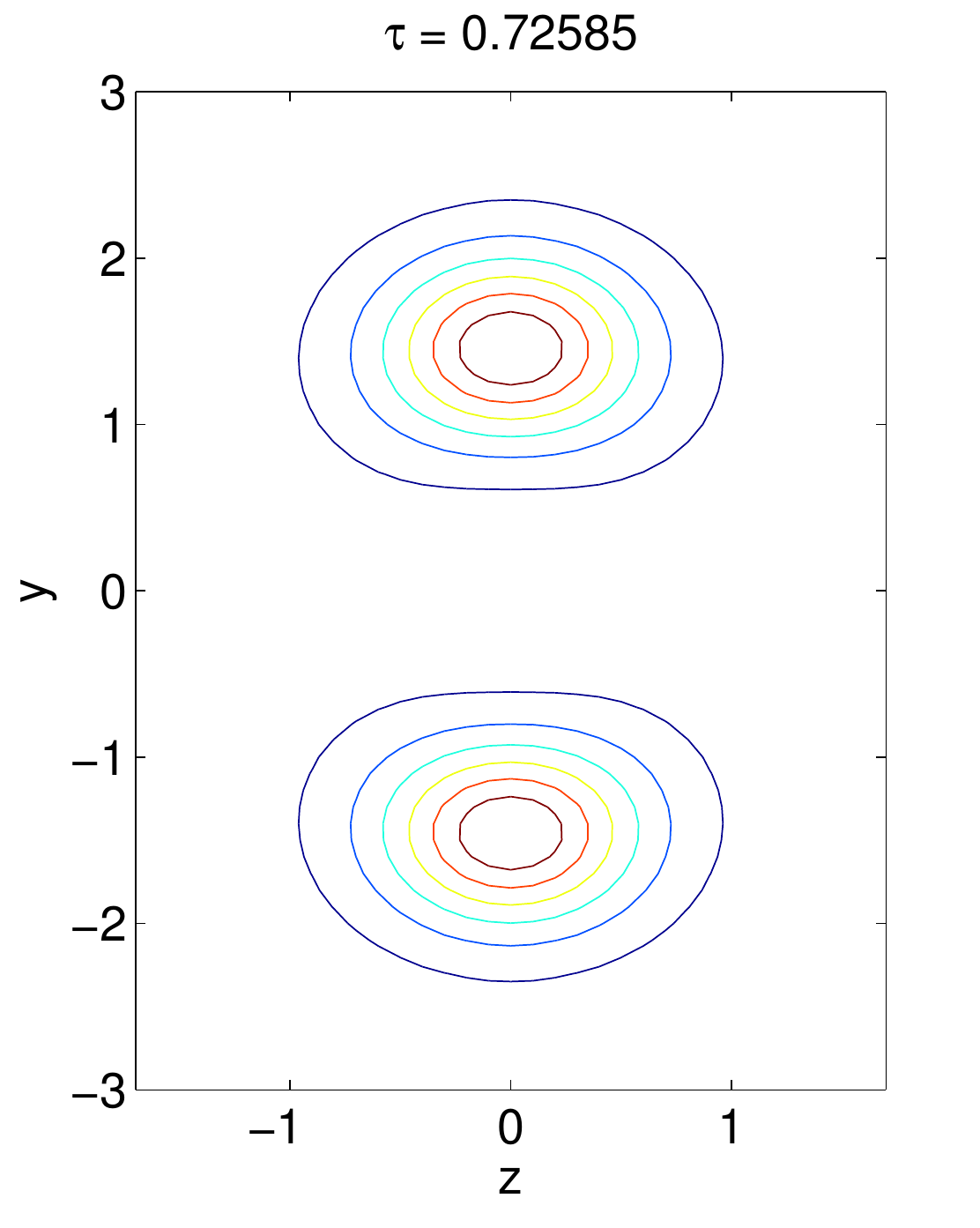}}
\subfigure{\includegraphics[width=0.22\linewidth]{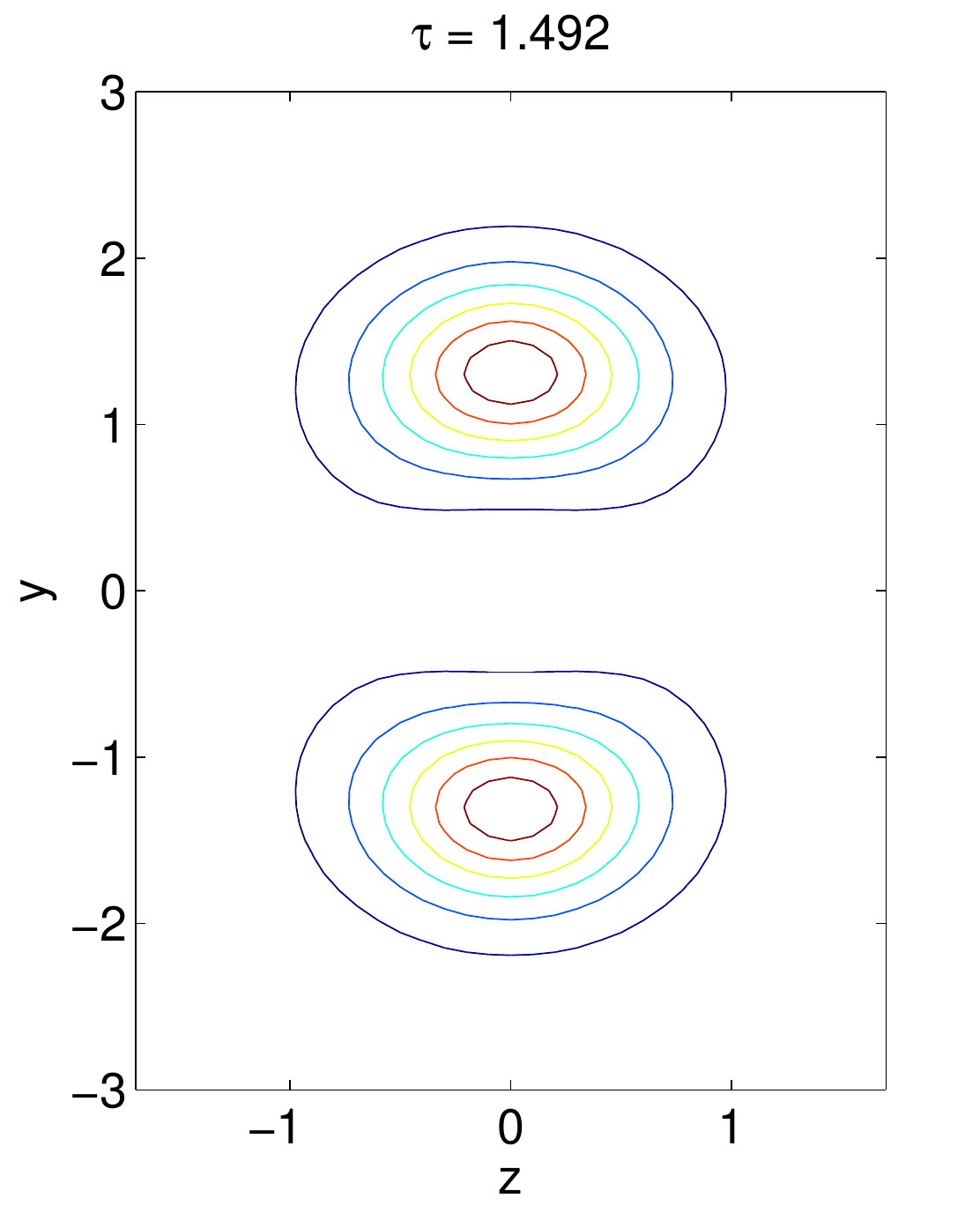}}
\subfigure{\includegraphics[width=0.22\linewidth]{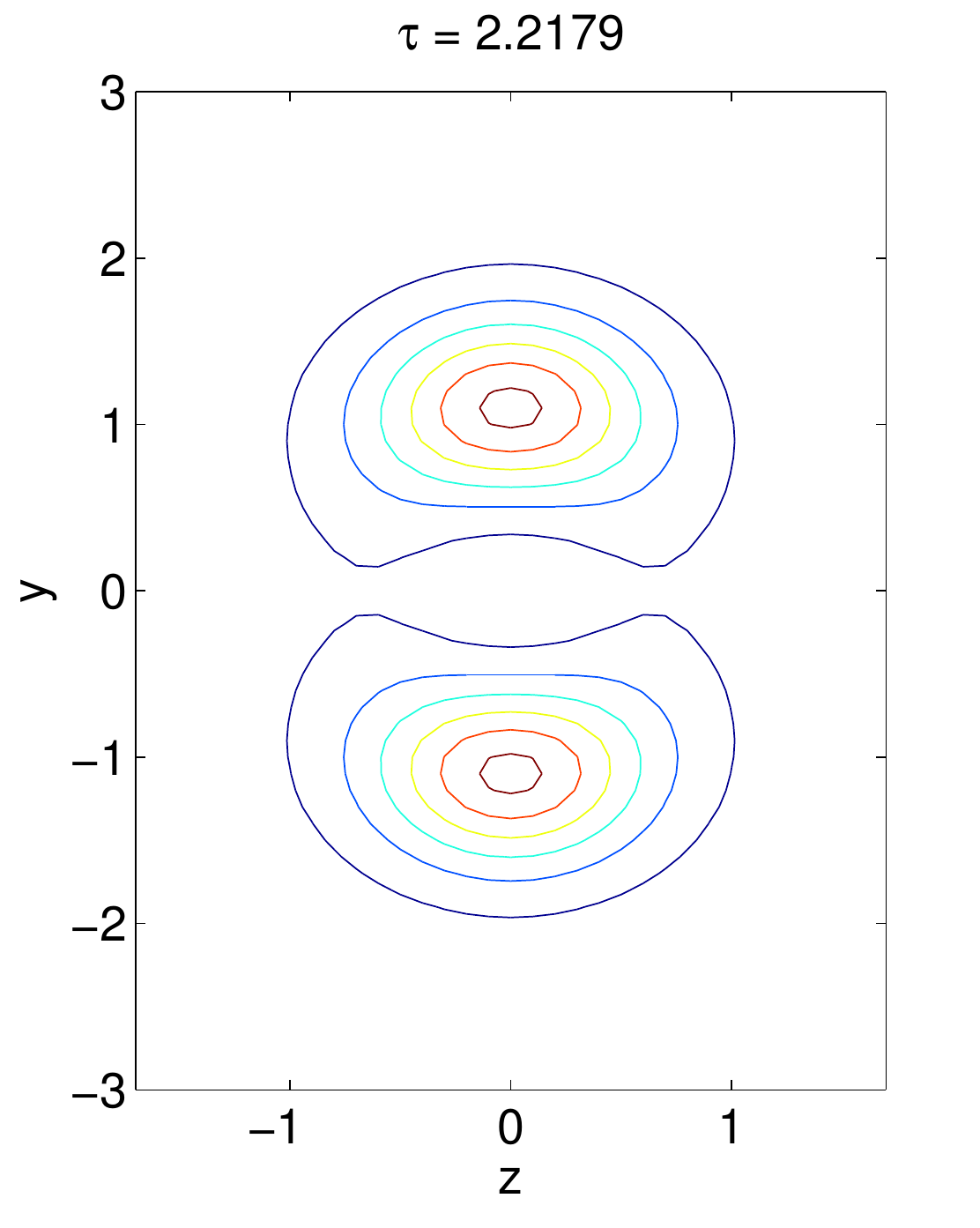}}}
\mbox{\subfigure{\includegraphics[width=0.22\linewidth]{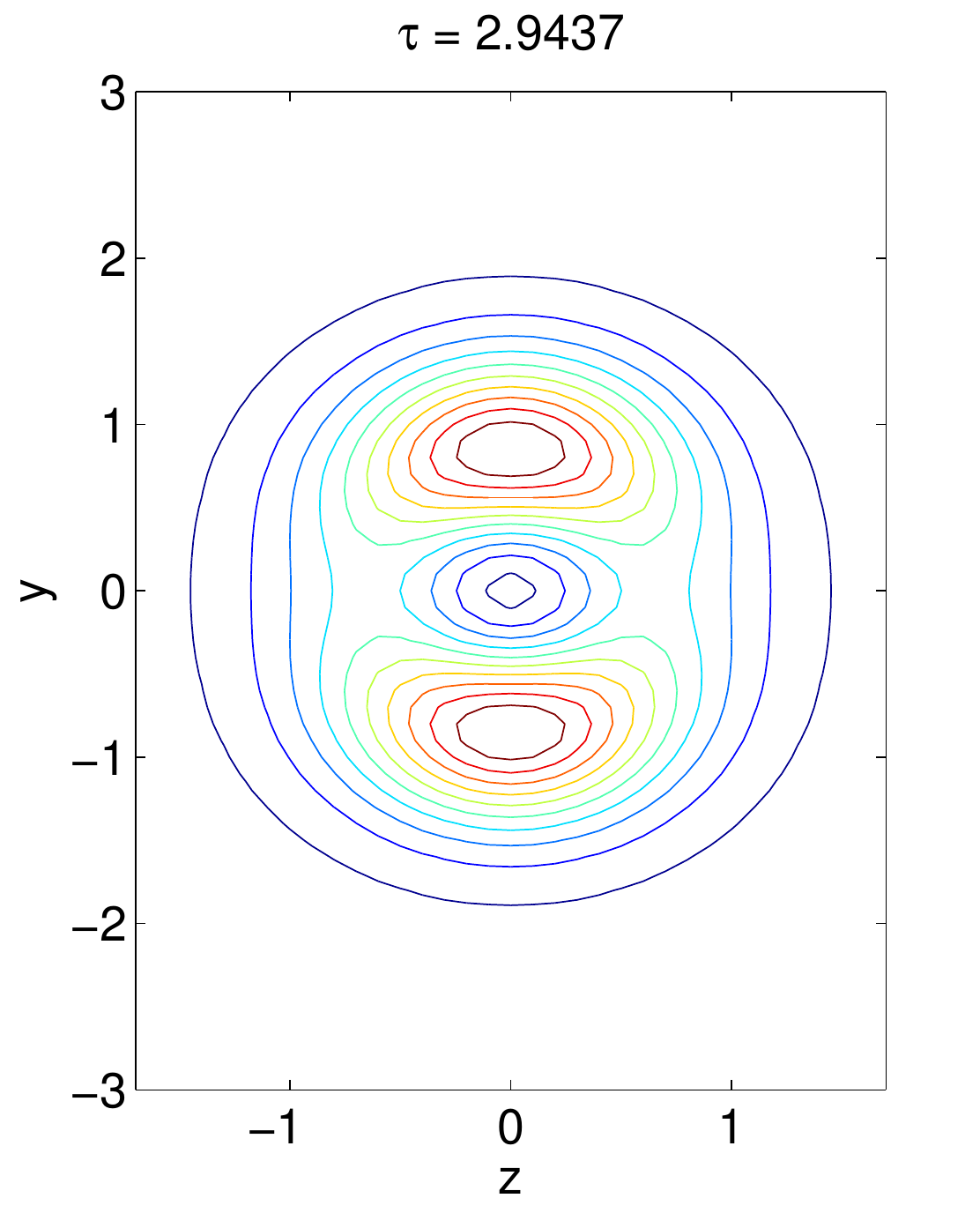}}
\subfigure{\includegraphics[width=0.22\linewidth]{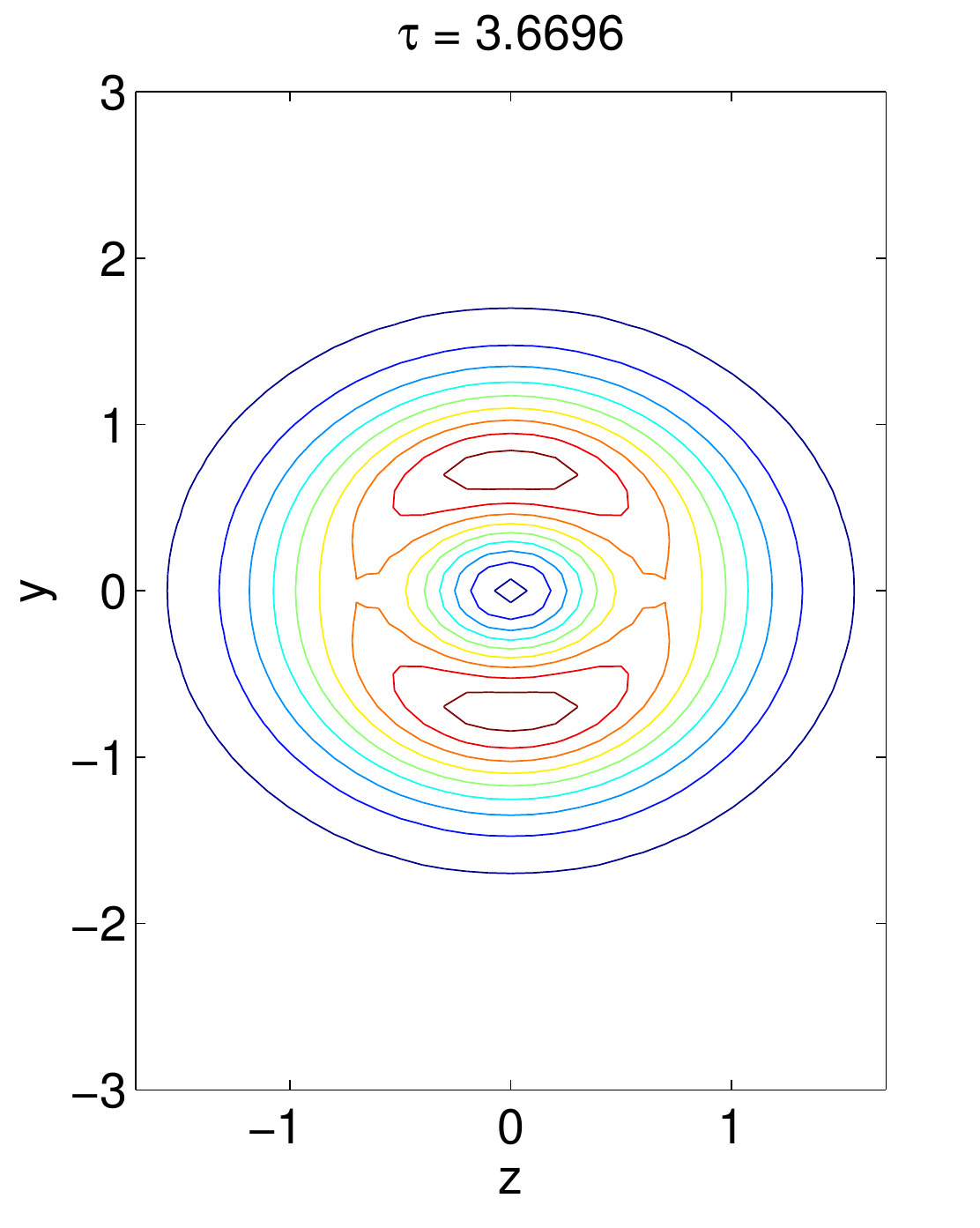}}
\subfigure{\includegraphics[width=0.22\linewidth]{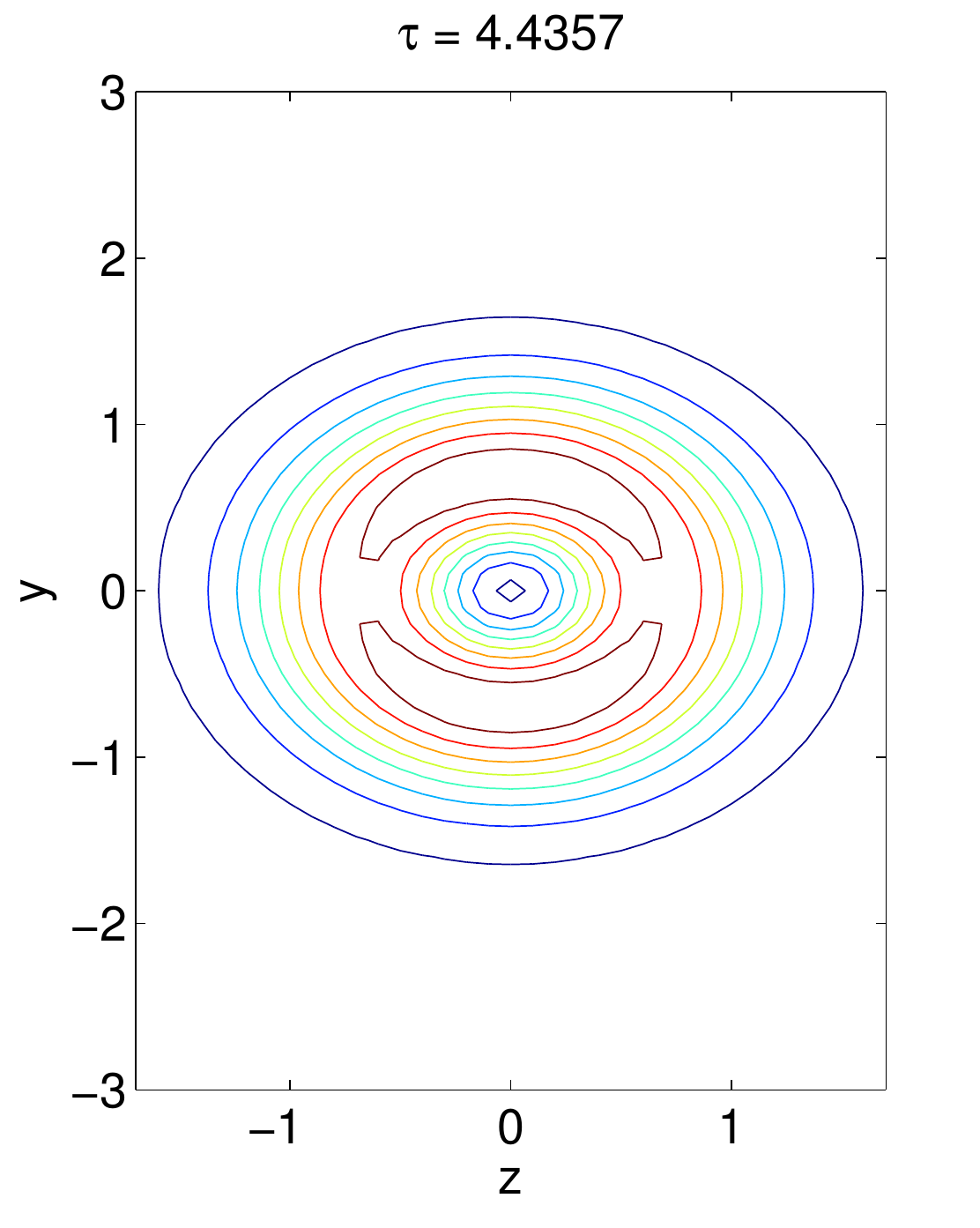}}
\subfigure{\includegraphics[width=0.22\linewidth]{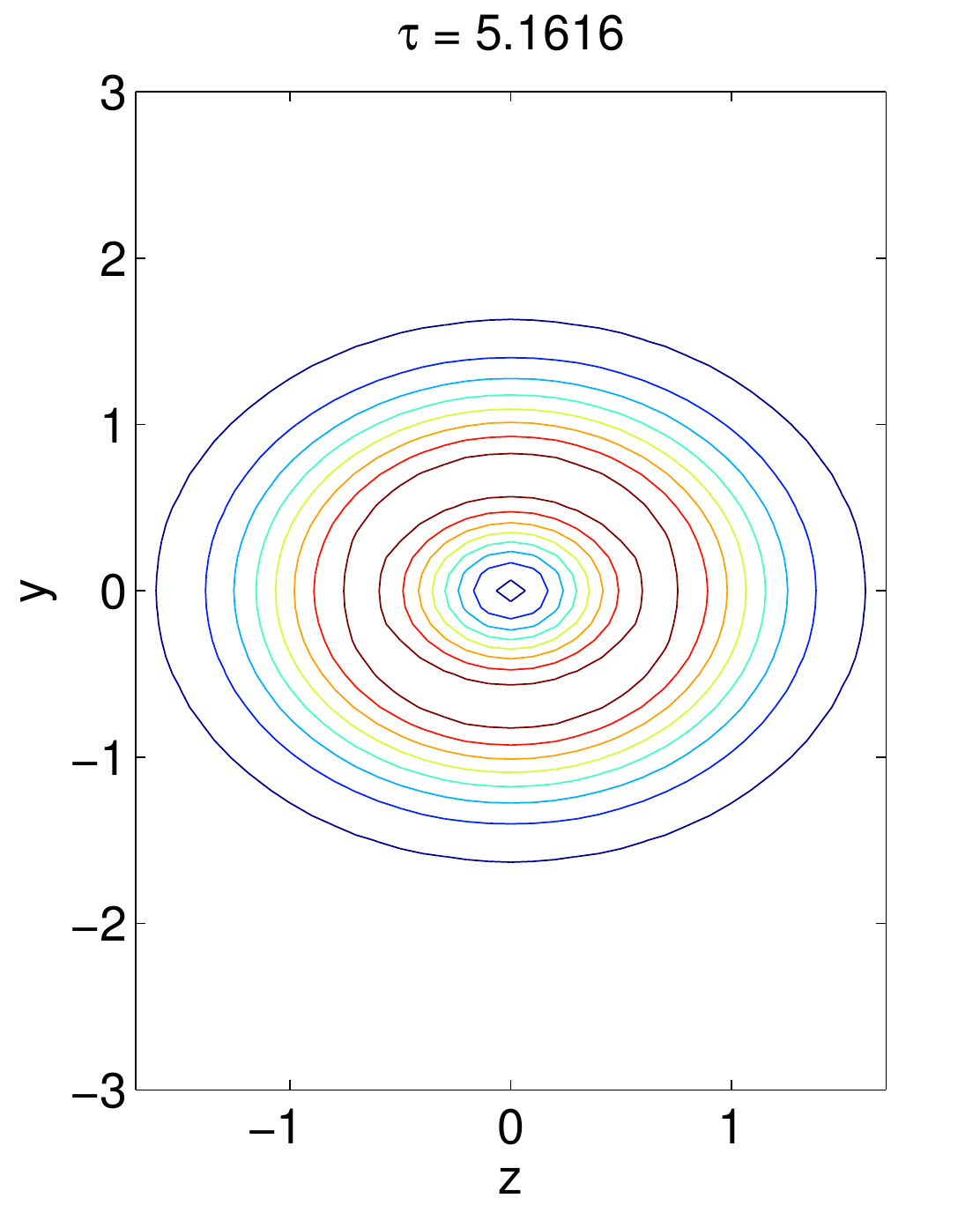}}}
\caption{Contours of constant energy in the $yz$-plane of the
  double-winding baby-Skyrmion on a wall (at $x=0$) in a linear 
  potential as function of relaxation time $\tau$. It is observed that
  the two separated baby-Skyrmions attract each other and find a
  minimum of the energy in a ring-like solution. This calculation was
  done on an $81^3$ cubic lattice and the potential used is
  \eqref{eq:Vlinear_babysk} with $m_4=4,m_3=2$. }
\label{fig:relaxation}
\end{center}
\end{figure}

In fig.~\ref{fig:relaxation} we show a series of energy contours of
the configuration in the middle of the domain wall (at $x=0$) as
function of relaxation time $\tau$ until the equations of motion are
satisfied to the required accuracy level. We provide this series of
configurations in order to show that the initial guess is not a
symmetric state that relaxes to a symmetric configuration. We start
off with a very asymmetric state and find a ring-like solution. 

\begin{thebibliography}{99}

\bibitem{Skyrme:1961vq} 
  T.~H.~R.~Skyrme,
  ``A Nonlinear field theory,''
  Proc.\ Roy.\ Soc.\ Lond.\ A {\bf 260}, 127 (1961);
  ``A Unified Field Theory of Mesons and Baryons,''
  Nucl.\ Phys.\  {\bf 31}, 556 (1962).

\bibitem{'tHooft:1973jz} 
  G.~'t Hooft,
  ``A Planar Diagram Theory for Strong Interactions,''
  Nucl.\ Phys.\ B {\bf 72}, 461 (1974);
  E.~Witten,
  ``Baryons in the 1/n Expansion,''
  Nucl.\ Phys.\ B {\bf 160}, 57 (1979);
  E.~Witten,
  ``Current Algebra, Baryons, and Quark Confinement,''
  Nucl.\ Phys.\ B {\bf 223}, 433 (1983).

\bibitem{Piette:1994ug} 
  B.~M.~A.~G.~Piette, B.~J.~Schroers and W.~J.~Zakrzewski,
  ``Multi - solitons in a two-dimensional Skyrme model,''
  Z.\ Phys.\ C {\bf 65}, 165 (1995)
  [hep-th/9406160].

\bibitem{Piette:1994mh}
   B.~M.~A.~Piette, B.~J.~Schroers and W.~J.~Zakrzewski,
   ``Dynamics of baby skyrmions,''
   Nucl.\ Phys.\  B {\bf 439}, 205 (1995)
  [arXiv:hep-ph/9410256].

\bibitem{Kudryavtsev:1997nw} 
  A.~E.~Kudryavtsev, B.~M.~A.~G.~Piette and W.~J.~Zakrzewski,
  ``Skyrmions and domain walls in (2+1)-dimensions,''
  Nonlinearity {\bf 11}, 783 (1998)
  [hep-th/9709187];

\bibitem{Weidig:1998ii} 
  T.~Weidig,
  ``The Baby skyrme models and their multiskyrmions,''
  Nonlinearity {\bf 12}, 1489 (1999)
  [hep-th/9811238].

\bibitem{Skyrme:1961vr} 
  T.~H.~R.~Skyrme,
  ``Particle states of a quantized meson field,''
  Proc.\ Roy.\ Soc.\ Lond.\ A {\bf 262}, 237 (1961).

\bibitem{Derrick:1964ww} 
  G.~H.~Derrick,
  ``Comments on nonlinear wave equations as models for elementary
  particles,''  J.\ Math.\ Phys.\  {\bf 5}, 1252 (1964).  

 \bibitem{Nitta:2012xq} 
   M.~Nitta,
   ``Josephson vortices and the Atiyah-Manton construction,''
   Phys.\ Rev.\ D {\bf 86}, 125004 (2012)
   [arXiv:1207.6958 [hep-th]].

\bibitem{Nitta:2012wi} 
   M.~Nitta,
   ``Correspondence between Skyrmions in 2+1 and 3+1 Dimensions,''
   Phys.\ Rev.\ D {\bf 87}, 025013 (2013)
   [arXiv:1210.2233 [hep-th]].

\bibitem{Nitta:2012rq} 
  M.~Nitta,
  ``Matryoshka Skyrmions,''
  Nucl.\ Phys.\ B {\bf 872}, 62 (2013)
  [arXiv:1211.4916 [hep-th]].

 \bibitem{Piette:1997ce} 
   B.~M.~A.~G.~Piette and W.~J.~Zakrzewski,
   ``Skyrmions and domain walls,''
   In *Kingston 1997, Solitons* 187-190
   [hep-th/9710011];
  A.~E.~Kudryavtsev, B.~M.~A.~G.~Piette and W.~J.~Zakrzewski,
  ``On the interactions of skyrmions with domain walls,''
  Phys.\ Rev.\ D {\bf 61}, 025016 (2000)
  [hep-th/9907197].

\bibitem{Jennings:2013aea} 
   P.~Jennings and P.~Sutcliffe,
   ``The dynamics of domain wall Skyrmions,''
   J.\ Phys.\ A {\bf 46}, 465401 (2013)
   [arXiv:1305.2869 [hep-th]].

\bibitem{Eslami:2000tj} 
  P.~Eslami, W.~J.~Zakrzewski and M.~Sarbishaei,
  ``Baby Skyrme models for a class of potentials,''
  Nonlinearity {\bf 13}, 1867 (2000)
  [hep-th/0001153].

\bibitem{baby-Skyrme-lattice}
  O.~Schwindt and N.~R.~Walet,
  ``Towards a phase diagram of the 2-D Skyrme model,''
  Europhys.\ Lett.\  {\bf 55}, 633 (2001)
  [hep-ph/0104229]; 
  R.~S.~Ward,
  ``Planar skyrmions at high and low density,''
  Nonlinearity 17, 1033 (2004) [hep-th/0307036];
  I.~Hen and M.~Karliner,
  ``Hexagonal structure of baby Skyrmion lattices,''
  Phys.\ Rev.\ D {\bf 77}, 054009 (2008)
  [arXiv:0711.2387 [hep-th]].

\bibitem{Kobayashi:2013ju} 
  M.~Kobayashi and M.~Nitta,
  ``Jewels on a wall ring,''
  Phys.\ Rev.\ D {\bf 87}, 085003 (2013)
  [arXiv:1302.0989 [hep-th]].

\bibitem{Jaykka:2010bq} 
  J.~Jaykka and M.~Speight,
  ``Easy plane baby skyrmions,''
  Phys.\ Rev.\ D {\bf 82}, 125030 (2010)
  [arXiv:1010.2217 [hep-th]].

\bibitem{Kobayashi:2013aja} 
  M.~Kobayashi and M.~Nitta,
  ``Fractional vortex molecules and vortex polygons in a baby Skyrme model,''
  Phys.\ Rev.\ D {\bf 87}, 125013 (2013)
  [arXiv:1307.0242 [hep-th]]; 
  M.~Kobayashi and M.~Nitta,
  ``Vortex polygons and their stabilities in Bose-Einstein condensates and field theory,''
  J.\ Low Temp.\ Phys. {\bf 175}, 208-215 (2014) 
  [arXiv:1307.1345 [cond-mat.quant-gas]].

\bibitem{Jaykka:2011ic} 
  J.~Jaykka, M.~Speight and P.~Sutcliffe,
  ``Broken Baby Skyrmions,''
  Proc.\ Roy.\ Soc.\ Lond.\ A {\bf 468}, 1085 (2012)
  [arXiv:1106.1125 [hep-th]];
  T.~Delsate, M.~Hayasaka and N.~Sawado,
  ``Non-axisymmetric baby-skyrmion branes,''
  Phys.\ Rev.\ D {\bf 86}, 125009 (2012)
  [arXiv:1208.6341 [hep-th]];
  P.~Jennings and T.~Winyard,
  ``Broken planar Skyrmions - statics and dynamics,''
  JHEP {\bf 1401}, 122 (2014)
  [arXiv:1306.5935 [hep-th]];
  N.~Sawado and Y.~Tamaki,
  ``Integrable, molecular-type solutions of the extended Skyrme-Faddeev model,''
  arXiv:1309.6004 [hep-th];

\bibitem{Gudnason:2013qba} 
  S.~B.~Gudnason and M.~Nitta,
  ``Baryonic sphere: a spherical domain wall carrying baryon number,''
  Phys.\ Rev.\ D {\bf 89}, 025012 (2014)
  [arXiv:1311.4454 [hep-th]].

\end{thebibliography}
\end{document}